\documentclass[11pt,preprint]{article}
\pdfoutput=1
\usepackage{jheppub}
\usepackage[usenames,dvipsnames]{xcolor}
\usepackage{mathrsfs}

\usepackage{tikz}
\usetikzlibrary{decorations.pathmorphing,decorations.pathreplacing,calligraphy,calc,snakes,cd,external}

\usepackage[export]{adjustbox}
\usepackage{hyperref}
\usepackage{bbold}
\usepackage{mathtools}
\usepackage{empheq}
\usepackage[most]{tcolorbox}

\hypersetup{
    pdfencoding=unicode,
	colorlinks=true,
	urlcolor=Maroon,
	linkcolor=RoyalBlue,
	citecolor=Maroon,
	pdftitle={What can be measured asymptotically?},
	pdfauthor={Simon Caron-Huot, Mathieu Giroux, Holmfridur S. Hannesdottir, Sebastian Mizera},
	pdfdisplaydoctitle=true,
	pdfstartview=FitH,
	linktocpage=true
}

\def\>{\rangle}
\def\<{\langle}
\def\vacR{|0\>}
\def\vac#1{\<0|#1|0\>}
\def\adag{a^\dagger}
\def\bdag{b^\dagger}
\def\cM{\mathcal{M}}
\def\cC{\mathcal{C}}

\def\.{{\cdot}}

\def\d{\mathrm{d}}
\def\be{\begin{equation}}
\def\ee{\end{equation}}

\def\Cut{\mathrm{Cut}}

\def\e{\mathrm{e}}

\def\eps{\varepsilon}

\def\ot{\leftarrow}
\def\gammaE{\gamma_{\text{E}}}
\usepackage{cancel}
\usepackage{nccmath}

\def\GeikA{G^{{\rm eik},A}}

\usetikzlibrary{decorations.markings}
\usetikzlibrary{decorations.pathmorphing}
\tikzset{->-/.style={decoration={
  markings,
  mark=at position #1 with {\arrow{>}}},postaction={decorate}}}
\tikzset{-latex-/.style={decoration={
  markings,
  mark=at position #1 with {\arrow{latex}}},postaction={decorate}}}

\usetikzlibrary{shapes.misc}
\tikzset{cross/.style={cross out, draw=black, very thick, minimum size=2*(#1-\pgflinewidth), inner sep=0pt, outer sep=0pt},
cross/.default={1pt}}

\tikzset{
    vector/.style={
        decoration={snake, aspect=0.75, mirror, segment length=2mm},
        decorate
    }
}
  
\def\D{\mathrm{D}}
\def\rd{\mathrm{d}}

\def\GL{\mathrm{GL}}

\def\leq{\leqslant}
\def\geq{\geqslant}

\def\cut{\mathrm{Cut}}
\def\bary{\bar{y}}
\def\barv{\bar{v}}
\def\barp{\bar{p}}
\def\barm{\bar{m}}
\def\barw{\bar{w}}
\def\Exp{\mathrm{Exp}}
\newcommand{\dbar}{\mathrm{d}\hspace*{-0.08em}\bar{}\hspace*{0.1em}}
\usepackage{bm}
\def\bdelta{\bm{\delta}}
\def\Dden{\mathcal{D}}
\def\cC{\mathcal{C}}
\def\cO{\mathcal{O}}
\def\cR{\mathcal{R}}
\def\cT{\mathcal{T}}
\def\cTbar{\overline{\mathcal{T}}}
\def\rone{\mathrm{I}} 
\def\rtwo{\mathrm{II}}
\def\rthree{\mathrm{III}}
\def\rdiff{{(\rone-\rtwo)}}
\def\ravg{\mathrm{avg}}
\def\GN{G}

\DeclareMathOperator*{\sumint}{%
\mathchoice%
  {\ooalign{$\displaystyle\sum$\cr\hidewidth$\displaystyle\int$\hidewidth\cr}}
  {\ooalign{\raisebox{.14\height}{\scalebox{.7}{$\textstyle\sum$}}\cr\hidewidth$\textstyle\int$\hidewidth\cr}}
  {\ooalign{\raisebox{.2\height}{\scalebox{.6}{$\scriptstyle\sum$}}\cr$\scriptstyle\int$\cr}}
  {\ooalign{\raisebox{.2\height}{\scalebox{.6}{$\scriptstyle\sum$}}\cr$\scriptstyle\int$\cr}}
}

\definecolor{darkgreen}{rgb}{0.0, 0.4, 0.0}
\definecolor{darkmagenta}{rgb}{0.55, 0.0, 0.55}

\usepackage[titles]{tocloft}
\usepackage{dirtytalk}
\usepackage{changepage}
\usepackage{subcaption}
\usepackage{float}
\usepackage{enumitem}  
\usepackage{stackengine}
\usepackage{scalerel}

\newcommand{\Lim}[1]{\raisebox{0.5ex}{\scalebox{0.8}{$\displaystyle \lim_{#1}\;$}}}

 \DeclarePairedDelimiterXPP\EV[1]{E}(){}{
 
 #1}
 \DeclarePairedDelimiterXPP\Var[1]{V}(){}{
 
 #1}

\DeclareRobustCommand
\Compactcdots{\mathinner{\cdotp\mkern-2mu\cdotp\mkern-2mu\cdotp}}

 \tikzset{gradGtoB/.style={
    postaction={
        decorate,
        decoration={
            markings,
            mark=at position \pgfdecoratedpathlength-0.5pt with {\arrow[ForestGreen,line width=#1] {}; },
            mark=between positions 0 and \pgfdecoratedpathlength-0pt step 0.5pt with {
                \pgfmathsetmacro\myval{multiply(divide(
                    \pgfkeysvalueof{/pgf/decoration/mark info/distance from start}, \pgfdecoratedpathlength),100)};
                \pgfsetfillcolor{RoyalBlue!\myval!ForestGreen!};
                \pgfpathcircle{\pgfpointorigin}{#1};
                \pgfusepath{fill};}
}}}}
\tikzset{gradRtoG/.style={
    postaction={
        decorate,
        decoration={
            markings,
            mark=at position \pgfdecoratedpathlength-0.5pt with {\arrow[Maroon,line width=#1] {}; },
            mark=between positions 0 and \pgfdecoratedpathlength-0pt step 0.5pt with {
                \pgfmathsetmacro\myval{multiply(divide(
                    \pgfkeysvalueof{/pgf/decoration/mark info/distance from start}, \pgfdecoratedpathlength),100)};
                \pgfsetfillcolor{ForestGreen!\myval!Maroon!};
                \pgfpathcircle{\pgfpointorigin}{#1};
                \pgfusepath{fill};}
}}}}
\tikzset{gradBlackToR/.style={
    postaction={
        decorate,
        decoration={
            markings,
            mark=at position \pgfdecoratedpathlength-0.5pt with {\arrow[black,line width=#1] {}; },
            mark=between positions 0 and \pgfdecoratedpathlength-0pt step 0.5pt with {
                \pgfmathsetmacro\myval{multiply(divide(
                    \pgfkeysvalueof{/pgf/decoration/mark info/distance from start}, \pgfdecoratedpathlength),100)};
                \pgfsetfillcolor{Maroon!\myval!black!};
                \pgfpathcircle{\pgfpointorigin}{#1};
                \pgfusepath{fill};}
}}}}

\tikzset{gradBtoBlack/.style={
    postaction={
        decorate,
        decoration={
            markings,
            mark=at position \pgfdecoratedpathlength-0.5pt with {\arrow[blue,line width=#1] {}; },
            mark=between positions 0 and \pgfdecoratedpathlength-0pt step 0.5pt with {
                \pgfmathsetmacro\myval{multiply(divide(
                    \pgfkeysvalueof{/pgf/decoration/mark info/distance from start}, \pgfdecoratedpathlength),100)};
                \pgfsetfillcolor{black!\myval!RoyalBlue!};
                \pgfpathcircle{\pgfpointorigin}{#1};
                \pgfusepath{fill};}
}}}}

\tikzset{gradRtoB/.style={
    postaction={
        decorate,
        decoration={
            markings,
            mark=at position \pgfdecoratedpathlength-0.5pt with {\arrow[RoyalBlue,line width=#1] {}; },
            mark=between positions 0 and \pgfdecoratedpathlength-0pt step 0.5pt with {
                \pgfmathsetmacro\myval{multiply(divide(
                    \pgfkeysvalueof{/pgf/decoration/mark info/distance from start}, \pgfdecoratedpathlength),100)};
                \pgfsetfillcolor{RoyalBlue!\myval!Maroon!};
                \pgfpathcircle{\pgfpointorigin}{#1};
                \pgfusepath{fill};}
}}}}

\newcommand{\tikzxmark}{%
\tikz[scale=5.23] {
    \draw[line width=0.7,line cap=round] (0,0) to [bend left=6] (1,1);
    \draw[line width=0.7,line cap=round] (0.2,0.95) to [bend right=3] (0.8,0.05);
}}

\definecolor{darkorange}{HTML}{E28413}
\definecolor{choral}{HTML}{E09891}
\definecolor{darkred}{HTML}{6B0F1A}
\tikzset{
    partial ellipse/.style args={#1:#2:#3}{
        insert path={+ (#1:#3) arc (#1:#2:#3)}
    }
}
\def\centerarc[#1](#2)(#3:#4:#5)%
    { \draw[#1] ($(#2)+({#5*cos(#3)},{#5*sin(#3)})$) arc (#3:#4:#5); }
\tikzset{
    vector/.style={
        decoration={snake, aspect=0.75, mirror, segment length=2mm},
        decorate
    },
	photon/.style={decorate, decoration={snake, amplitude=1pt, segment length=6pt}
	}
}

\newcommand{\cloud}{\tikzset{every picture/.style={line width=0.75pt}}  
\begin{tikzpicture}[x=0.75pt,y=0.75pt,yscale=-0.5,xscale=0.5]
\draw[color=charcoal,fill=white  ,fill opacity=1 ]   (107.28,126.75) .. controls (106.64,121.92) and (108.75,117.13) .. (112.73,114.42) .. controls (116.7,111.71) and (121.85,111.56) .. (125.97,114.03) .. controls (127.43,111.21) and (130.1,109.27) .. (133.18,108.79) .. controls (136.26,108.3) and (139.38,109.33) .. (141.6,111.57) .. controls (142.84,109.02) and (145.29,107.31) .. (148.06,107.04) .. controls (150.84,106.77) and (153.55,107.98) .. (155.24,110.25) .. controls (157.49,107.55) and (161.07,106.41) .. (164.42,107.33) .. controls (167.78,108.25) and (170.32,111.06) .. (170.93,114.55) .. controls (173.69,115.31) and (175.98,117.27) .. (177.22,119.9) .. controls (178.47,122.53) and (178.53,125.58) .. (177.41,128.27) .. controls (180.12,131.88) and (180.75,136.68) .. (179.07,140.89) .. controls (177.39,145.11) and (173.65,148.09) .. (169.24,148.74) .. controls (169.21,152.69) and (167.09,156.32) .. (163.7,158.22) .. controls (160.3,160.12) and (156.16,160.01) .. (152.88,157.91) .. controls (151.48,162.65) and (147.54,166.13) .. (142.76,166.86) .. controls (137.99,167.59) and (133.23,165.43) .. (130.55,161.32) .. controls (127.26,163.34) and (123.31,163.93) .. (119.59,162.94) .. controls (115.88,161.95) and (112.71,159.46) .. (110.8,156.05) .. controls (107.44,156.45) and (104.19,154.67) .. (102.67,151.59) .. controls (101.14,148.51) and (101.66,144.79) .. (103.98,142.27) .. controls (100.98,140.47) and (99.45,136.89) .. (100.18,133.4) .. controls (100.92,129.92) and (103.76,127.31) .. (107.21,126.94) ; \draw[color=charcoal]   (103.98,142.27) .. controls (105.39,143.13) and (107.03,143.51) .. (108.66,143.38)(110.8,156.05) .. controls (111.51,155.96) and (112.2,155.79) .. (112.85,155.52)(130.55,161.32) .. controls (130.05,160.56) and (129.64,159.75) .. (129.31,158.9)(152.88,157.91) .. controls (153.13,157.05) and (153.3,156.16) .. (153.37,155.26)(169.24,148.74) .. controls (169.28,144.53) and (166.94,140.67) .. (163.23,138.83)(177.41,128.27) .. controls (176.81,129.7) and (175.89,130.97) .. (174.73,131.98)(170.93,114.55) .. controls (171.04,115.12) and (171.08,115.71) .. (171.07,116.3)(155.24,110.25) .. controls (154.68,110.92) and (154.22,111.68) .. (153.87,112.49)(141.6,111.57) .. controls (141.3,112.18) and (141.08,112.83) .. (140.93,113.5)(125.97,114.03) .. controls (126.84,114.55) and (127.65,115.18) .. (128.37,115.9)(107.28,126.75) .. controls (107.37,127.42) and (107.51,128.08) .. (107.7,128.72) ;
\end{tikzpicture}}

\definecolor{charcoal}{HTML}{343837}
\definecolor{yellowish}{rgb}{0.880722,0.611041,0.142051}

\title{What can be measured asymptotically?}
\usepackage{orcidlink}
\author[\!a,b,\orcidlink{0000-0002-7005-9652}]{Simon Caron-Huot,}\emailAdd{schuot@physics.mcgill.ca}
\author[\!a,\orcidlink{0000-0002-2672-634X}]{Mathieu Giroux,}\emailAdd{mathieu.giroux2@mail.mcgill.ca}
\author[\!b,\orcidlink{0000-0002-5440-2086}]{Holmfridur S. Hannesdottir,}\emailAdd{hofie@ias.edu}
\author[\,b,\orcidlink{0000-0002-8066-5891}]{\\ Sebastian Mizera}\emailAdd{smizera@ias.edu}

\affiliation{$^a$Department of Physics, McGill University, 3600 Rue University, Montr\'eal, H3A 2T8, QC Canada}
\affiliation{$^b$Institute for Advanced Study, Einstein Drive, Princeton, NJ 08540, USA}

\abstract{We consider asymptotic observables in quantum field theories in which the S-matrix makes sense.
We argue that in addition to scattering amplitudes, a whole compendium of inclusive observables exists where the time-ordering is relaxed. These include expectation values of electromagnetic or gravitational radiation fields as well as out-of-time-order amplitudes.
We explain how to calculate them in two ways: by relating them to amplitudes and products of amplitudes, and by using a generalization of the LSZ reduction formula. As an application, we discuss one-loop master integrals contributing to gravitational radiation in the post-Minkowski expansion, emphasizing the role of classical cut contributions and highlighting the different infrared physics of in-in observables.
}

\begin{document} 

\maketitle

\setcounter{page}{2}

\newpage
\section{Introduction}

Scattering amplitudes describe the quantum mechanical evolution
between ``in'' and ``out'' states of asymptotically non-interacting particles.
Amplitudes are closely related to measurable quantities, such as
differential cross-sections measured at colliders
that can be computed by squaring amplitudes,
so much so that amplitudes are often themselves called ``observable.''
The aim of this paper is to explore more general amplitude-like quantities
that describe measurements done in the asymptotic past and future.

The distinction between amplitudes and cross-sections is in fact not entirely sharp,
as made explicit by the optical theorem.
To be concrete, the total cross-section for proton-proton scattering through the strong interaction
for example
is given by the imaginary part of the proton-proton elastic ($2\to 2$) amplitude in the forward limit:
\be\label{eq:ppCrossSection}
 \sigma^{\rm tot}_{pp}(s) = \frac{1}{\sqrt{s(s-4m_p^2)}} {\rm Im}\,\cM_{pp\to pp}(s,t=0)\,.
\ee
On the other hand, the total cross-section could be computed by summing the squared amplitudes for two protons to evolve into any possible final states $X$:
\be
 \sigma^{\rm tot}_{pp}(s) \propto \sumint_X \cM_{pp\to X}\cM^\dag_{X\to pp}\,.  \label{unitarity_intro_2}
\ee
This relation suggests that many physical observables thought of as quadratic in amplitudes
can also be viewed as linear in some kind of generalized amplitudes.
Indeed, such a device is often used in perturbative cross-section calculations in a number of
quantum field theory or particle physics textbooks, wherein one draws ``doubled'' diagrams that combine the amplitudes and complex-conjugate amplitudes on the two sides of a cut.

Another interesting class of asymptotic observables includes (momentum-space) radiation waveforms.
Waveforms answer questions of the following general type:
what is the expectation value of some field, like the electromagnetic field, after the collision of two particles?  According to the Kosower-Maybee-O'Connell (KMOC) formalism \cite{Kosower:2018adc}, they can be
expressed in terms of creation-annihilation operators as \cite{Cristofoli:2021vyo}:
\be
{}_{\rm in}\<1'2'| \, a_3^{\rm out}\,|12\>_{\rm in}=
\<0|\,a_{1'}^{\rm in} a_{2'}^{\rm in}\, a_3^{\rm out}\,\adag_1{}^{\rm in} \adag_2{}^{\rm in}\,|0\>\, ,
\label{KMOC_intro}
\ee
where $a_3$ absorbs a photon with some on-shell momentum $p_3$.
Physically, the photon is on-shell because one is interested in the radiation field far in the future of the collision. The observable \eqref{KMOC_intro} differs from a conventional scattering amplitude only in that the bra and ket are both ``in'' states prepared in the past (and eventually to be integrated against a known initial wavefunction and its complex conjugate, omitted here).
It makes physical sense that the collided particles do not appear in any ``out'' operator
since the question does not specify what happens to them after the collision, nor even if they survive it.
The evaluation of \eqref{KMOC_intro}, as reviewed below, features the familiar $2\to 3$ amplitudes as well as additional ``cut'' terms.

In some simple situations, such as classical or non-relativistic collisions of charged particles, tree-level waveforms reproduce familiar results from classical field theory such as the far-field limit of retarded (Li\'enard-Wiechert) potentials and its corresponding radiated power \cite{Cristofoli:2021vyo}.
In a gravitational context, \eqref{KMOC_intro} is very interesting due to its relation to the
gravitational waveform as measured in a classical detector such as LIGO \cite{LIGOScientific:2016aoc,LIGOScientific:2017vwq,LIGOScientific:2017ync}.
We stress that waveforms are linear in the observed field:
they are examples of physical observables that are fundamentally not quadratic in amplitudes.

The main idea of this paper is that \eqref{unitarity_intro_2} and \eqref{KMOC_intro} are simply two examples of non time-ordered scattering amplitudes.
Thinking of in and out $a$ and $\adag$ operators as measurements done
in the asymptotic past and future, we would like to understand all possible ways to order and combine them.
What other options are possible, what do they mean physically, and how to compute them?
For example, what generalization of the LSZ reduction formula relates out-of-time-order correlation functions to out-of-time-order amplitudes?
How can the techniques used to compute time-ordered amplitudes be adapted to these new objects?

In the context of conformal field theory and the AdS/CFT correspondence,
out-of-time-ordered correlators (OTOCs) have found a number of applications.
For example, four-point OTOCs providing a diagnosis of chaos \cite{Shenker:2013pqa, Maldacena:2015waa}.
Four-point correlators with a distinct ordering (also not time-ordered)
can also describe ``radar-like'' experiments,
wherein one launches a test particle and bounces photons off it to track its trajectory,
which can be used to probe the local geometry of the bulk spacetime \cite{Caron-Huot:2022lff}.
We anticipate that out-of-time-order scattering amplitudes could be physically interesting in a similar way.

\begin{figure}[t]
    \centering
    \begin{tikzpicture}[line width=1,draw=charcoal, scale=0.5]
  \draw[->] (-6.5, 0) -- (6.5, 0);
  \draw[->] (0, -0.3) -- (0, 4.5);
  \node[] at (6,4.0) {$s$};
  \draw[shift={(6,4)},scale=0.7] (-0.5,0.5) -- (-0.5,-0.5) -- (0.5,-0.5);
  \centerarc[<->,Maroon,thick](0,-1.5)(35:145:5);
\node[] at (-4.5,0.7) {$\cM_{1\bar{3} \to \bar{2}4}^\dag$};
\node[] at (4.5,0.7) {$\cM_{12 \to 34}$};
\end{tikzpicture}
\qquad
\begin{tikzpicture}[line width=1,draw=charcoal, scale=0.5]
  \draw[->] (-6.5, 0) -- (6.5, 0);
  \draw[->] (0, -0.3) -- (0, 4.5);
  \centerarc[<->,Maroon,thick](0,-1.5)(35:145:5);
\node[] at (-4.5,0.7) {$?$};
\node[] at (4.5,0.7) {$\cM_{12 \to 345}$};
\end{tikzpicture}
    \caption{Asymptotic observables can be connected through new version of crossing symmetry, made precise by analytic continuation. \textbf{Left:} Well-established relation between the exclusive amplitude $\cM_{12\to34}$ and its crossed process $\cM_{1\bar{3} \to \bar{2}4}^\dag$. \textbf{Right:} What is the result of the analytic continuation of $\cM_{12\to 345}$?}
    \label{fig:introcross}
\end{figure}
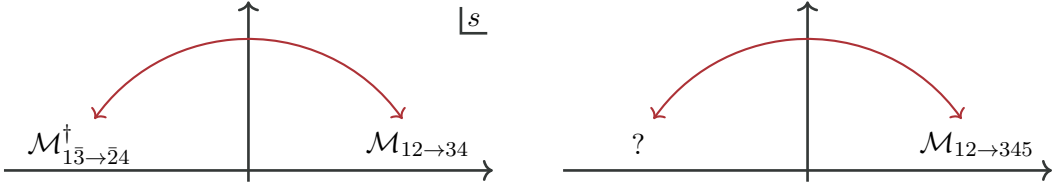

Another motivation to study arbitrary-ordered amplitudes is to better understand microcausality
at the level of the $S$-matrix. This is the statement that operators spacelike-commute:
\be
 [A^\dag(x),A(y)]=0\quad\mbox{for $x{-}y$ spacelike}\,. \label{microcausality}
\ee
At a most basic level, the asymptotic limit of this relation ought to connect amplitudes with different operator orderings.  Concretely, just like consistency of \eqref{microcausality} in the vacuum is known to imply that for each particle there exists an antiparticle with the opposite quantum numbers, \eqref{microcausality} should imply crossing relations for scattering amplitudes.
For $2\to 2$ scattering, this is the well-known analytic continuation in the upper-half $s$-plane that relates the amplitude for the process $12 \to 34$ to the complex conjugate of the amplitude for
$1\bar{3} \to \bar{2}4$ \cite{Bros:1965kbd}, as depicted in Fig.~\ref{fig:introcross}.
The \emph{anti-time-ordered} amplitude $\cM^\dag$ appears because the continuation lands on the ``wrong'' side of the cut.  In a forthcoming paper, we will propose that a certain analytic continuation of the five-point conjugated scattering amplitude $\cM^\dag_{12\to 345}$ yields precisely the inclusive expectation value \eqref{KMOC_intro} \cite{crossing}, see Fig.~\ref{fig:introcross}. A satisfactory account of the analyticity properties of amplitudes seems to require us to consider non-standard orderings.

We would like to stress that the generalized amplitudes we will consider, and their calculation using various alternative $i\varepsilon$ prescriptions, do not constitute in any way a modification of the theory.
This idea is conceptually similar to the use of \emph{reverse unitarity} to relate the phase-space integrals in inclusive cross sections to ordinary Feynman integrals on different contours \cite{Anastasiou:2002yz}.
All our asymptotic observables will be ultimately equivalent to products of the standard $S$ and $S^\dag$ matrix elements summed over unobserved intermediate states.
We find it natural to look for more intrinsic ways to define and calculate them.

\paragraph{Outline} This paper is organized as follows. In Sec.~\ref{sec:asympt_measurements}, we introduce the algebra of asymptotic creation and annihilation operators and use it to give an account of various asymptotic observables that can be constructed out of them.  A blob notation is described in Sec.~\ref{sec:blobology}, where the importance of disconnected terms is emphasized. In Sec.~\ref{sec:localFields}, we generalize the LSZ reduction formulae to out-of-time-ordered observables and explain their relation to Schwinger-Keldysh contours. This technology is put to practical use in Sec.~\ref{sec:waveforms}, where we compute all master integrals needed for the one-loop computation of gravitational waveforms (with details of the differential equation relegated to App.~\ref{sec:eikonalDiffEq}), emphasizing in particular which terms survive in the classical limit. Finally, we conclude with a list of open questions in Sec.~\ref{sec:conclusions}.
\def\ot{{\leftarrow}}

\section{Asymptotic measurements}
\label{sec:asympt_measurements}

In this section, we review the rules of the game in S-matrix theory in a flat spacetime: axioms that abstract the properties of measurements made from asymptotically large distances.  We will find that scattering amplitudes are just one out of many interesting physical observables.
Our setup can be applied to any theory in which an S-matrix can be computed, e.g., pion scattering in four dimensions, perturbative QCD in
dimensional regularization or $\mathcal{N}=4$ super Yang-Mills.

\subsection{S-matrix theory and foundational axioms}

The basic assumption is that any finite energy excitation decays at late time to a finite set of stable particles, which separate from each other and effectively become free.
We assume the following set of axioms:
\begin{enumerate}[label=(\roman*)]
\item The algebra of asymptotic measurements in the \emph{far past} is generated by \emph{creation} and \emph{annihilation operators} of stable particles.
They satisfy the canonical relation
\begin{equation} \label{canonical}
 [a_1,a^\dagger_{2}] =
\delta_{1,2}\ 2p_1^{0}\ (2\pi)^{\D-1} \delta^{\D-1}(\vec{p}_1-\vec{p}_2)\, ,
\end{equation}
where $p_i^{0}$ is the (positive) energy of the $i^{\text{th}}$ particle and
$\delta_{i,j}$ is a Kronecker delta on flavor and spin indices.%
\footnote{
To avoid clutter, we treat all
particles as bosons and lump momentum, spin and flavor into a single subscript $i$. With fermions, one should replace certain commutators by anticommutators where appropriate.
}

\item[(i')] There is an equivalent algebra of ``out'' measurements in the \emph{far future}, which we denote with $b$ and $\bdag$'s.
\item These operators act on equivalent Hilbert spaces and are related by a unitary \emph{evolution operator} $S$:
\begin{equation} \label{b from a}
    b_i = S^\dagger a_i S, \qquad
    b_i^\dagger = S^\dagger a_i^\dagger S\,.
\end{equation}
\item There exists a \emph{time-invariant vacuum} which does not contain particles. That is,
\begin{equation} \label{Svac}
    a_i\vacR=b_i\vacR=0,\qquad S\vacR=\vacR\,.
\end{equation}
\item One-particle states evolve trivially. In other words, they are \emph{stable}
\begin{equation}
    \bdag_i \vacR = S^\dagger\adag_i \vacR = \adag_i \vacR\,. \label{stability}
\end{equation}
In particular, this implies $S a_i^\dag \vacR = a_i^\dag \vacR$ thanks to the unitarity condition $S^\dagger S=\mathbb{1}$.
\end{enumerate}
We emphasize that it would be hard to picture a flat spacetime scattering theory that does not include some version of these principles.
Throughout this paper, we will additionally assume Poincar\'e invariance.
We also stress that all the above assertions apply for real momenta $p_i$. Typically, one would like to supplement them with analyticity properties at complex $p_i$, which requires some notion of causality. We will study this in detail in \cite{crossing}.

While our considerations could conceivably be generalized in various ways
that are excluded by the above axioms, for example to deal with long-range topological order\footnote{
Physical setups where asymptotic states carry nontrivial topology include anyons in three-dimensional Chern-Simons theories or dyons in four-dimensional U(1) gauge theories, see \cite{Jain:2014nza,Mehta:2022lgq,Csaki:2022tvb} and references therein.} or nontrivial infrared dynamics, we will focus in this paper on the basic historical setup where the infrared dynamics is trivial.

Translated to textbook conventions, the above creation and annihilation operators correspond to
\be
a_i \equiv a_i^{\mathrm{in}} \quad \text{and} \quad b_i \equiv a_i^{\mathrm{out}}\, .
\ee
Following common nomenclature,
we call \say{\emph{in}} and \say{\emph{out}} states those prepared using exclusively operations in the past or future, namely
\begin{equation}
    |m\cdots 1\>_{\rm in} \equiv  
      \adag_m \cdots \adag_1 |0\> \quad \text{and} \quad 
    |m\cdots 1\>_{\rm out} \equiv  
      \bdag_m \cdots \bdag_1 |0\>\,.
\end{equation}
The overlap between \say{in} and \say{out} states is given by matrix elements of $S$. That is,
\begin{equation}\begin{aligned}\label{S from ab}
     {}_{\rm out}\< m{+}n \cdots m{+}1 |m \cdots 1\>_{\rm in} &=
     \vac{b_{m+n}\cdots b_{m+1}\
     \adag_m \cdots \adag_1}
\\    &= \vac{a_{m+n}\cdots a_{m+1}\ S\
     \adag_{m} \cdots \adag_{1}}
\\ &\equiv
     \< m{+}n \cdots m{+}1| S |m \cdots 1 \>\, ,
\end{aligned}\end{equation}
where we have used both rules \eqref{b from a} and \eqref{Svac}.
In equations like the last line, a state $|\ldots\>$ without a subscript is conventionally
taken to be an ``in'' state by default.

\subsection{A compendium of observables} \label{ssec:asympt}

Let us now consider an arbitrary measurement involving asymptotic operators in the past and/or future.  The $m\to n$ amplitude in \eqref{S from ab} is one natural instance. We are keen to ask: \emph{are there other options}?

\paragraph{Four particles}
Let us first answer this question in the case of four particles, where we can consider an arbitrary product of four $a$, $\adag$, $b$, or $\bdag$'s.  A priori, there are $4^4=256$ options for such products. However, many trivially vanish due to \eqref{Svac} and its Hermitian conjugate.  Many others are trivially reducible by virtue of
\begin{equation}
  a_2 \adag_1\vacR
  =[a_2,\adag_1]\ \vacR \propto \vacR \quad \text{and} \quad
  b_2 \adag_1 \vacR
  =[b_2,\bdag_1]\ \vacR\propto \vacR\, ,
\end{equation}
where in the second case we used the stability relation $\bdag_i\vacR=\adag_i\vacR$.
Since these commutators are $c$-numbers (see \eqref{canonical}), 
there is no new information to be learned from correlators that contain them.  Similar comments apply to
$a\bdag$ and $b\bdag$ products acting on the vacuum.
Thus, in any nontrivial correlator, we can assume that the rightmost two operators are creation operators.
Furthermore, using again the stability condition, we have 
\be\label{eq:ab-aa-bb}
a_1^\dag b_2^\dag \vacR = \adag_2\bdag_1\vacR =\adag_2\adag_1\vacR,
\qquad
b_1^\dag a_2^\dag \vacR = \bdag_2\adag_1\vacR = \bdag_2\bdag_1\vacR\,,
\ee
such that we can assume that both are of the same type. That is,
\begin{equation}
 \adag_2 \adag_1|0\> \quad \text{and} \quad 
 \bdag_2\bdag_1|0\>
\quad
\mbox{are the \emph{only} non-trivial two-particle states}\,.
\label{two particles}
\end{equation}
Applying the same logic to the two leftmost operators,  we conclude that any vacuum expectation value of four asymptotic operators takes one of the following four forms
\begin{subequations}\begin{gather}
\vac{b_{4}b_{3} \adag_2 \adag_1} =
\< 43|S|21\>\, , \qquad  \vac{a_{4}a_{3} \bdag_2 \bdag_1} =
\<43|S^\dagger|21\>\,,
\\
\vac{a_{4}a_{3} \adag_2 \adag_1} =
 \<43|\mathbb{1}|21\>
 =\vac{b_{4}b_{3} \bdag_2 \bdag_1}\, .
\end{gather}\end{subequations}
The cases in the bottom line give disconnected products of two-point functions.  Focusing on connected terms, this analysis confirms that there are
really just two quantities one can measure with four particles:
the amplitude and its complex conjugate.

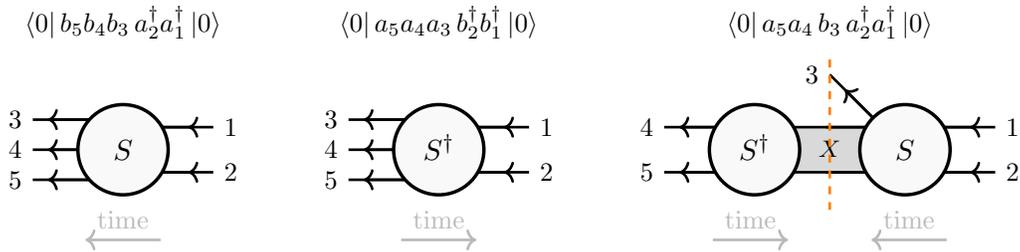
\begin{figure}[t]
\centering
\adjustbox{valign=t}{\begin{tikzpicture}[line width=1]
\draw[] (0,1.7) node {\small$\vac{\,b_5 b_4 b_3\, a_2^\dag a_1^\dag\,}$};
\draw[->, color=gray!55] (0.5,-1.2) -- (-0.5,-1.2) node[above,midway]{\small{time}};
\draw[] (1.2,0.3) node[right] {\small$1$} -- (0,0.3);
\draw[] (0,0.4) -- (-1.2,0.4) node[left] {\small$3$};
\draw[>->] (0,0.4) -- (-1,0.4);
\draw[>->] (1,0.3) -- (0,0.3);
\draw[] (0,0) node[right] {} -- (-1.2,0) node[left] {\small$4$};
\draw[] (1.2,-0.3) node[right] {\small$2$} -- (0,-0.3) node[left] {};
\draw[>->] (0,0) -- (-1,0);
\draw[>->] (1,-0.3) -- (0,-0.3);
\draw[] (0,-0.4) node[right] {} -- (-1.2,-0.4) node[left] {\small$5$};
\draw[>->] (0,-0.4) -- (-1,-0.4);
\filldraw[fill=gray!5, line width=1.3](0,0) circle (0.6) node {$S$};
\end{tikzpicture}}
\qquad
\adjustbox{valign=t}{\begin{tikzpicture}[line width=1]
\draw[] (0,1.7) node {\small$\vac{\,a_5 a_4 a_3\, b_2^\dag b_1^\dag\,}$};
\draw[<-, color=gray!55] (0.5,-1.2) -- (-0.5,-1.2) node[above,midway]{\small{time}};
\draw[] (1.2,0.3) node[right] {\small$1$} -- (0,0.3);
\draw[] (0,0.4) -- (-1.2,0.4) node[left] {\small$3$};
\draw[->] (0,0.4) -- (-1,0.4);
\draw[>-] (1,0.3) -- (0,0.3);
\draw[] (0,0) node[right] {} -- (-1.2,0) node[left] {\small$4$};
\draw[] (1.2,-0.3) node[right] {\small$2$} -- (0,-0.3) node[left] {};
\draw[->] (0,0) -- (-1,0);
\draw[>-] (1,-0.3) -- (0,-0.3);
\draw[] (0,-0.4) node[right] {} -- (-1.2,-0.4) node[left] {\small$5$};
\draw[->] (0,-0.4) -- (-1,-0.4);
\filldraw[fill=gray!5, line width=1.3](0,0) circle (0.6) node[yshift=1] {$S^\dag$};
\end{tikzpicture}}
\qquad
\adjustbox{valign=t}{\begin{tikzpicture}[line width=1]
\draw[] (1,1.7) node {\small$\vac{\,a_5a_4\, b_3\, \adag_2\adag_1\,}$};
\draw[->, color=gray!55,xshift=30pt] (1.5,-1.2) -- (0.5,-1.2) node[above,midway]{\small{time}};
\draw[<-, color=gray!55,xshift=-30pt] (1.5,-1.2) -- (0.5,-1.2) node[above,midway]{\small{time}};
\draw[] (0,0.3) -- (-1.2,0.3) node[left] {\small$4$};
\draw[->] (0,0.3) -- (-1,0.3);
\draw[] (0,-0.3) -- (-1.2,-0.3) node[left] {\small$5$};
\draw[->] (0,-0.3) -- (-1,-0.3);
\draw[] (2,0.3) -- (3.2,0.3) node[right] {\small$1$};
\draw[-<] (2,0.3) -- (3,0.3);
\draw[] (2,-0.3) -- (3.2,-0.3) node[right] {\small$2$};
\draw[-<] (2,-0.3) -- (3,-0.3);
\draw[] (2,0) -- (1,1) node[left] {\small$3$};
\draw[->] (2,0) -- (1.2,0.8);
\filldraw[fill=gray!30](0,-0.3) rectangle (2,0.3);
\draw (2,0.3) -- (0.95,0.3);
\draw (2,-0.3) -- (0.95,-0.3);
\draw[] (1,0) node {$\medmath{X}$};
\filldraw[fill=gray!5, line width=1.3](0,0) circle (0.6) node[yshift=1] {$S^\dag$};
\filldraw[fill=gray!5, line width=1.3](2,0) circle (0.6) node {$S$};
\draw[dashed,orange] (1,1.2) -- (1,-0.8);
\end{tikzpicture}}
\caption{\label{fig:5pt-examples}Example asymptotic measurements with five particles,
corresponding to the first three terms in \eqref{five point correlators}.
Placement of the external legs follows the bra-ket notation, which does not necessarily coincide with the flow of time (indicated with arrows).
The third figure derives from inserting a complete basis of intermediate states as in \eqref{inclusive 5}.
}
\end{figure}

\paragraph{Five particles} The story becomes more interesting with five asymptotic measurements.  Accounting for \eqref{two particles} on both the leftmost and rightmost pairs,
and discarding anything reducible via commutators, we find the following options
\begin{equation} \label{five point correlators}
    \vac{b_5b_4b_3 \adag_2\adag_1}\, ,\quad
    \vac{a_5a_4a_3 \bdag_2\bdag_1}\, ,\quad
    \vac{a_5a_4b_3 \adag_2\adag_1}\, ,\quad
    \vac{b_5b_4a_3 \bdag_2\bdag_1}\, ,
\end{equation}
together with their Hermitian conjugates.
It will be convenient to refer to them with shorthands:
\begin{equation}\label{eq:5pt-measurements}
  S_{543\ot 21}\, ,\quad S^\dagger_{543\ot 21}\, ,\quad
  \text{Exp}_3\equiv {}_{\rm in}\<54|b_3|21\>_{\rm in}\, , \quad
  {}_{\rm out}\<54|a_3|21\>_{\rm out}\, .
\end{equation}
The last two are new objects:
\emph{inclusive amplitudes} or \emph{expectation values}. We have labeled the S-matrix consistently with the operator ordering.
Using \eqref{Svac} to eliminate $b$'s
we can relate them to conventional amplitudes. For example,
\begin{equation}
    \text{Exp}_3 =  \vac{a_5a_4S^\dagger a_3 S\adag_2\adag_1}
= \sumint_X \<54|S^\dagger |X\> \<X3|S|21\>\,, \label{inclusive 5}
\end{equation}
which is shown on the third panel of Fig.~\ref{fig:5pt-examples}.
We call those ``inclusive'' because of the infinite sum over unobserved final states $X$: all details of the final state are marginalized over, except for what is measured by $b_3$.

A natural physical instance of \eqref{inclusive 5} occurs when $b_3$ absorbs a quantum of a classical field,
as mentioned in the introduction.  For example, one could collide electrically charged particles and measure with $b_3$ the expectation value of the electromagnetic field at future (null) infinity after the collision.
In this context one would typically integrate the state $|12\rangle_{\mathrm{in}}$ against some wavefunction,
and integrate ${}_{\mathrm{in}}\<45|$ against the complex conjugate of the \emph{same} wavefunction.
Note that they are both \say{in} states because we are computing an \emph{expectation value} $\Exp_{3}$ in a definite state prepared before the collision.  Thus this measures the electromagnetic field after the collision irrespective of other properties of the outcome. Similarly, if $b_3$ is a graviton operator, this yields the gravitational waveform as measured by a classical detector such as LIGO \cite{Kosower:2018adc,Cristofoli:2021vyo}, which will serve as a concrete example below in Sec.~\ref{sec:waveforms}. This physical interpretation justifies our notation choice, $\text{Exp}_3$, in \eqref{eq:5pt-measurements}.

Note that in this paper we adopt the (unusual!) convention where \emph{time flows from right to left}. This is to streamline the passage from bra-ket formulas to pictures.
Pictures as in Fig.~\ref{fig:5pt-examples} can always derived simply by eliminating all $b$'s using the identity $b_i= S^\dagger a_i S$, then inserting complete bases of states between $S$ factors, similarly to \eqref{inclusive 5}.  Their relations to connected matrix elements $\cM$ are discussed in Sec.~\ref{sec:blobology}.

It is worth noting that \eqref{inclusive 5} does \emph{not} measure the number of radiated photons/gravitons, nor the radiated power: these would be quadratic in the field.
Rather, it is linear in the field and is a complex number with a physically meaningful phase. Notably, through the Fourier transform,
the frequency dependence of the phase determines the waveform shape in the time domain (see \cite{Cristofoli:2021vyo}).
In contrast, the radiated energy $\propto b_3^\dagger b_3$ in a two-body collision would be
an example of a 6-point (generalized) amplitude.

\paragraph{Six particles} Let us now describe possible asymptotic measurements involving six on-shell momenta,
besides the familiar $2\to 4$ and $3\to 3$ exclusive amplitudes.
One is an inclusive measurement of the sort just mentioned, namely
\begin{equation}
 \vac{\,a_6a_5\, \bdag_4 b_3\, \adag_2\adag_1\,} =
 \sumint_X \< 65|S^\dagger|4X\> \<X3|S|21\>\,. \label{6pt N}
\end{equation}
In the multiple forward limit ($6,5,4\to 1,2,3$), this reduces to
the expectation value of the number of particles of type $3$ produced in a collision. In other words, to the inclusive differential cross-section for the process $12\to 3X$ where $X$ stands for an arbitrary set of unobserved particles:
\begin{equation}
    \frac{\text{d}\sigma}{\text{d}p_3} \propto \sumint_X \big| \mathcal{M}_{12\to 3X}|^2\,.
\end{equation}
Simple instances include the fully inclusive distribution of $Z$ bosons created at LEP or of Higgs bosons created at LHC (see \cite{Anastasiou:2015vya,Anastasiou:2016cez,Dreyer:2016oyx,Chen:2021isd} for recent examples, some of which use the method of reverse unitarity mentioned above).

\begin{figure}
    \centering
    \begin{equation*}
    {\adjustbox{valign=c}{\input{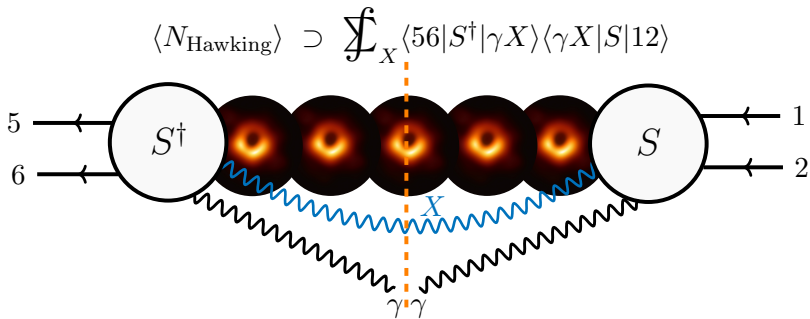}}}
    \end{equation*}
    \caption{The inclusive observable \eqref{6pt N} probing Hawking radiation from a black hole produced by a two-body collision. For supermassive black holes (like M87 above \cite{EventHorizonTelescope:2019dse}), in addition to the (very cold) Hawking radiation, the inclusive spectrum defined by this observable 
    contains extra radiation from, e.g., an accretion disk.}
    \label{fig:HawkingRadiation0}
\end{figure}

In a gravitational context, one could also imagine that particles $1$ and $2$ collide and form a black hole,
and $\bdag_4 b_3$ measures the inclusive spectrum of Hawking radiation (see Fig.~\ref{fig:HawkingRadiation0}). Note that non-forward measurements with $p_4\neq p_3$ additionally retain information about the arrival time of this radiation, so the 6-point ``amplitude'' \eqref{6pt N} includes knowledge of the late-time radiation spectrum. Yet another application of \eqref{6pt N} to momentum impulse computations will be discussed in Sec.~\ref{sec:impulse}.

Another 6-point observable is a one-point function between two- and three-particle states
\begin{equation}
 \vac{\, a_6a_5\, b_4\, \adag_3\adag_2\adag_1\,} =
 {}_{\rm in}\<65|b_4|321\>_{\rm in} \,. \label{6pt radar}
\end{equation}
This can be interpreted in different ways.
For example, it could capture an interference term for the one-point function of $b_4$ similar to \eqref{inclusive 5}, but between two and three-particle initial states.
Perhaps most concretely, this could capture
coherence effects between the one- and two-parton wavefunctions in double-parton scattering (see for example \cite{Diehl:2017wew}).

Alternatively, \eqref{6pt radar} could describe a ``radar''-type experiment: in the background of a $X\ot 12$ scattering event, we use $a_3^\dag$ to send in a pulse of light, whose reflection is recorded at a later point with $b_4$. If $1$ and $6$ create and absorb a macroscopic object like a black hole, and $2$ and $5$ create and absorb a test particle moving in its background,
this could very accurately measure the black hole's metric as discussed recently in \cite{Caron-Huot:2022lff}. This is somewhat similar to probing the Solar System's metric by tracking the time it takes for light to travel to a spacecraft and back.

All observables so far involve at most two factors of $S$ and $S^\dag$.
Starting from 6-points, new options arise that are more radically out-of-time-order. An example is
\begin{equation} \label{OTOC example}
{}_{\rm in}\<6|\,b_5\,a_4\,\bdag_3\,\adag_2\,|1\>_{\rm in}=\vac{\,b_6 b_5\,a_4\,\bdag_3\,\adag_2\adag_1\,}
=\vac{\,a_6a_5\,S\,a_4\,S^\dag\,\adag_3\,S\,\adag_2\adag_1\,}\,.
\end{equation}
where we used the stability condition in the first equality.
In the leftmost expression, we think of 1 and 6 as providing a background state and interpret
this amplitude as a four-point out-of-time-order correlator in this background. 
Fourier transforming to the time domain and applying a large time translation to $\bdag_3$ and $b_5$, this
can measure a Lyapunov exponent characterizing the chaotic growth of small perturbations \cite{Shenker:2013pqa, Maldacena:2015waa}. As discussed in \cite{Caron-Huot:2022lff}, such correlators using a sufficiently large energy can also provide particularly detailed probes of the spacetime geometry of the background state.

\begin{figure}
\centering
\adjustbox{valign=t}{\begin{tikzpicture}[line width=1,scale=1]
\draw[] (0,1.7) node {\small$\vac{\,b_6 b_5 b_4\, a_3^\dag a_2^\dag a_1^\dag\,}$};
\coordinate (a) at (0,0);
\coordinate (a0) at ($(a)+(180:1.2)$);
\coordinate (a1) at ($(a)+(0:1.2)$);
\coordinate (as0) at ($(a)+(180:1)$);
\coordinate (as1) at ($(a)+(0:1)$);
\coordinate (b) at ($(a)+(90:0.4)$);
\coordinate (b0) at ($(b)+(180:1.2)$);
\coordinate (b1) at ($(b)+(0:1.2)$);
\coordinate (bs0) at ($(b)+(180:1)$);
\coordinate (bs1) at ($(b)+(0:1)$);
\coordinate (c) at ($(a)+(-90:0.4)$);
\coordinate (c0) at ($(c)+(180:1.2)$);
\coordinate (c1) at ($(c)+(0:1.2)$);
\coordinate (cs0) at ($(c)+(180:1)$);
\coordinate (cs1) at ($(c)+(0:1)$);
\draw[] (a1) node[right] {\small$2$} -- (a0) node[left] {\small$5$};
\draw[] (b1) node[right] {\small$1$} -- (b0) node[left] {\small$4$};
\draw[] (c1) node[right] {\small$3$} -- (c0) node[left] {\small$6$};
\draw[>->] (as1) -- (as0);
\draw[>->] (bs1) -- (bs0);
\draw[>->] (cs1) -- (cs0);
\filldraw[fill=gray!5, line width=1.3](a) circle (0.6) node {$S$};
\end{tikzpicture}}
\quad \vspace{0.7cm}
\adjustbox{valign=t}{\begin{tikzpicture}[line width=1,scale=0.9]
\draw[] (1,1.7) node {\small$\vac{\,a_6a_5\, \bdag_4 b_3\, \adag_2\adag_1\,}$};
\draw[] (0,0.3) -- (-1.2,0.3) node[left] {\small$5$};
\draw[->] (0,0.3) -- (-1,0.3);
\draw[] (0,-0.3) -- (-1.2,-0.3) node[left] {\small$6$};
\draw[->] (0,-0.3) -- (-1,-0.3);
\draw[] (2,0.3) -- (3.2,0.3) node[right] {\small$1$};
\draw[-<] (2,0.3) -- (3,0.3);
\draw[] (2,-0.3) -- (3.2,-0.3) node[right] {\small$2$};
\draw[-<] (2,-0.3) -- (3,-0.3);
\draw[] (2,0) -- (1,-1) node[left] {\small$3$};
\draw[->] (2,0) -- (1.2,-0.8);
\draw[] (1,1) node[right] {\small$4$} -- (0,0);
\draw[>-] (0.8,0.8) -- (0,0);
\filldraw[fill=gray!30](0,-0.3) rectangle (2,0.3);
\draw (2,0.3) -- (0.95,0.3);
\draw (2,-0.3) -- (0.95,-0.3);
\draw[] (1,0) node {$\medmath{X}$};
\filldraw[fill=gray!5, line width=1.3](0,0) circle (0.6) node[yshift=1] {$S^\dag$};
\filldraw[fill=gray!5, line width=1.3](2,0) circle (0.6) node {$S$};
\draw[dashed,orange] (1,1.2) -- (1,-1.2);
\end{tikzpicture}}
\quad \vspace{-0.7cm}
\adjustbox{valign=t}{
\begin{tikzpicture}[line width=1,scale=0.9]
\draw[] (1.5,1.7) node {\small$\vac{\,b_6 b_5\, a_4\, \bdag_3\, \adag_2\adag_1\,}$};
\draw[] (0,0.3) -- (-1.2,0.3) node[left] {\small$5$};
\draw[->] (0,0.3) -- (-1,0.3);
\draw[] (0,-0.3) -- (-1.2,-0.3) node[left] {\small$6$};
\draw[->] (0,-0.3) -- (-1,-0.3);
\draw[] (3,0.3) -- (4.2,0.3) node[right] {\small$1$};
\draw[-<] (3,0.3) -- (4,0.3);
\draw[] (3,-0.3) -- (4.2,-0.3) node[right] {\small$2$};
\draw[-<] (3,-0.3) -- (4,-0.3);
\draw[] (1.5,0) -- (2.5,-1) node[right] {\small$3$};
\draw[-<] (1.5,0) -- (2.3,-0.8);
\draw[] (0.5,1) node[left] {\small$4$} -- (1.5,0);
\draw[<-] (0.7,0.8) -- (1.5,0);
\filldraw[fill=gray!30](0,-0.3) rectangle (1.5,0.3);
\filldraw[fill=gray!30](1.5,-0.3) rectangle (3,0.3);
\filldraw[fill=gray!5, line width=1.3](0,0) circle (0.6) node {$S$};
\filldraw[fill=gray!5, line width=1.3](1.5,0) circle (0.6) node[yshift=1] {$S^\dag$};
\filldraw[fill=gray!5, line width=1.3](3,0) circle (0.6) node {$S$};
\draw (2.25,0.3) -- (2.26,0.3);
\draw (2.25,-0.3) -- (2.26,-0.3);
\draw (0.75,0.3) -- (0.76,0.3);
\draw (0.75,-0.3) -- (0.76,-0.3);
\draw[dashed,orange] (0.75,1.2) -- (0.75,-1.2);
\draw[dashed,orange] (2.25,1.2) -- (2.25,-1.2);
\draw[] (0.75,0) node {\small $\medmath{X}$};
\draw[] (2.25,0) node {\small $\medmath{Y}$};
\end{tikzpicture}}
\caption{\label{fig:6pt-examples}Example measurements with six particles:
exclusive $3\ot 3$ scattering amplitude; inclusive two-point function in a collision \eqref{6pt N};
out-of-time-order correlator \eqref{OTOC example}.
}
\end{figure}
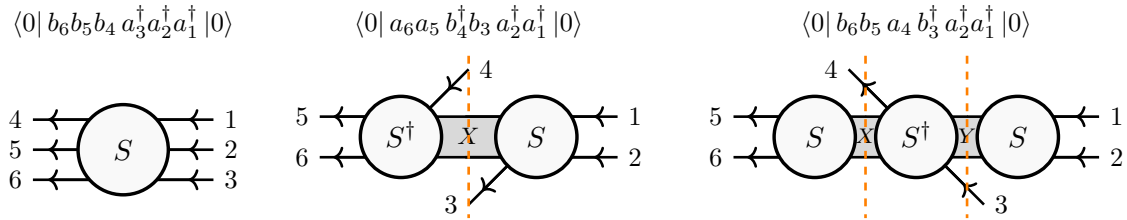

\paragraph{Summary}
Amplitudes with the same momenta but
different operator orderings answer very different yet natural physical questions.
The distinctions are perhaps most evident when macroscopic objects which carry a large entropy $\mathcal{S}$, such as black holes, appear as intermediate states.
Exclusive amplitudes with a finite number of external particles are then likely to be exponentially small $\sim \e^{-\mathcal{S}/2}$ by general statistical considerations. On the other hand, inclusive amplitudes like \eqref{6pt N}---which captures, among other information, the spectrum of Hawking radiation---are generally not exponentially small.  Eventually, we would like to understand all the possible relations between inclusive and exclusive amplitudes.

\subsection{\label{sec:classification}Enumerating asymptotic measurements}

It will be useful to systematically enumerate the different measurements
involving $n$ asymptotic operators, and to introduce a uniform notation for them.

To address this question systematically, let us first fix a given ordering of the operators $a$, $a^\dag$, $b$ and $b^\dag$, thus yielding $4^n$ options for potential asymptotic measurements.  We will assume in this discussion
that the momenta in each $a^\dag$ is different from those in $a$'s, and that the momenta in each $b^\dag$ is different
from those in $b$'s, such that we can ignore $c$-number commutators like \eqref{canonical}.

The first step is to eliminate all $b$ and $b^\dag$'s  in favor of $a$ and $a^\dag$'s via \eqref{b from a},
so as to get a string of just $a$, $\adag$, $S$ and $S^\dag$'s.
As before, we also use properties of the vacuum \eqref{Svac} and stability \eqref{stability} to eliminate
$S$ or $S^\dag$'s acting on the vacuum or on one-particle states.
Finally, commutators allow us to move all $a$'s to the right of $a^\dag$'s.
As a result, all measurements can be brought into the form:
\be\label{eq:measurement-type}
\stackon[7pt]{\hspace{0.7cm}\scalebox{1}{\ensuremath{\langle 0| \underbracket[0.4pt]{a \cdots a}_{k_{2s}} S \underbracket[0.4pt]{a^\dag \cdots a^\dag}_{k_{2s-1}} \underbracket[0.4pt]{a \cdots a}_{k_{2s-2}}
S^\dag \cdots
S^\dag \underbracket[0.4pt]{a^\dag \cdots a^\dag}_{k_3} \underbracket[0.4pt]{a \cdots a}_{k_2} S \underbracket[0.4pt]{a^\dag \cdots a^\dag}_{k_1} | 0\rangle}}\,,}
{ \adjustbox{valign=c}{\input{tikz/gobs}}}
\ee
together with Hermitian conjugate cases where the rightmost operator is an $S^\dag$.
The particle labels are suppressed for clarity and the numbers $k_i$ denote the length of the chains of $a$'s and $a^\dag$'s with $\sum_{i=1}^{2s} k_i = n$.
The total of $n$ operators $a$ and $a^\dag$ are separated by $s$ operators $S$ and $S^\dag$, which appear alternatively. As a shorthand, we will denote \eqref{eq:measurement-type} as
\be
S_{k_{2s} \ot k_{2s-1} | \cdots| k_4\ot k_3|k_2\ot k_1}\, ,
\ee
when $k_1$ is nonempty.
Between every pair of $S$ and $S^\dag$, there has to be at least one operator, and the endpoints have to have at least two each. In other words,
\be
k_1, k_{2s} \geq 2\,, \qquad k_{2i} + k_{2i+1} \geq 1 \quad\mathrm{for}\quad i=1,2,\ldots,s-1\,.
\ee
The possible range is thus $s \in \{1,2,\ldots,n-3\}$. The special cases with $s=1$ are the connected scattering amplitudes
\be
\langle 0| \underbracket[0.4pt]{a \cdots a}_{k_{2}} S \underbracket[0.4pt]{a^\dag \cdots a^\dag}_{k_1} | 0\rangle =
S_{k_2 \ot k_1}\,.
\ee
Those with $s=2$ are the inclusive measurements
\be
\langle 0| \underbracket[0.4pt]{a \cdots a}_{k_{4}} S \underbracket[0.4pt]{a^\dag \cdots a^\dag}_{k_{3}} \underbracket[0.4pt]{a \cdots a}_{k_{2}} S^\dagger \underbracket[0.4pt]{a^\dag \cdots a^\dag}_{k_1} | 0\rangle = \sumint_{X} S^\dag_{k_4\ot X,k_3} S_{X,k_2\ot k_1}\,,
\ee
where we inserted a complete basis of states between $k_2$ and $k_3$, and so on.

Finally, to account for cases where the rightmost evolution operator is $S^\dag$ instead of $S$
(formally corresponding to $k_1$ and $k_2$ both empty), we will use the notation
$S_{k_{2s} \ot k_{2s-1} | \cdots| k_4\ot k_3|k_2\ot k_1}^\dag$ 
to denote the process of \eqref{eq:measurement-type}
where every $S \leftrightarrow S^\dagger$.
Throughout this paper, the operator ${}^\dag$ will indicate the use of anti-time-ordered amplitude (closely related, but not identical, to complex conjugation), see Sec.~\ref{sec:3.1} for more details.

Let's now ask how many distinct types of measurements $T_n$ one can make at a given multiplicity. Again, we omit disconnected measurements from the counting, for instance for $n=5$ we count only \eqref{eq:5pt-measurements} and its Hermitian conjugates.
We find that $T_n$ satisfies the recursion relation
\begin{equation}
    T_n = 4T_{n-1} - 2T_{n-2}\, ,
\end{equation}
and hence grows asymptotically as
\begin{equation}
    T_n \sim \frac{1}{\sqrt{2}} (2+ \sqrt{2})^{n-3}\,.
\end{equation}
This means that exclusive scattering amplitudes, whose number grows only linearly in $n$, are of \emph{measure zero} among all possible measurements one can construct as $n\to \infty$. The first few $T_n$'s are recorded in  the first row of Tab.~\ref{tab:measurements}.

\begin{table}[t]
    \centering
    \begin{tabular}{r|c|c|c|c|c|c|c}
$n$             & 4 & 5 & 6 & 7 & 8 & 9 & 10\\\hline
$\boldsymbol{\#}$ \textbf{measurement types} & 2 & 8 & 28 & 96 & 328 & 1120 & 3824\\\hline
$\boldsymbol{\#}$ \textbf{measurements}     & 6 & 80 & 1370 & 27\ 692 & 639\ 310 & 16\ 601\ 832 & 479\ 026\ 722
    \end{tabular}
    \caption{\textbf{Top:} Number of asymptotic measurement types one can make with $n$ particles.  \textbf{Bottom:} Number of all inequivalent measurements including permutations of the external legs and accounting for $\textrm{CPT}$ relations.}
    \label{tab:measurements}
\end{table}

At this stage, we can further enumerate all distinct permutations of the
measurements by taking into account particle labels of $a$ and $a^\dag$ operators. For each measurement type \eqref{eq:measurement-type}, we have $\binom{n}{k_1\, k_2\, \cdots\, k_{2s}}$ distinct ways of sprinkling the particle labels. Once this is taken into account, using CPT invariance we can also identify the measurements with one labeling to those with the opposite labeling and complex conjugate. More precisely, in the $S$-free notation,
CPT equates observables that differ by simultaneously reflecting the order of operators and exchanging
\be
a_m \leftrightarrow b_{\bar{m}}^\dagger \quad \text{and} \quad b_m \leftrightarrow a_{\bar{m}}^\dagger\,,
\qquad S \leftrightarrow S^\dagger
\ee
where $\bar{m}$ denotes the anti-particle of $m$.
A simple example would be
\be
S_{34 \ot 12} = \vac{\,b_4 b_3\, a_2^\dag a_1^\dag\,} \stackrel{\mathrm{CPT}}{=} \vac{\,b_{\bar{1}} b_{\bar{2}}\, a_{\bar{3}}^\dag a_{\bar{4}}^\dag\,} = S_{\bar{1}\bar{2} \ot \bar{3}\bar{4}}\,.
\ee
We stress that CPT symmetry has nothing to do with crossing symmetry.
It is further discussed in the context of reduction formulae in Sec.~\ref{sec:CPT}.

In particular, if we do not distinguish between particles and anti-particles, the total number of distinct exclusive (i.e., conventional) amplitudes for any $n$ is
\be
\sum_{k_1 = 2}^{n-2} \binom{n}{k_1} = 2^{n} - 2n - 2\,,
\ee
which includes the amplitudes $\cM$ and their complex conjugates $\cM^\dag$. For example, for $n=4$ we get three amplitudes corresponding to scattering in $s$-, $t$-, and $u$-channels, plus three more complex conjugates. The \emph{total} number of measurements grows much faster and is tabulated in the second row of Tab.~\ref{tab:measurements}.

\newcommand{\OneFourPTSblob}[1]{
\adjustbox{valign=c}{
\begin{tikzpicture}[line width=1]
     \draw[] (0,0.3) -- (-1.2,0.3);
     \draw[->] (0,0.3) -- (-1,0.3);
     \draw[] (0,-0.3) -- (-1.2,-0.3);
     \draw[->] (0,-0.3) -- (-1,-0.3);
     \draw[] (0,0.3) -- (1.2,0.3);
     \draw[-<] (0,0.3) -- (1,0.3);
     \draw[] (0,-0.3) -- (1.2,-0.3);
     \draw[-<] (0,-0.3) -- (1,-0.3);
     \filldraw[fill=gray!5, line width=1.2, line width=1.3pt](0,0) circle (0.6) node {$#1$};
\end{tikzpicture}}
}

\newcommand{\TwoFourPTSblob}[2]{\adjustbox{valign=c}{\begin{tikzpicture}[line width=1]
\draw[] (0,0.3) -- (-1.2,0.3) ;
\draw[->] (0,0.3) -- (-1,0.3);
\draw[] (0,-0.3) -- (-1.2,-0.3) ;
\draw[->] (0,-0.3) -- (-1,-0.3);
\draw[] (2,0.3) -- (3.2,0.3) ;
\draw[-<] (2,0.3) -- (3,0.3);
\draw[] (2,-0.3) -- (3.2,-0.3);
\draw[-<] (2,-0.3) -- (3,-0.3);
\filldraw[fill=gray!30](0,-0.3) rectangle (2,0.3);
\draw (2,0.3) -- (0.95,0.3);
\draw (2,-0.3) -- (0.95,-0.3);
\draw[] (1,0) node {$X$};
\filldraw[fill=gray!5, line width=1.2, line width=1.3pt](0,0) circle (0.6) node[yshift=1] {$#1$};
\filldraw[fill=gray!5, line width=1.2, line width=1.3pt](2,0) circle (0.6) node {$#2$};
\draw[dashed,orange] (1,0.5) -- (1,-0.5);
\end{tikzpicture}}}

\allowdisplaybreaks

\section{Introduction to blobology}\label{sec:blobology}

In this section, we introduce useful diagrammatic tools to organize computations of asymptotic observables:
the $S$ and $S^\dag$ blobs, already depicted in the preceding section.
We will see how the blobs allow us to decompose observables into products of connected amplitudes
times appropriate phase space integrals, and how this can be applied diagram-by-diagram
in perturbation theory by matching \say{blob patterns.}

\subsection{Connected versus disconnected subamplitudes}
\label{sec:3.1}

To find the connected components of asymptotic observables, it is useful to separate out the interacting parts of $S$. For $2 \ot 2$ scattering, we have
\begin{equation}\label{eq:2to2}
    S_{2 \ot 2} = \mathbb{1} +  i T_{2\ot 2} \,.
\end{equation}
The matrix element $\mathcal{M}$  of $T$ between some final and initial states $\< 34 |$ and $|12\>$ is then defined as usual by factoring out the overall momentum-conserving $\delta$-function,
\begin{equation}
    \< 34 | T | 12 \> =  \, \mathcal{M}\bdelta_{4} \quad \text{with} \quad  \bdelta_{m} \equiv \left(2\pi\right)^\D \delta^\D \big(\textstyle \sum_{i=1}^{m} p_i \big) \,.
\end{equation}
We will come back to the general $m \ot n$ case shortly.
Diagrammatically, \eqref{eq:2to2} gives an expansion of $S$ blobs:
\begin{equation}
\begin{split}
\adjustbox{valign=c}{\begin{tikzpicture}[line width=1]
    \draw[] (0,0.3) -- (-1.2,0.3) node[above] {\small$3$};
    \draw[->] (0,0.3) -- (-1,0.3);
    \draw[] (0,-0.3) -- (-1.2,-0.3) node[below] {\small$4$};
    \draw[->] (0,-0.3) -- (-1,-0.3);
    \draw[] (0,0.3) -- (1.2,0.3) node[above] {\small$1$};
    \draw[-<] (0,0.3) -- (1,0.3);
    \draw[] (0,-0.3) -- (1.2,-0.3) node[below] {\small$2$};
    \draw[-<] (0,-0.3) -- (1,-0.3);
    \filldraw[fill=gray!5, line width=1.2, line width=1.3pt](0,0) circle (0.6) node {$S$};
\end{tikzpicture}}
&
=
\adjustbox{valign=c}{\begin{tikzpicture}[line width=1]
\begin{scope}[xshift=-220]
    \draw[] (0,0.3) -- (-1.2,0.3) node[above] {\small$3$};
    \draw[->] (0,0.3) -- (-1,0.3);
    \draw[] (0,-0.3) -- (-1.2,-0.3) node[below] {\small$4$};
    \draw[->] (0,-0.3) -- (-1,-0.3);
    \draw[] (0,0.3) -- (1.2,0.3) node[above] {\small$1$};
    \draw[-<] (0,0.3) -- (1,0.3);
    \draw[] (0,-0.3) -- (1.2,-0.3) node[below] {\small$2$};
    \draw[-<] (0,-0.3) -- (1,-0.3);
\end{scope}
\begin{scope}[xshift=-165]
        \node[] at (0,0) {$+$};
\end{scope}
\begin{scope}[xshift=-110]
      \draw[] (1.2,0.3) node[above] {\small$1$} -- (-1.2,-0.3) node[below] {\small$4$};
    \draw[>->] (1,0.25) -- (-1,-0.25);
    \path (1.2,0.3) -- node[fill=white,minimum width=0.3cm,minimum height=0.3cm,inner sep=2pt]{} (-1.2,-0.3);
    \draw[] (1.2,-0.3) node[below] {\small$2$} -- (-1.2,0.3) node[above] {\small$3$};
    \draw[>->] (1,-0.25) -- (-1,0.25);
\end{scope}
\begin{scope}[xshift=-55]
        \node[] at (0,0) {$+$};
\end{scope}
    \draw[] (0,0.3) -- (-1.2,0.3) node[above] {\small$3$};
    \draw[->] (0,0.3) -- (-1,0.3);
    \draw[] (0,-0.3) -- (-1.2,-0.3) node[below] {\small$4$};
    \draw[->] (0,-0.3) -- (-1,-0.3);
    \draw[] (0,0.3) -- (1.2,0.3) node[above] {\small$1$};
    \draw[-<] (0,0.3) -- (1,0.3);
    \draw[] (0,-0.3) -- (1.2,-0.3) node[below] {\small$2$};
    \draw[-<] (0,-0.3) -- (1,-0.3);
    \filldraw[fill=gray!5, line width=1.2, line width=1.3pt](0,0) circle (0.6) node {$i\mathcal{M}$};
\end{tikzpicture}}\label{eq:dec4pts}
\end{split}\,.
\end{equation}
Above, all the factors are explicit except for an overall momentum-conserving $\bdelta$-function,
which comes with each $i\cM$ blob.\footnote{Note that some of the literature from the last century (see, e.g., Refs.~\cite{Eden:1966dnq,Stapp:1976mx,Iagolnitzer:1994xv}) used a similar but slightly different notation, in which the blobs for $i \mathcal{M}$ were denoted with a \say{$+$} and the blobs for $i \mathcal{M}^\dag$ were denoted with a \say{$-$} and had an overall minus sign.  Our notation makes all factors as explicit as possible.} The expansion for $S^\dag$ is simply obtained by replacing the $i\cM$ blob by a $-i\cM^\dag$ one, where $\cM^\dag$ denotes the matrix elements of $T^\dag$ in the same fashion $\cM$ does for $T$.

Furthermore, in terms of blobs, the unitarity condition $S^\dag S-\mathbb{1}=0$ sandwiched between two-particle states translates into
\begin{subequations}
\begin{align}
0&=\TwoFourPTSblob{S^\dag}{S} - \OneFourPTSblob{\mathbb{1}}\\&=\OneFourPTSblob{i\cM}~+\OneFourPTSblob{-i\cM^\dag}~+~\TwoFourPTSblob{-i\cM^\dag}{i\cM}\label{eq:unitarityBlobs}\,,
\end{align}
\end{subequations}
where the non-interacting $\mathbb{1}$ blob corresponds to the first two terms on the right-hand side of \eqref{eq:dec4pts}. 

Throughout this paper, the inclusive sum over the states $X$ (denoted diagrammatically by the cut through $X$ in \eqref{eq:unitarityBlobs}) corresponds to the on-shell phase space integrals,
\begin{equation}
    \sumint_X \equiv \sum_{X_j} \prod_{i \in X_j} \int \frac{\rd^{\D}q_i }{(2\pi)^{\D-1}} \delta_+ (q_i^2+m_i^2)\,,
    \label{eq:inclsum}
\end{equation}
where $m_i$ is the mass of particle $i$, and each $X_j$ consists of a possible set of particles that can be exchanged. The subscript \say{$+$} on the $\delta$-function indicates the presence of an Heaviside step function $\Theta(\pm q^0)$ enforcing that positive energy flows from the right to the left of the diagram.
Finally, it is understood that only states that are inequivalent under permutations are integrated over
(equivalently, one could integrate over the full phase space including permutations at the cost of additional $1/|X_j|!$ symmetry factors). In particular, this guarantees that $\mathbb{1} \cdot \mathbb{1} = \mathbb{1}$.

Note that complex conjugation works just as in quantum mechanics, namely
\begin{equation}\label{eq:MstarToMdag}
(\cM_{A\leftarrow B})^\ast=\<A|S|B\>^\ast=\<B|S^\dag|A\>=\cM^\dag_{B\leftarrow A}\,.
\end{equation}
In the case of $A\leftarrow A$ scattering, this makes the product in \eqref{eq:unitarityBlobs} real and positive, since 
\begin{equation}
 \underbracket[0.4pt]{\adjustbox{valign=c}{\begin{tikzpicture}[line width=1]
\draw[] (0,0.3) -- (-1.2,0.3) node[midway,below,xshift=-25pt,yshift=9pt] {\small$A \textcolor{gray!20}{\adjustbox{valign=c}{\resizebox{0.5cm}{!}{\{}}}$}; 
\draw[->] (0,0.3) -- (-1,0.3);
\draw[] (0,-0.3) -- (-1.2,-0.3); 
\draw[->] (0,-0.3) -- (-1,-0.3);
\draw[] (2,0.3) -- (3.2,0.3) node[midway,below,xshift=25pt,yshift=9pt] {\small$\textcolor{gray!20}{\adjustbox{valign=c}{\resizebox{0.5cm}{!}{\}}}}A$}; 
\draw[-<] (2,0.3) -- (3,0.3);
\draw[] (2,-0.3) -- (3.2,-0.3); 
\draw[-<] (2,-0.3) -- (3,-0.3);
\filldraw[fill=gray!30](0,-0.3) rectangle (2,0.3);
\draw (2,0.3) -- (0.95,0.3);
\draw (2,-0.3) -- (0.95,-0.3);
\draw[] (1,0) node {$X$};
\filldraw[fill=gray!5, line width=1.2, line width=1.3pt](0,0) circle (0.6) node[yshift=1] {$-i\mathcal{M}^\dag$};
\filldraw[fill=gray!5, line width=1.2, line width=1.3pt](2,0) circle (0.6) node {$i\mathcal{M}$};
\draw[dashed,orange] (1,0.5) -- (1,-0.5);
\end{tikzpicture}}}_{\cM_{A \ot X}^\dag \cM_{X \ot A}}
= \Bigg| \underbracket[0.4pt]{\adjustbox{valign=c}{\begin{tikzpicture}[line width=1]
    \draw[] (0,0.3) -- (-1.2,0.3) node[midway,below,xshift=-25pt,yshift=9pt] {\small$X \textcolor{gray!20}{\adjustbox{valign=c}{\resizebox{0.5cm}{!}{\{}}}$};
    \draw[->] (0,0.3) -- (-1,0.3) node[left,below,yshift=6pt, xshift=-3pt] {$\vdots$};
    \draw[] (0,-0.3) -- (-1.2,-0.3);
    \draw[->] (0,-0.3) -- (-1,-0.3);
    \draw[] (0,0.3) -- (1.2,0.3) node[midway,below,xshift=25pt,yshift=9pt] {\small$\textcolor{gray!20}{\adjustbox{valign=c}{\resizebox{0.5cm}{!}{\}}}}A$};
    \draw[-<] (0,0.3) -- (1,0.3);
    \draw[] (0,-0.3) -- (1.2,-0.3) ;
    \draw[-<] (0,-0.3) -- (1,-0.3);
    \filldraw[fill=gray!5, line width=1.2, line width=1.3pt](0,0) circle (0.6) node {$i\mathcal{M}$};
\end{tikzpicture}}}_{\cM_{X \ot A}}\Bigg|^2\geq0\,.
\end{equation}

Let us add that, in even spacetime dimensions, CPT guarantees that amplitudes are symmetrical under transposition:
$\cM_{B\ot A}=\cM_{A\ot B}$. Consequently, $\cM^\dag$ is equivalent to the complex conjugate
$\cM^*$, which is a notation often used in textbooks (see, e.g., \cite{Peskin:1995ev}).
A general discussion of CPT including odd spacetime dimensions will be given in Sec.~\ref{sec:CPT} below.

In all calculations in this paper, we will assume that \emph{external} legs are non-forward,
or more generally that no subset of external momenta sums up to zero, so that there is a single overall connected component.
Thus, when computing $2\ot 2$ scattering amplitudes, we drop the forward terms corresponding to the identity in \eqref{eq:2to2}.
However, these terms can still be important within subamplitudes.

Let us illustrate this on the simplest inclusive observable: the five-point inclusive measurement $\text{Exp}_k= \sumint_X \<54|S^\dagger |X\> \<X3|S|21\>$.
We can break each $S^\dag$ and $S$ into its connected components.  The disconnected part of $S$ gives rise
to forward terms that only have support when $p_3$ is forward to $p_1$ or $p_2$, which we discard.
However, the disconnected part of $S^\dag$ still gives rise to amplitudes that are overall connected, which we must thus retain:
\begin{equation}
    \adjustbox{valign=c}{\begin{tikzpicture}[line width=1]
\begin{scope}[xshift=0]
\draw[] (0,0.3) -- (-1.2,0.3) node[above] {\small$4$};
\draw[->] (0,0.3) -- (-1,0.3);
\draw[] (0,-0.3) -- (-1.2,-0.3) node[below] {\small$5$};
\draw[->] (0,-0.3) -- (-1,-0.3);
\draw[] (2,0.3) -- (3.2,0.3) node[above] {\small$1$};
\draw[-<] (2,0.3) -- (3,0.3);
\draw[] (2,-0.3) -- (3.2,-0.3) node[below] {\small$2$};
\draw[-<] (2,-0.3) -- (3,-0.3);
\draw[] (2,0) -- (1,1) node[left] {\small$3$};
\draw[->] (2,0) -- (1.2,0.8);
\filldraw[fill=gray!30](0,-0.3) rectangle (2,0.3);
\draw (2,0.3) -- (0.95,0.3);
\draw (2,-0.3) -- (0.95,-0.3);
\draw[] (1,0) node {$X$};
\filldraw[fill=gray!5, line width=1.2, line width=1.3pt](0,0) circle (0.6) node[yshift=1] {$S^\dag$};
\filldraw[fill=gray!5, line width=1.2, line width=1.3pt](2,0) circle (0.6) node {$S$};
\draw[dashed,orange] (1,1.2) -- (1,-0.8);
\end{scope}
\begin{scope}[xshift=107]
    \node[] at (0,0) {$=$};
\end{scope}
\begin{scope}[xshift=100]
\draw[] (2,0.3) -- (0.8,0.3) node[above] {\small$4$};
\draw[->] (2,0.3) -- (1,0.3);
\draw[] (2,-0.3) -- (0.8,-0.3) node[below] {\small$5$};
\draw[->] (2,-0.3) -- (1,-0.3);
\draw[] (2,0.3) -- (3.2,0.3) node[above] {\small$1$};
\draw[-<] (2,0.3) -- (3,0.3);
\draw[] (2,-0.3) -- (3.2,-0.3) node[below] {\small$2$};
\draw[-<] (2,-0.3) -- (3,-0.3);
\draw[] (2,0) -- (1,1) node[left] {\small$3$};
\draw[->] (2,0) -- (1.2,0.8);
\draw (2,0.3) -- (0.95,0.3);
\draw (2,-0.3) -- (0.95,-0.3);
\filldraw[fill=gray!5, line width=1.2, line width=1.3pt](2,0) circle (0.6) node {$i \mathcal{M}$};
\draw[dashed,orange] (1,1.2) -- (1,-0.8);
\end{scope}
\begin{scope}[xshift=206]
    \node[] at (0,0) {$+$};
\end{scope}
\begin{scope}[xshift=255]
\draw[] (0,0.3) -- (-1.2,0.3) node[above] {\small$4$};
\draw[->] (0,0.3) -- (-1,0.3);
\draw[] (0,-0.3) -- (-1.2,-0.3) node[below] {\small$5$};
\draw[->] (0,-0.3) -- (-1,-0.3);
\draw[] (2,0.3) -- (3.2,0.3) node[above] {\small$1$};
\draw[-<] (2,0.3) -- (3,0.3);
\draw[] (2,-0.3) -- (3.2,-0.3) node[below] {\small$2$};
\draw[-<] (2,-0.3) -- (3,-0.3);
\draw[] (2,0) -- (1,1) node[left] {\small$3$};
\draw[->] (2,0) -- (1.2,0.8);
\filldraw[fill=gray!30](0,-0.3) rectangle (2,0.3);
\draw (2,0.3) -- (0.95,0.3);
\draw (2,-0.3) -- (0.95,-0.3);
\draw[] (1,0) node {$X$};
\filldraw[fill=gray!5, line width=1.2, line width=1.3pt](0,0) circle (0.6) node {$- i \mathcal{M}^\dag$};
\filldraw[fill=gray!5, line width=1.2, line width=1.3pt](2,0) circle (0.6) node {$ i \mathcal{M}$};
\draw[dashed,orange] (1,1.2) -- (1,-0.8);
\end{scope}
\end{tikzpicture}}\,.
    \label{eq:5ptdecomp}
\end{equation}

If a diagram contains two (or more) $i\cM$ or $-i\cM^\dag$ blobs strung together, there will be separate momentum-conserving $\delta$-functions for each blob, so after factoring out the overall one,
in a string of $N$ blobs we will be left with $(N-1)$ $\delta$-functions to be soaked up by the integrals \eqref{eq:inclsum}.

For observables with more external legs, the decomposition into amplitudes and their conjugates proceeds in the same way. Note that in general $S\neq \mathbb{1}+iT$; here is a (hopefully) self-explanatory example with
four incoming and four outgoing particles:
\begin{equation}\label{eq:dec8pts}
    \adjustbox{valign=c}{
    \begin{tikzpicture}[line width=1,scale=0.9]
    \newcommand\coords{
        \coordinate (z) at (0,0);
        \coordinate (a) at (0,-0.15);
        \coordinate (a0) at ($(a)+(180:1.2)$);
        \coordinate (a1) at ($(a)+(0:1.2)$);
        \coordinate (as0) at ($(a)+(180:1)$);
        \coordinate (as1) at ($(a)+(0:1)$);
        \coordinate (b) at ($(a)+(90:0.3)$);
        \coordinate (b0) at ($(b)+(180:1.2)$);
        \coordinate (b1) at ($(b)+(0:1.2)$);
        \coordinate (bs0) at ($(b)+(180:1)$);
        \coordinate (bs1) at ($(b)+(0:1)$);
        \coordinate (c) at ($(a)+(-90:0.3)$);
        \coordinate (c0) at ($(c)+(180:1.2)$);
        \coordinate (c1) at ($(c)+(0:1.2)$);
        \coordinate (cs0) at ($(c)+(180:1)$);
        \coordinate (cs1) at ($(c)+(0:1)$);
        \coordinate (d) at ($(a)+(90:0.6)$);
        \coordinate (d0) at ($(d)+(180:1.2)$);
        \coordinate (d1) at ($(d)+(0:1.2)$);
        \coordinate (ds0) at ($(d)+(180:1)$);
        \coordinate (ds1) at ($(d)+(0:1)$);
    }
    \newcommand\skeleton{
        \draw[] (a1) -- (a0) ;
        \draw[] (b1) -- (b0) ;
        \draw[] (c1) -- (c0) ;
        \draw[] (d1) -- (d0) ;
        \draw[{>[scale=0.7]}-{>[scale=0.7]}] (as1) -- (as0) ;
        \draw[{>[scale=0.7]}-{>[scale=0.7]}] (bs1) -- (bs0) ;
        \draw[{>[scale=0.7]}-{>[scale=0.7]}] (cs1) -- (cs0) ;
        \draw[{>[scale=0.7]}-{>[scale=0.7]}] (ds1) -- (ds0) ;
    }
    \begin{scope}[xshift=0]
        \coords
        \skeleton
        \filldraw[fill=gray!5, line width=1.2](z) circle (0.6) node {$S$};
    \end{scope}
    \begin{scope}[xshift=45]
        \node[] at (0,0) {$=$};
    \end{scope}
    \begin{scope}[xshift=90]
        \coords
        \skeleton
        \filldraw[fill=gray!5, line width=1.2] (z) circle (0.6) node {$i \mathcal{M}$};
    \end{scope}
    \begin{scope}[xshift=135]
        \node[] at (0,0) {$+$};
    \end{scope}
    \begin{scope}[xshift=180]
        \coords
        \skeleton
        \filldraw[fill=gray!5, line width=1.2] (0,0.15) circle (0.5) node {$i \mathcal{M}$};
    \end{scope}
    \begin{scope}[xshift=225]
        \node[] at (0,0) {$+$};
    \end{scope}
    \begin{scope}[xshift=270]
        \coords
        \skeleton
        \filldraw[fill=gray!5, line width=1.2] (0,0.35) circle (0.3) node {\scalebox{0.65}{$i \mathcal{M}$}};
    \end{scope}
    \begin{scope}[xshift=315]
        \node[] at (0,0) {$+$};
    \end{scope}
    \begin{scope}[xshift=-45]
    \begin{scope}[xshift=315,yshift=-50]
        \node[] at (0,0) {$+$};
    \end{scope}
    \begin{scope}[xshift=270,yshift=-50]
        \coords
        \skeleton
        \filldraw[fill=gray!5, line width=1.2] (0,0.35) circle (0.3) node {\scalebox{0.65}{$i \mathcal{M}$}};
        \filldraw[fill=gray!5, line width=1.2] (0,-0.35) circle (0.3) node {\scalebox{0.65}{$i \mathcal{M}$}};
    \end{scope}
    \begin{scope}[xshift=225,yshift=-50]
        \node[] at (0,0) {$+$};
    \end{scope}
    \begin{scope}[xshift=360,yshift=-50]
        \coords
        \skeleton
    \end{scope}
    \begin{scope}[xshift=425,yshift=-50]
        \node[] at (0,0) {$+ \text{ perms\,.}$};
    \end{scope}
    \end{scope}
    \end{tikzpicture}
    }
\end{equation}
If we were computing a non-forward scattering amplitude, we would only retain the first term,
but all terms are potentially important when the $S$ blob appears as a subamplitude of a more general observable,
since the non-forward assumption only enforces that the \emph{full} diagram is connected.
An example of an overall non-connected diagram would be
\be
\adjustbox{valign=c}{\begin{tikzpicture}[line width=1,scale=0.9]
\begin{scope}
\coordinate (a) at (0,0);
\coordinate (a0) at ($(a)+(180:1.2)$);
\coordinate (a1) at ($(a)+(0:3.2)$);
\coordinate (as0) at ($(a)+(180:1)$);
\coordinate (as1) at ($(a)+(0:3)$);
\coordinate (b) at ($(a)+(90:0.4)$);
\coordinate (b0) at ($(b)+(180:1.2)$);
\coordinate (b1) at ($(b)+(0:3.2)$);
\coordinate (bs0) at ($(b)+(180:1)$);
\coordinate (bs1) at ($(b)+(0:3)$);
\coordinate (c) at ($(a)+(-90:0.4)$);
\coordinate (c0) at ($(c)+(180:1.2)$);
\coordinate (c1) at ($(c)+(0:3.2)$);
\coordinate (cs0) at ($(c)+(180:1)$);
\coordinate (cs1) at ($(c)+(0:3)$);
\draw[] (a1) node[right] {\small$4$} -- (a0) node[left] {\small$10$};
\draw[] (b1) node[right] {\small$3$} -- (b0) node[left] {\small$9$};
\draw[] (c1) node[right] {\small$5$} -- (c0) node[left] {\small$11$};
\draw[>->] (as1) -- (as0);
\draw[>->] (bs1) -- (bs0);
\draw[>->] (cs1) -- (cs0);
\filldraw[fill=gray!30](0,-0.4) rectangle (2,0);
\filldraw[fill=gray!5, line width=1.2](2,0) circle (0.6) node {$S$};
\filldraw[fill=gray!5, line width=1.2](0,-0.2) circle (0.4) node[yshift=1] {$S^\dag$};
\draw[] (1,-0.2) node {$\medmath{Y}$};
\end{scope}
\begin{scope}[yshift=45]
\draw[] (0,0.3) -- (-1.2,0.3) node[left] {\small$7$};
\draw[->] (0,0.3) -- (-1,0.3);
\draw[] (0,-0.3) -- (-1.2,-0.3) node[left] {\small$8$};
\draw[->] (0,-0.3) -- (-1,-0.3);
\draw[] (2,0.3) -- (3.2,0.3) node[right] {\small$1$};
\draw[-<] (2,0.3) -- (3,0.3);
\draw[] (2,-0.3) -- (3.2,-0.3) node[right] {\small$2$};
\draw[-<] (2,-0.3) -- (3,-0.3);
\draw[] (2,0) -- (1,1) node[left] {\small$6$};
\draw[->] (2,0) -- (1.2,0.8);
\filldraw[fill=gray!30](0,-0.3) rectangle (2,0.3);
\draw (2,0.3) -- (0.95,0.3);
\draw (2,-0.3) -- (0.95,-0.3);
\draw[] (1,0) node {$X$};
\filldraw[fill=gray!5, line width=1.2](0,0) circle (0.6) node[yshift=1] {$S^\dag$};
\filldraw[fill=gray!5, line width=1.2](2,0) circle (0.6) node {$S$};
\draw[dashed,orange] (1,1.2) -- (1,-2.5);
\end{scope}
\end{tikzpicture}}\,,
\ee
\noindent
since it only has support on the forward kinematics $p_{3459,10,11}=0$ and $p_{12678}=0$ in the notation $p_I = \sum_{i\in I} p_i$. 
Physically, we discard such disconnected clusters for the usual reason that
they describe experiments that do not interfere with each other (one could be taking place in Andromeda for all we know).   Adding any single line between the two clusters above would make it connected.

\subsection{Perturbative blobology}
\label{sec:pert_blobology}
Our discussion so far has been entirely at an axiomatic level, assuming that an operator $S$ containing the full non-perturbative dynamics has been supplied to us.
It is therefore natural to ask whether we can, just like for scattering amplitudes, compute inclusive observables 
in perturbation theory. We will see in this subsection that this is indeed possible, even on a diagram-by-diagram or topology-by-topology basis.

The fact that the formulas must make sense diagram-by-diagram can be seen using a trick: the inclusive observables are defined in \textit{any} quantum field theory, so we can engineer a Lagrangian that reproduces any desired diagram. As a simple example, consider the observable $\Exp_3$ from \eqref{inclusive 5} for the following topology:
\begin{equation}\label{eq:M_12_345}
\begin{gathered}
\begin{tikzpicture}[baseline= {($(current bounding box.base)+(10pt,10pt)$)},line width=1, scale=0.7,yscale=-1]
\coordinate (a) at (0,0) ;
\coordinate (b) at (1,0) ;
\coordinate (c) at ($(b)+(-40:1)$);
\draw[] (a) to [in=120,out=60](b);
\draw[] (a) to [in=-120,out=-60](b);
\draw[] (b) to (c);
\draw[RoyalBlue] (b) -- ++ (30:1) node[right] {\footnotesize$2$};
\draw[Maroon] (c) -- ++ (-150:1) node[left]{\footnotesize$3$};
\draw[RoyalBlue] (c) -- ++ (-30:1) node[right] {\footnotesize$1$};
\draw[RoyalBlue] (a) -- (-150:1) node[left] {\footnotesize$4$};
\draw[RoyalBlue] (a) -- (150:1) node[left] {\footnotesize$5$};
\node[] at (1.85,-0.05) {\footnotesize{$6$}};
\node[] at (0.5,-0.55) {\footnotesize{$7$}};
\node[] at (0.5,0.65) {\footnotesize{$8$}};
\fill[black,thick] (a) circle (0.07);
\fill[black,thick] (b) circle (0.07);
\fill[black,thick] (c) circle (0.07);
\end{tikzpicture}
\end{gathered} \,.
\end{equation}
To isolate its scalar contribution, we design the Lagrangian
\begin{equation}
    \mathcal{L} = \sum_{i=1}^8 \frac{1}{2} \left( \phi_i \Box \phi_i - m_i^2 \right)- g_1 \phi_1 \phi_3 \phi_6 - g_2 \phi_2 \phi_6 \phi_7 \phi_8 - g_3 \phi_4 \phi_5 \phi_7 \phi_8 \,,
\end{equation}
that precisely reproduces only this diagram to order $\mathcal{O}(g_1g_2g_3)$ in the coupling constants. (In gauge theories one might need to be a bit more careful; we can either work in a physical gauge or add up the relevant contributions that render the amplitudes gauge invariant.) Since we can always play this game, we run the arguments from the previous sections, then expand the observables in perturbation theory with the designed Lagrangian. This allows us to compute the inclusive observables order-by-order as products of amplitudes connected by phase-space integrals.

To show how such perturbation-theory computations of the observables work, let us compute the contribution
to $\Exp_3=\vac{a_5a_4 b_3 \adag_2\adag_1}$ with the same graph topology as the diagram in \eqref{eq:M_12_345}.
We simply have to sum over all ways of fitting the diagram inside blobs shown \eqref{eq:measurement-type}, where each factor $i\cM$ or $-i\cM^\dag$ must be fully connected:
\begin{equation}
\begin{tikzpicture}[baseline= {($(current bounding box.base)+(10pt,-2pt)$)},line width=1,scale=0.9]
\begin{scope}[xshift=-180,xshift=120,baseline= {($(current bounding box.base)+(10pt,-2pt)$)},scale=0.7]
    \node[] at (0,0) {$\textcolor{white}{=}$};
\end{scope}
\begin{scope}[xshift=-180,xshift=120,baseline= {($(current bounding box.base)+(10pt,-2pt)$)},scale=0.7, yshift=-70]
    \node[] at (-0.75,0) {$\Exp_3 =$};
\end{scope}
\end{tikzpicture}
\underbracket[0.4pt]{
\begin{tikzpicture}[baseline= {($(current bounding box.base)+(10pt,-2pt)$)},line width=1,scale=0.9]
\draw[] (0,0.3) -- (-1.2,0.3) node[left] {\small$4$};
\draw[->] (0,0.3) -- (-1,0.3);
\draw[] (0,-0.3) -- (-1.2,-0.3) node[left] {\small$5$};
\draw[->] (0,-0.3) -- (-1,-0.3);
\draw[] (2,0.3) -- (3.2,0.3) node[right] {\small$1$};
\draw[-<] (2,0.3) -- (3,0.3);
\draw[] (2,-0.3) -- (3.2,-0.3) node[right] {\small$2$};
\draw[-<] (2,-0.3) -- (3,-0.3);
\draw[] (2,0) -- (1,1) node[left] {\small$3$};
\draw[->] (2,0) -- (1.2,0.8);
\filldraw[fill=gray!20](0,-0.3) rectangle (2,0.3);
\draw (2,0.3) -- (0.95,0.3);
\draw (2,-0.3) -- (0.95,-0.3);
\filldraw[fill=gray!5, line width=1.2](0,0) circle (0.6) node {\scalebox{1}{$\mathbb{1}$}};
\filldraw[fill=gray!5, line width=1.2](2,0) circle (0.6) node {\scalebox{1}{$i \mathcal{M}$}};
\draw[dashed,orange] (1,1.2) -- (1,-2.2);
\begin{scope}[xshift=35,yshift=-50,line width=1, scale=0.6,yscale=-1,baseline= {($(current bounding box.base)+(10pt,-2pt)$)}]
\coordinate (a) at (0,0) ;
\coordinate (b) at (1,0) ;
\coordinate (c) at ($(b)+(-40:1)$);
\draw[] (a) to [in=120,out=60](b);
\draw[] (a) to [in=-120,out=-60](b);
\draw[] (b) to (c);
\draw[RoyalBlue] (b) -- ++ (30:1) node[right] {\footnotesize$2$};
\draw[Maroon] (c) -- ++ (-150:1) node[above]{\footnotesize$3$};
\draw[RoyalBlue] (c) -- ++ (-30:1) node[right] {\footnotesize$1$};
\draw[RoyalBlue] (a) -- (-150:1) node[left] {\footnotesize$4$};
\draw[RoyalBlue] (a) -- (150:1) node[left] {\footnotesize$5$};
\fill[black,thick] (a) circle (0.07);
\fill[black,thick] (b) circle (0.07);
\fill[black,thick] (c) circle (0.07);
\end{scope}
\end{tikzpicture}}_{\text{term }1}
\quad
\begin{tikzpicture}[baseline= {($(current bounding box.base)+(10pt,-2pt)$)},line width=1,scale=0.9]
\begin{scope}[xshift=-180,xshift=120,baseline= {($(current bounding box.base)+(10pt,-2pt)$)},scale=0.7]
    \node[] at (0,0) {$\textcolor{white}{+}$};
\end{scope}
\begin{scope}[xshift=-180,xshift=120,baseline= {($(current bounding box.base)+(10pt,-2pt)$)},scale=0.7, yshift=-70]
    \node[] at (0,0) {$+$};
\end{scope}
\end{tikzpicture}
\quad
\underbracket[0.4pt]{
\begin{tikzpicture}[line width=1,baseline= {($(current bounding box.base)+(10pt,-2pt)$)},scale=0.9,xshift=-40]
\draw[] (0,0.3) -- (-1.2,0.3) node[left] {\small$4$};
\draw[->] (0,0.3) -- (-1,0.3);
\draw[] (0,-0.3) -- (-1.2,-0.3) node[left] {\small$5$};
\draw[->] (0,-0.3) -- (-1,-0.3);
\draw[] (2,0.3) -- (3.2,0.3) node[right] {\small$1$};
\draw[-<] (2,0.3) -- (3,0.3);
\draw[] (2,-0.3) -- (3.2,-0.3) node[right] {\small$2$};
\draw[-<] (2,-0.3) -- (3,-0.3);
\draw[] (2,0) -- (1,1) node[left] {\small$3$};
\draw[->] (2,0) -- (1.2,0.8);
\filldraw[fill=gray!30](0,-0.3) rectangle (2,0.3);
\draw (2,0.3) -- (0.95,0.3);
\draw (2,-0.3) -- (0.95,-0.3);
\filldraw[fill=gray!5, line width=1.2](0,0) circle (0.6) node {\scalebox{1}{$-i \mathcal{M}^\dag$}};
\filldraw[fill=gray!5, line width=1.2](2,0) circle (0.6) node {\scalebox{1}{$i \mathcal{M}$}};
\draw[dashed,orange] (1,1.2) -- (1,-2.2);
\begin{scope}[xshift=20,yshift=-50,line width=1, scale=0.6,yscale=-1,baseline= {($(current bounding box.base)+(10pt,-2pt)$)}]
\coordinate (a) at (0,0) ;
\coordinate (b) at (1,0) ;
\coordinate (c) at ($(b)+(-40:1)$);
\draw[] (a) to [in=120,out=60](b);
\draw[] (a) to [in=-120,out=-60](b);
\draw[] (b) to (c);
\draw[RoyalBlue] (b) -- ++ (30:1) node[right] {\footnotesize$2$};
\draw[Maroon] (c) -- ++ (-150:1) node[above]{\footnotesize$3$};
\draw[RoyalBlue] (c) -- ++ (-30:1) node[right] {\footnotesize$1$};
\draw[RoyalBlue] (a) -- (-150:1) node[left] {\footnotesize$4$};
\draw[RoyalBlue] (a) -- (150:1) node[left] {\footnotesize$5$};
\fill[black] (a) circle (0.07);
\fill[black] (b) circle (0.07);
\fill[black] (c) circle (0.07);
\end{scope}
\end{tikzpicture}
}_{\text{term }2}\,.
\end{equation}
The term on the left corresponds to the usual connected scattering amplitude $i \mathcal{M}_{345 \ot 12}$, while the term on the right represents two on-shell amplitudes $-i \mathcal{M}^\dag_{45 \ot 78}$ and $i \mathcal{M}_{378 \ot 12}$, connected via an inclusive sum over the phase space (see \eqref{eq:inclsum}),
and a $\delta$-function in one of the blobs.
Of course, this construction is neither new nor mysterious: in perturbation theory we can compute the cut blobs simply as \emph{unitarity \emph{(}Cutkosky\emph{)} cuts}~\cite{Cutkosky:1960sp}, where particles that flow across the cut are taken to be on-shell with positive energy. In perturbation theory, it simply amounts to replacing each propagator with a $\delta$-function with positive-energy flow across the cut,
\begin{equation}
    \frac{-i}{q_i^2+m_i^2 - i\varepsilon} \;\to\;  2 \pi \delta_+(q_i^2+m_i^2)\,,
\end{equation}
and conjugating the part of the diagram contained in $-i \mathcal{M}^\dag$.

Continuing our example in \eqref{eq:M_12_345}, assuming that the masses of all the scalars are equal for simplicity, we can write the contribution to the observable $\Exp_k$ from the first term as,
\begin{equation}
    \left. \Exp_3 \right\vert_{\text{(term 1)}} = \frac{-ig_1 g_2 g_3}{-s_{13}+m^2} \underbracket[0.4pt]{\int \frac{\d^\D \ell }{i(2\pi)^{\D}}
    \frac{1}{[\ell^2+m^2-i\varepsilon] [(p_{45}-\ell)^2+m^2-i\varepsilon]}}_{\Sigma(s_{45})}\,,
    \label{eq:term1}
\end{equation}
with $s_{ij} \equiv -p_{ij}^2$, and we have defined the self-energy loop integral as $\Sigma(s_{45})$ for later convenience. Note that since $s_{13}<0$ in this example, we have dropped the $i\varepsilon$ in its propagator.
For the cut term, on the other hand, we add phase space integrals for particles 7 and 8, as well as a momentum-conserving $\delta$-function,
\begin{subequations}
\begin{align}
    \left. \Exp_3 \right\vert_{\text{(term 2)}} & = \frac{-g_1 g_2 g_3}{-s_{13}+m^2} \int \frac{\d^\D q_7}{(2\pi)^{\D-1}} \frac{\d^\D q_8}{(2\pi)^{\D-1}} 
    \delta_+(q_7^2+m^2)
    \delta_+(q_8^2+m^2) \bdelta_{45 \ot 78} 
    \\ & =  \frac{-g_1 g_2 g_3}{-s_{13}+m^2} 
    \int\frac{\d^\D \ell }{(2\pi)^{\D}}
    (2\pi) \delta_+(\ell^2+m^2)
    (2\pi) \delta_+\big[(p_{45}-\ell)^2+m^2 \big]\,.
\end{align}
\end{subequations}
This term is now directly written as a Cutkosky cut, where we have replaced propagators in the amplitude \eqref{eq:term1} with $\delta$-functions with positive-energy flow, and conjugated the vertex in $-i \mathcal{M}^\dag$. More generally, the momentum-conserving $\delta$-functions for each blob will collapse one of the phase-space integrals, to reduce them into ordinary cut integrals.
Of course, in this particular example, the cut integral can be recognized as (twice) the imaginary part of the self-energy, so the transition from amplitude to inclusive observable is rather simple:
\begin{equation}
    i\cM\supset \frac{-ig_1g_2g_3}{-s_{13}+m^2}\Sigma(s_{45})
    \quad\Rightarrow\quad
    \Exp_3 \supset \frac{-ig_1g_2g_3}{-s_{13}+m^2}\underbracket[0.4pt]{\Big[\Sigma(s_{45}) -2i{\rm Im}\Sigma(s_{45})\Big]}_{\Sigma(s_{45})^*} \label{cut from Im Sigma}\,,
\end{equation}
for $s_{45}$ real. We warn the reader that such a simple substitution may not always be available.

As the final point in this subsection, we note that when fitting diagrams into the blobs in \eqref{eq:measurement-type}, it is crucial to respect the order at which particles are absorbed or emitted.
In Sec.~\ref{sec:waveforms}, we will encounter an example that relies heavily on this fact:
the following four pentagon topologies for the inclusive observable $\Exp_3$ (leg 3 being the emitted graviton):
\begin{equation}\label{ABCD topos}
\stackrel{\adjustbox{valign=c,scale={0.75}{0.75}}{\input{tikz/I6}}}{A}
\qquad
\stackrel{\adjustbox{valign=c,scale={0.75}{0.75}}{\input{tikz/IB}}}{B}
\qquad
\stackrel{\adjustbox{valign=c,scale={0.75}{0.75}}{\input{tikz/IC}}}{C}
\qquad
\stackrel{\adjustbox{valign=c,scale={0.75}{0.75}}{\input{tikz/IDsource}}}{D}\,.
\end{equation}
We take the thick-line particles to have constant masses along each thick line and the wavy lines to be massless, which makes many cuts vanish by kinematic considerations.
To compute $\Exp_3$, we fit these diagrams in all different ways into the blob pattern in \eqref{eq:5ptdecomp}. However, cuts through $A$ of the form
\begin{equation}
    \Exp_3\not\supset \underbracket[0.4pt]{\Bigg\{\adjustbox{valign=c,scale={0.75}{0.75}}{\input{tikz/I6cut1}}\,,
    \quad \adjustbox{valign=c,scale={0.75}{0.75}}{\input{tikz/I6cut2}}\,,
    \quad \adjustbox{valign=c,scale={0.75}{0.75}}{\input{tikz/I6cut3}}\Bigg\}}_{ \text{(not matching blobs pattern)}}\,,
\end{equation}
do not contribute, since the particle $3$ is only emitted after the cuts. Similarly, cuts of $B$ and $C$ of the form
\begin{equation}
\Exp_3\not\supset \underbracket[0.4pt]{\Bigg\{\adjustbox{valign=c,scale={0.75}{0.75}}{\input{tikz/IBnotAcut0}}\,,\adjustbox{valign=c,scale={0.75}{0.75}}{\input{tikz/ICnotAcut0}}\Bigg\}}_{\text{(not unitarity cuts)}}\,,
\end{equation}
are disallowed, since they do not put all particles through the cut on-shell. Additionally, while cuts of the form
\begin{equation}
    \Exp_3\supset\underbracket[0.4pt]{\Bigg\{\adjustbox{valign=c,scale={0.75}{0.75}}{\input{tikz/IBnotAcut}}\,,\adjustbox{valign=c,scale={0.75}{0.75}}{\input{tikz/ICnotAcut}}\,,\adjustbox{valign=c,scale={0.75}{0.75}}{\input{tikz/IBcutCorner}}\Bigg\}}_{\text{(vanish)}}\,,
\end{equation}
are valid unitary cuts fitting the blobs pattern, they vanish by momentum conservation for equal-mass thick lines: the particle of mass $m$ cannot decay into a particle of mass $m$ and a massless particle, and by time reversal, two particles cannot combine into one with mass $m$.

Finally, let us record all non-vanishing contributions to $\Exp_3$ from the topologies \eqref{ABCD topos} obtained by matching the blobs pattern
\begin{equation}
\Exp_3\supset \underbracket[0.4pt]{\Bigg\{\adjustbox{valign=c,scale={0.75}{0.75}}{\input{tikz/nontrivial2}}\,,\adjustbox{valign=c,scale={0.75}{0.75}}{\input{tikz/nontrivial3}}\,,\adjustbox{valign=c,scale={0.75}{0.75}}{\input{tikz/nontrivial4}}\,,\adjustbox{valign=c,scale={0.75}{0.75}}{\input{tikz/nontrivial0}}\,,\adjustbox{valign=c,scale={0.75}{0.75}}{\input{tikz/nontrivial1}}\Bigg\}}_{ \text{(non-trivial)}}\,.
\end{equation}
To be clear, all the diagrams above are conventional time-ordered amplitudes, with the exception of the last one, which is cut.

\subsection{Cuts through tree-level propagators and narrow-width approximation}
\label{sec:treelevel}

When computing observables for diagrams with tree-level propagators, it is important to include cuts through those as well. Tree-level cuts can be nonzero in physical kinematics when and only when the exchanged particle is unstable: ``if it can be produced, it can decay.'' If one were to resum self-energies, one would find a propagator of the approximate form
\begin{equation}
\frac{-i}{q_i^2-\Sigma(q_i^2)} \approx  \frac{-i}{q_i^2+M^2 - i M \Gamma} \, ,
\end{equation}
where $M$ is the mass after the resummation, and $\Gamma \geq 0$ is the decay width. In the \emph{narrow-width approximation} $\Gamma \to 0^+$, such a propagator behaves just like that of a stable particle with $\Gamma$ playing the role of $\varepsilon$.  In the $S$-matrix context, one would never include an unstable particle
directly in a cut but rather one would include its decay products, proportional to the imaginary part of the self-energy, but these are effectively equivalent due to the relation
\be
\frac{2M\Gamma}{(q_i^2+M^2)^2+M^2\Gamma^2} = 2{\rm Re} \frac{-i}{q_i^2+M^2 - i M\Gamma}
 \approx 2\pi\delta(q_i^2+M^2) + \mathcal{O}(\Gamma)\, ,
\ee
which can be understood from the familiar distributional identity
$\frac{1}{x+i\varepsilon} - \frac{1}{x-i\varepsilon} = -2 \pi i \delta(x)$.
Thus, in the narrow-width approximation (or, order-by-order in perturbation theory), it is correct to
simply treat unstable particles as if they were ordinary particles and allow cuts through their propagators.
This phenomenon is familiar in calculations of cross-sections involving unstable but long-lived particles \cite{Peskin:1995ev,Campbell:2017hsr} and here we are just saying that it also applies to products of blobs.

As a simple example, if we look at a tree-level contribution to the process from \eqref{eq:M_12_345} with a massive propagator,
\begin{equation}\label{eq:M_12_345tree}
\begin{gathered}
\begin{tikzpicture}[baseline= {($(current bounding box.base)+(10pt,10pt)$)},line width=1, scale=0.7,yscale=-1]
\coordinate (a) at (0,0) ;
\coordinate (b) at (1,0) ;
\coordinate (c) at ($(b)+(0:1)$);
\draw[] (a) -- (b);
\draw[] (b) -- (c);
\draw[RoyalBlue] (c) -- ++ (30:1) node[right] {\footnotesize$2$};
\draw[Maroon] (b) -- ++ (-150:1) node[left,yshift=2]{\footnotesize$3$};
\draw[RoyalBlue] (c) -- ++ (-30:1) node[right] {\footnotesize$1$};
\draw[RoyalBlue] (a) -- (-150:1) node[left] {\footnotesize$4$};
\draw[RoyalBlue] (a) -- (150:1) node[left] {\footnotesize$5$};
\fill[black,thick] (a) circle (0.07);
\fill[black,thick] (b) circle (0.07);
\fill[black,thick] (c) circle (0.07);
\end{tikzpicture}
\end{gathered}\,,
\end{equation}
we can write the contribution to the observable $\Exp_3$ from this diagram in the narrow-width approximation as
\begin{subequations}
\begin{align}
\Exp_3 & =
\begin{tikzpicture}[baseline= {($(current bounding box.base)+(10pt,-2pt)$)},line width=1, scale=0.7,yscale=-1]
\begin{scope}
\coordinate (a) at (0,0) ;
\coordinate (b) at (1,0) ;
\coordinate (c) at ($(b)+(0:1)$);
\draw[] (a) -- (b);
\draw[] (b) -- (c);
\draw[RoyalBlue] (c) -- ++ (30:1) node[right] {\footnotesize$2$};
\draw[Maroon] (b) -- ++ (-150:1) node[left,yshift=2]{\footnotesize$3$};
\draw[RoyalBlue] (c) -- ++ (-30:1) node[right] {\footnotesize$1$};
\draw[RoyalBlue] (a) -- (-150:1) node[left] {\footnotesize$4$};
\draw[RoyalBlue] (a) -- (150:1) node[left] {\footnotesize$5$};
\fill[black,thick] (a) circle (0.07);
\fill[black,thick] (b) circle (0.07);
\fill[black,thick] (c) circle (0.07);
\end{scope}
\begin{scope}[xshift=110]
    \node[] at (0,0) {$+$};
\end{scope}
\begin{scope}[xshift=160]
\coordinate (a) at (0,0) ;
\coordinate (b) at (1,0) ;
\coordinate (c) at ($(b)+(0:1)$);
\draw[] (a) -- (b);
\draw[] (b) -- (c);
\draw[RoyalBlue] (c) -- ++ (30:1) node[right] {\footnotesize$2$};
\draw[Maroon] (b) -- ++ (-150:1) node[left,yshift=2]{\footnotesize$3$};
\draw[RoyalBlue] (c) -- ++ (-30:1) node[right] {\footnotesize$1$};
\draw[RoyalBlue] (a) -- (-150:1) node[left] {\footnotesize$4$};
\draw[RoyalBlue] (a) -- (150:1) node[left] {\footnotesize$5$};
\draw[orange,dashed] ($(0.5,0)+(0,0.75)$) -- ($(0.5,0)+(0,-0.75)$) ;
\fill[black,thick] (a) circle (0.07);
\fill[black,thick] (b) circle (0.07);
\fill[black,thick] (c) circle (0.07);
\end{scope}
\end{tikzpicture}
\\
& =
\frac{-i g_1 g_2 g_3}{(-s_{45}^2+M^2-i\eps)(-s_{12}+M^2 - i \eps)} + \frac{-g_1 g_2 g_3}{-s_{12}+M^2 - i \eps} 2 \pi \delta_+ (s_{45}-M^2)
\\ & =
\frac{-i g_1 g_2 g_3}{(-s_{45}+M^2+i\eps)(-s_{12}+M^2- i \eps)} 
\label{eq:Exp3res}
\,,
\end{align}
\end{subequations}
where we have defined $s_{ij} = - p_{ij}^2$, and in the last line we used the distributional identity $\frac{1}{x+i\varepsilon} - \frac{1}{x-i\varepsilon} = -2 \pi i \delta(x)$ and the fact that $\delta_+(s_{45} - M^2)=\delta(s_{45} - M^2)$ due to positive energy flowing through the cut.

We end this section by making a remark about a curious feature of \eqref{cut from Im Sigma} and \eqref{eq:Exp3res}:
adding the cut terms to the amplitude only had the effect of changing the $i \varepsilon$ in the $s_{45}$ channel, so the observable $\Exp_3$ could have been obtained from the conventional amplitude by an analytic continuation.
This observation is not a coincidence. It turns out that this is a simple instance of a much richer story: inclusive observables are actually related to amplitudes and complex-conjugated amplitudes via analytic continuations in the kinematic invariants. We explore this correspondence in \cite{crossing}.
Physically, the reversal of $i\varepsilon$ in that channel makes sense because particles $4$ and $5$ are absorbed in the past and so must be propagated backward, unlike in a conventional scattering amplitude.

\section{Reduction formulas for out-of-time-order correlators}
\label{sec:localFields}

In local quantum field theory, \emph{reduction formulas} express scattering amplitudes in terms of the on-shell limit of amputated Green's functions.  They are useful both for proving abstract properties of the amplitudes as well as for practical calculations, based for example on perturbation theory.  In this section, we explain how applying the LSZ procedure to generic out-of-time-ordered correlators leads to the asymptotic observables from Sec.~\ref{sec:asympt_measurements}.  As a simple test, the Feynman rules obtained by applying the \emph{Schwinger-Keldysh formalism} to these correlators will be found to coincide with those just obtained by blob-matching.
\subsection{Out-of-time-order correlators and Schwinger-Keldysh formalism}

Let us first motivate the set of out-of-time ordered correlators we would like to reduce.
They involve products of \emph{time-ordered products} $\mathcal{T}$ of the form
\be
 \< 0| \cT\{ \phi_1\cdots \phi_{i_1}\} \cT\{\phi_{i_1+1}\cdots \phi_{i_2}\} \cdots
 \cT\{\phi_{i_{n-1}+1}\cdots \phi_{i_n}\} |0\>\,. \label{many T products}
\ee
This type of correlators has been introduced in \cite{ruelle1961,araki1960properties}.
Linear combinations of them contain other familiar objects such as \emph{anti-time-ordered products} $\cTbar\{\cdots\}$ or \emph{retarded products} $\cR_i\{\cdots\}$, as can be verified by formally manipulating step functions. For example, at two points one has
\begin{subequations}
\begin{align}
 \bar{\cT}\{\phi_1\phi_2\} &\equiv \phi_1 \phi_2\, \theta(x_1^0 < x_2^0) + \phi_2 \phi_1\, \theta(x_2^0 < x_1^0) \\
 &= -\cT\{\phi_1\phi_2\} + \cT\{\phi_1\}\cT\{\phi_2\} + \cT\{\phi_2\}\cT\{\phi_1\}\,,
 \end{align}
 \end{subequations}
 and 
 \begin{subequations} \label{ret versus T}
 \begin{align}
 \cR_1\{\phi_1\phi_2\} &\equiv [\phi_1,\phi_2]\,\theta(x_1^0>x_2^0)\label{eq:RPDef}\\
 &= \cT\{\phi_1\phi_2\} - \cT\{\phi_2\}\cT\{\phi_1\}\,,
\end{align}
\end{subequations}
where the subscript on $\mathcal{R}$ indicates the (necessarily future-most) field being measured. Such relations generalize to any number of points and are invertible (this should be clear below after we introduce the largest time equation \eqref{largest time equation}).
Thus, replacing $\cT$ by $\bar{\cT}$ or $\cR$ in \eqref{many T products} would generate the same linear span of correlators.  

In order to better understand these objects and their calculation, it proves useful to turn to a path integral representation. The \emph{Schwinger-Keldysh formalism} provides an insightful perspective in these regards. The idea behind it is quite simple: if the usual path integral computes time-ordered products, anti-time-ordered products can be computed by a path integral running backward in time. This approach allows us to write down any combination of anti- and time-ordered products by linking the corresponding Lorentzian (real) time axes by infinitesimal Euclidean (imaginary) time shifts. Ultimately, this results in a path integral featuring numerous time-folds oscillating between the past and the future.

To illustrate this in the context of correlators, we can consider the following seven-field correlator represented by a path integral over a Schwinger-Keldysh contour with three time-folds (I, II, and III):
\begin{align}
   \<0|\, \cT\{FG\}\, \overline{\cT}\{ CDE\}\, \cT\{AB\}\,|0\>
\equiv \<0| \cC\{ A^{\rone}B^{\rone} C^{\rtwo}D^{\rtwo}E^{\rtwo} F^{\rthree}G^{\rthree}\}|0\> 
\nonumber\\ = \adjustbox{valign=c}{\input{tikz/SKcontour}}.
\label{example 3fold}
\end{align}
The \emph{contour-ordering symbol} $\cC$ exemplified here is a natural generalization of the time-ordering symbol. The superscript on an operator labels on which time-fold it is inserted. More generally, one can view the fields either as a single function on the $k$-fold contour or, equivalently, as $k$ copies of the usual spacetime fields, with endpoint identifications (i.e., the boundary condition $\lim_{t\to+\infty}\phi^\rone-\phi^\rtwo=0$).
The total action includes the contribution from past-directed branches with a minus sign (in our convention, the even ones: $\rtwo$, $\mathrm{IV}$, \ldots) since $\d t<0$, such that $\e^{iS_{\mathcal{C}}}=\e^{iS^{\rone}-iS^{\rtwo}+iS^{\rthree}-iS^{\mathrm{IV}}+\ldots}$.

Historically, the contour considered by Schwinger and by Keldysh contained two time-folds. It is commonly used in finite temperature field theory, where the two endpoints are identified following an imaginary excursion of $-i\beta$, rather than extending to $\pm i\infty$, in order to compute averages in the Gibbs ensembles $\e^{-\beta H}$, see for example \cite{Das:1997gg}.  At zero-temperature, the formalism simplifies and naturally reproduces Cutkosky-like cutting rules \cite{Meltzer:2020qbr}, as will become clear in the calculations below.
Multiple time-fold contours have been used to calculate out-of-time-ordered correlators and described more generally in \cite{Stanford:2015owe,Haehl:2017qfl}, and were used to derive certain finite-temperature cutting rules in \cite{Caron-Huot:2007zhp}.

Based on the comments below \eqref{ret versus T}, the set of operators spanned by multi-time-folds Schwinger-Keldysh correlators is equivalent to the set identified in \eqref{many T products}.

\subsection{Reduction formulas}

We will now extend the familiar LSZ reduction formula to the correlators just introduced; to our knowledge, this construction is novel. It is useful to introduce ``currents'' which abstract the amputation procedure:
\be\label{eq:A1}
 j(x) = i(-\partial_x^2 + m^2)\phi(x)\,,
\ee
where $\phi(x)$ is a local operator.
Note that the parenthesis is simply $(p^2+m^2)$ in Fourier space. To avoid clutter, we focus on a real scalar theory with a single field $\phi$, which is normalized in such a way that its two-point function features a canonically normalized pole at $p^2+m^2=0$.

\begin{figure}
    \centering
     \adjustbox{valign=c}{\input{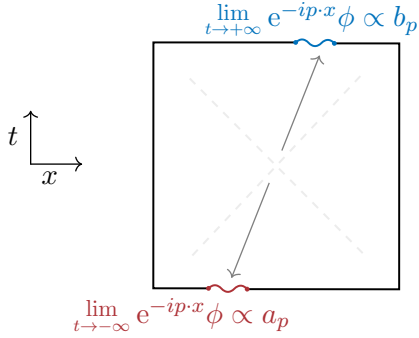}\qquad\qquad}
    \caption{Asymptotic measurements made in the far past and future ($a$ and $b$, respectively) and how they follow from the constructive interference of $\e^{-ip{\cdot}x}\phi$ at late times along the trajectory of particles;
    the product oscillates rapidly along other directions. The dotted lines are meant to represent light-cone axes centered, say, around a collision point in the bulk.}
    \label{fig:jAsympt}
\end{figure}

To derive reduction formulas, one exploits that the on-shell limit of the current is a total derivative, namely
\be \label{total derivative}
j(p)\equiv\int \text{d}^\D x\, \e^{-ip{\cdot} x} j(x)
\xrightarrow{\stackrel{\text{on-shell:}}{p^2\to -m^2}}  \int \text{d}^\D x \frac{\partial}{\partial x^\mu}
\left[ \e^{-ip{\cdot} x}\ (-i\partial^\mu_x+p^\mu) \phi(x)\right]\,.
\ee
Note that this is simply an identity about the Fourier transform of a distribution and equations of motion were not used.
Dynamics enters the picture through the assumption that the product $\e^{-ip{\cdot}x}\phi(x)$
is rapidly oscillatory as $x$ approaches infinity \emph{except} potentially along the paths $x^\mu \propto \pm p^\mu$ of particles going to infinity.
Thus, after integrating against any smooth test function of the (on-shell) momentum $p$,
the on-shell currents reduce to surface terms along the direction of particles, where constructive interference may occur (see Fig. \ref{fig:jAsympt}).
  It is natural to interpret these surface terms as \emph{defining} the asymptotic creation and annihilation operators postulated in Sec.~\ref{sec:asympt_measurements}:
\be \label{j from a and b}
\lim_{p^2\to -m^2} j(p) \equiv \left\{\begin{array}{ll}
 \adag_{-p} - \bdag_{-p} \qquad & {\rm if }\quad p^0<0\quad {\rm (incoming)}\,,\\
 b_p - a_p \qquad & {\rm if }\quad p^0>0\quad {\rm (outgoing)}\,.
 \end{array}\right.
\ee
The above argument is familiar from textbook derivations of the LSZ reduction formula (see for example \cite{Srednicki:2007qs}).
Textbooks, as well as the original LSZ paper \cite{Lehmann:1954rq},
typically discard the $a_p$ and $\bdag_{-p}$ terms because they vanish for vacuum time-ordered correlators (i.e., fields in the far past only have negative frequency components). Our main claim is that keeping these extra terms suffices to extend the LSZ formula to the products \eqref{many T products}.
 
In \eqref{j from a and b}, the positive signs for $a^\dag_{-p}$ and $b_p$ match the standard textbook conventions, while the relative minus signs for the other terms are due to evaluating the total derivative \eqref{total derivative} at the other boundary.

The understanding of $a$ and $\adag$ operators as surface terms at past infinity, and of $b$ and $\bdag$ as surface terms at future infinity, lets us unambiguously define the action of the time-ordering symbol on them: $\cT$ simply moves all $\bdag$ and $b$'s to the left of $\adag$ and $a$'s.
Thus, for example, if particles 1 and 2 are both incoming,
we have (once on-shell)\footnote{
It is important that $\cT$ is defined from its operation on the underlying fields $\phi$, such that derivatives can act on $\theta$-functions as well: $\cT\{j(x)\cdots\} \equiv i(-\partial_x^2+m^2) \cT\{\phi(x)\cdots\}$.  This ensures that the total-derivative argument \eqref{total derivative} continues to hold for $\cT$ products. Physically, allowing the derivatives to act on $\theta$ is necessary to correctly account for interaction vertices at which two or more external legs meet; see \cite{Henning:2022xlj} for a recent discussion.}
\be \label{time ordered j}
\cT\{ j(-p_1)j(-p_2)\} = \cT\{(\adag_1-\bdag_1) (\adag_2-\bdag_2)\} =(\adag_1\adag_2) +(\bdag_1\bdag_2) -\bdag_2 \adag_1 - \bdag_1 \adag_2\,,
\ee
where we have grouped factors which commute with each other into parentheses.

Clearly, the extension of \eqref{time ordered j} to $n$-point contains $2^n$ terms.
We will see that in many situations, most terms vanish.
As a first simple example, consider a vacuum expectation value with $n=4$ points. Taking the pair $p_1,p_2$ to be incoming and the pair $p_3,p_4$ to be outgoing, and all momenta to be non-forward, we have
\begin{subequations}
\begin{align} \label{LSZ 4pt}
 \<0| \cT\left\{j(p_4)j(p_3)j(-p_2)j(-p_1)\right\}|0\>
 &\equiv
 \<0|\, \cT\{ (b_4-a_4)(b_3-a_3)(\adag_2-\bdag_2)(\adag_1-\bdag_1)\}\, |0\>
\\ & = \<0|\, b_3b_4\, \adag_1\adag_2\, |0\> = i\cM_{34\ot12}\bdelta_{4}\,.
\end{align}
\end{subequations}
This is, of course, an instance of the textbook \emph{LSZ reduction formula}. All terms with $a_i$ canceled because the $\cT$ operation moves them to the right where they annihilate the vacuum (up to commutators with other $\adag$'s, which vanish by assumption). Similarly, any $\bdag$ gets moved to the left, where it annihilates the vacuum. Using the same argument, one reproduces the general LSZ formula relating
$n\ot m$ scattering amplitudes to time-ordered correlators of currents.

In what follows, we will abbreviate products of currents as in \eqref{LSZ 4pt} as $\cT\{j_1j_2j_3j_4\}$. Below, we will discard forward terms and will return to them in Sec.~\ref{sec:impulse}.

\subsection{Example reductions, discontinuity formulas, and Steinmann relations}

It is instructive to consider other reduction formulas for $34\ot 12$ kinematics.
Using \eqref{eq:ab-aa-bb}, the following simplification
\begin{subequations}\begin{align}
\cT\{ j_1j_2\}\,|0\> &=
\left(\adag_1\adag_2-\bdag_1 \bdag_2\right)|0\>
\label{jj reduction1}\\&=(\mathbb{1}-S^\dag)|12\>\,, \label{jj reduction2}
\end{align}\end{subequations}
when \eqref{time ordered j} acts on the vacuum will be useful.
The factor $(\mathbb{1}-S^\dag)$ has a simple interpretation: it picks out the interacting part of $S$ and ensures that the result vanishes if the particles do not interact. Indeed, the product on the left manifestly vanishes in free theory by equations of motion.
The above identity can be applied in any state. For example,
\begin{subequations}
\begin{align}
 {}_{\rm out}\<34|\, \cT\{ j_1j_2\}\,|0\>&=
{}_{\rm out}\<34|(\mathbb{1}-S^\dag)|12\> 
\\ &= \<34|S(\mathbb{1}-S^\dag)|12\>
= i{\cM}_{34\ot 12}\bdelta_{4}\,,
\end{align}
\end{subequations}
where in passing to the second line we again dropped forward terms.

\paragraph{Discontinuity formulas}

Reduction formulas related to cutting rules can be naturally obtained by inserting a complete basis of states $|\Psi\>$ in the field theory. For example,
\be
\<0|\,\cTbar\{j_3j_4\}\,\cT\{j_1j_2\}\,|0\>
= \sumint_\Psi \<0| \,\cTbar\{j_3j_4\}\,|\Psi\>\,\<\Psi|\,\cT\{j_1j_2\}\,|0\>\,. \label{unitarity off shell}
\ee
Note that one could replace $\cTbar$ by $-\cT$ in the left factor without changing the present discussion in the context of $2\ot 2$ scattering (though $\cTbar$ would generalize more naturally to higher multiplicity).
In order to clarify this connection, we first apply \eqref{jj reduction1} to the left-hand side of \eqref{unitarity off shell} and obtain
\begin{subequations}
\begin{align} 
\<0|\,\cTbar\{j_3j_4\}\,\cT\{j_1j_2\}\,|0\>
&=
\<0| \big(a_4 a_3 - b_3 b_4\big)\big(\adag_1\adag_2-\bdag_1 \bdag_2\big)|0\>
\\
&= -\<0|\, b_3b_4\, \adag_1\adag_2\,|0\>-\<0|\,a_3a_4\,\bdag_1\bdag_2\,|0\> \\
&= -i\cM_{34\ot12}\bdelta_{4} + i\cM^\dag_{34\ot12}\bdelta_{4}\,,\label{2->2 cut example}
\end{align}
\end{subequations}
where, in the second line, we have discarded forward terms.  Next, applying \eqref{jj reduction2} to the right-hand side of \eqref{unitarity off shell} gives
\be \label{2->2 cut example B}
\<0|\,\cTbar\{j_3j_4\}\,\cT\{j_1j_2\}\,|0\>
= \sumint_\Psi\,\<34|(\mathbb{1} - S)|\Psi\>\,\<\Psi|(\mathbb{1}-S^\dag)|12\>\,.
\ee
Equating \eqref{2->2 cut example} and \eqref{2->2 cut example B} relates amplitudes to products of amplitudes.
Of course, a similar relation could have been obtained by inserting a complete basis of asymptotic states directly in the first line of \eqref{2->2 cut example}, giving
\begin{subequations}
\begin{align}
    \<0|\,\cTbar\{j_3j_4\}\,\cT\{j_1j_2\}\,|0\>
  &= \sumint_X\, \<34|(\mathbb{1} - S)|X\>\,\<X|(\mathbb{1}-S^\dag)|12\>
\\&= \sumint_X\, \cM_{34\ot X}^\dag\cM_{X\ot 12}\bdelta_{4}\,.\label{discontinuity formula simple}
\end{align}
\end{subequations}

Let us compare the preceding three formulas.
On one hand, the agreement between \eqref{2->2 cut example} and \eqref{discontinuity formula simple}  is precisely the unitarity relation encountered earlier in \eqref{eq:unitarityBlobs}.
Note that it coincides with the field theory cutting rule \eqref{2->2 cut example B} provided  that the field theory Hilbert space is spanned by asymptotic states. The product \eqref{unitarity off shell} was in some sense designed such that inserting a complete basis of asymptotic states in the first line of \eqref{2->2 cut example} would yield a product of \emph{connected} amplitudes.
On the other hand, the agreement between \eqref{2->2 cut example B} and 
\eqref{discontinuity formula simple} is less trivial. It amounts to an equivalence between
field theory and asymptotic Hilbert spaces, known as \emph{asymptotic completeness}.
Concretely, this property ensures that
\be
 \sumint_\Psi |\Psi \>\<\Psi| \,=\, \sumint_X |X\>_{\rm in}\,{}_{\rm in}\<X| \,=\, \sumint_X |X\>_{\rm out}\,{}_{\rm out}\<X|\,.
\ee
As will become clear below, this is trivially manifest in perturbation theory. This is because the Feynman rules for computing \eqref{discontinuity formula simple} in field theory involve the same cut propagators as those in $\sumint_X$ (defined in \eqref{eq:inclsum}).
In practice, one way that asymptotic completeness could fail is if one ``forgets'' to account for some bound state in the asymptotic Hilbert space.

\paragraph{Retarded products and their role in axiomatic field theory}

Focusing again on $34\ot 12$ kinematics, amplitudes can be written in various ways using retarded products. The following useful identity is a consequence of the definition in \eqref{eq:RPDef} combined with \eqref{j from a and b}:
\be \label{R jj}
\cR_2\{j_2 j_1\} =\cR_2\{(\adag_2{-}\bdag_2)\,(\adag_1{-}\bdag_1)\} = -[\bdag_2,\adag_1]\,.
\ee
The terms involving $\bdag_1$  vanish because the second operator must be in the past lightcone of $j_2$. Consequently, particle 1 can never reach future infinity. One can additionally check that the same result is also obtained by combining \eqref{ret versus T} with \eqref{jj reduction1}.\footnote{The identity \eqref{R jj} is a special case of Eq.~(37) in a second LSZ paper \cite{Lehmann:1957zz}.}
Putting everything together, we arrive at the reduction formula
\be \label{2to2 R}
{}_{\rm out}\<43| \cR_2\{j_2 j_1\}|0\> =
{}_{\rm out}\<43| \big( \adag_1\adag_2-\bdag_1\bdag_2\big)|0\>
= {}_{\rm out}\<43| 12\>_{\rm in}\,,
\ee
where in the first transition we used \eqref{eq:ab-aa-bb} and in the second one unitarity and non-forward kinematics.

The fact that the $\cR$-product is supported over $x_2^0>x_1^0$ guarantees that its Fourier transform is analytic in a certain domain, in which equal positive timelike imaginary parts are added to $p_2^\mu$ and $-p_1^\mu$. Naturally, one could also represent the amplitude using an advanced product in \eqref{2to2 R}. The equality of these two representations suggests analyticity in a larger domain, in which the timelike requirement can be relaxed.

This line of thought is central to the axiomatic proofs that scattering amplitudes are analytic in, for example, a neighborhood of the mass shell \cite{Bros:1964iho, Bros:1965kbd,Sommer:1970mr}.
The argument generalizes to $n\ot m$ scattering and has been used to show that amplitudes near the mass shell are generally equal to a finite sum of analytic functions \cite{Bros:1972jh} (with a single function sufficing in $2 \ot 2$ cases). At four points, \eqref{2to2 R} also implies analyticity in the so-called small \emph{Lehmann ellipse} in the momentum transfer plane  \cite[Sec. 4]{Sommer:1970mr}.

The existence of multiple reduction formulas for the same quantity has important consequences.
Here, our aim is very humble: we simply wish to emphasize that this situation is natural given the classification presented in Sec.~\ref{sec:classification}.  In particular, there exist only two distinct asymptotic observables at four points: $\cM$ and $\cM^\dag$, but a significantly greater number of correlation functions enjoying distinct analyticity properties can be considered.

Another reduction formula can be obtained similarly to \eqref{R jj} by inserting a retarded product between
one-particle states (again in $34\ot12$ kinematics), $\cR_3\{j_3 j_2\} = [b_3, a_2^\dag]$, we get:
\be
 {}_{\rm out}\<4| \cR_3 \{ j_3j_2 \} |1\>_{\rm in} =
 {}_{\rm out}\<4| b_3\adag_2|1\>_{\rm in} = 
{}_{\rm out}\<34|12\>_{\rm in}\,.
\ee
This formula is relevant to analyticity at fixed-$t$ and crossing symmetry through the upper-half $s$-plane \cite{Bros:1965kbd}.  When the energies of particles $2$ and $3$ are flipped, the same correlator reduces to $\cM^\dag$ instead of $\cM$ in accordance with the left endpoint of Fig.~\ref{fig:introcross}.

\paragraph{A five-point discontinuity and Steinmann-like relation}

Let us now describe a $345\ot 12$ reduction formula, which combines the last two examples,\footnote{
This could be written using fields only with the retarded product \eqref{ret3pt}:
$\cR_3\{j_3j_2\}\,|1\>=\cR_3\{j_1j_2j_3\}|0\>$.
} namely
\be\begin{aligned} \label{Steinmann reductions}
    -\<0|\, \cR_5\{j_5j_4\}\, \cR_3\{j_3j_2\}\,|1\>
    &= \vac{[b_5, a_4][b_3, a_2^\dag] a_1^\dag} =  \<0|\big(a_5 a_4-b_5 b_4\big)\, b_3\,\adag_2\adag_1|0 \>
    \\ &=
    \raisebox{-3em}{
\begin{tikzpicture}[line width=1]
\draw[line width = 1] (0,0.3) -- (-1.2,0.3) node[left] {\small$4$};
\draw[line width = 1] (0,-0.3) -- (-1.2,-0.3) node[left] {\small$5$};
\draw[line width = 1] (2,0.3) -- (3.2,0.3) node[right] {\small$2$};
\draw[line width = 1] (2,-0.3) -- (3.2,-0.3) node[right] {\small$1$};
\draw[line width = 1] (2,0) -- (1,-1) node[left] {\small$3$};
\filldraw[fill=gray!30, very thick](0,-0.3) rectangle (2,0.3);
\draw[] (1,0) node {$X$};
\filldraw[fill=gray!5, line width=1.2](0,0) circle (0.6) node {$-i\mathcal{M}^\dag$};
\filldraw[fill=gray!5, line width=1.2](2,0) circle (0.6) node {$i\mathcal{M}$};
\draw[dashed,Orange] (1,0.7) -- (1,-1.2);
\end{tikzpicture}}
    \equiv {\rm Cut}_{45}\bdelta_5\,.
\end{aligned}
\ee
This representation of the unitarity cut in terms of retarded commutators has an interesting feature.
To see it, consider the joint translation of the coordinates $x_3$ and $x_4$:
\be
 x_3 \mapsto x_3 + \xi \quad  \text{and} \quad x_4 \mapsto x_4  + \xi\,.
\ee
Upon performing the Fourier transforms in \eqref{Steinmann reductions}, we expect physically to find a singularity in the $s_{34}$ channel if and only if the integral over $\xi$ is allowed to reach infinity, either in the past or in the future
(since singularities in momentum space originate from the large-distance or large-time behavior of
correlators).  However, the $\xi$-range in \eqref{Steinmann reductions} is obstructed in the future by the condition that $x_4$ is in the past lightcone of $x_5$,
and it is obstructed in the past by the condition that $x_3$ be in the future lightcone of $x_2$, and therefore we do not expect any singularity in that channel.
This suggests that
\be
 {\rm Disc}_{s_{34}}\,{\rm Cut}_{45} = 0\,. \label{Steinmann schematic}
\ee
This is in essence the so-called \emph{Steinmann relations} \cite{Steinmann1960a,Steinmann1960b}:
``sequential discontinuities of discontinuities vanish in overlapping channels''
(see \cite{Bartels:2008ce,Caron-Huot:2016owq} for examples,
and \cite[App.~A]{Bartels:2014jya} for further historical references).
It would be interesting to more precisely connect \eqref{Steinmann schematic} with this statement. Notably, the relation between discontinuities and cuts should be clarified, as well as the precise implications from the support of Fourier transforms.
Massless thresholds in two-particle channels have proven to be particularly confusing in this context
(see \cite{Brower:2008nm}) and were avoided altogether in \cite{Caron-Huot:2016owq}.

By replacing $\cR_5$ by $\cR_4$ in \eqref{Steinmann reductions} one could similarly argue that ${\rm Cut}_{45}$ has no discontinuity in the $s_{35}$ channel.
As in the preceding examples, the existence of multiple reduction formulas for the same amplitude seems key to understanding its full analyticity properties.

\subsection{Retarded products and a reduction formula for waveforms}

A natural choice of operator ordering can typically be determined from the physical question being asked.  Let's consider the problem of determining the expectation value of a certain field (e.g., electromagnetic or gravitational) following a collision, regardless of any other properties of the final state, as discussed in the introduction.  This question will lead us to a reduction formula for the expectation value \eqref{inclusive 5}.

Since the setup is analogous to that covered by \emph{nonlinear response theory}, we expect the solution to involve retarded products. Let's briefly review the general setup of nonlinear response theory and the definition of $\cR$-products, following \cite{Chou:1984es}. We begin with a system in a predefined state $|\Psi\>$, perturb its Hamiltonian $H$ by coupling a field $\phi$ to a source $f(x)$ according to
\begin{equation}
    \delta H=\int f(x)\phi(x)\,,
\end{equation}
and subsequently measure an expectation value of say $\phi(x_0)$ at late time.
Formally, this involves computing
\be\label{eq:phiExp}
\<\phi_0\>_{H+\delta H}=
{}_{\rm in}\<\Psi|\,\cTbar \{\e^{i\int \d t (H+\delta H)} \}\,\phi_0\,
 \cT \{\e^{-i\int \d t (H+\delta H)}\}|\Psi\>_{\rm in}\,.
\ee
The task is simplified once one notices \eqref{eq:phiExp} can be expressed as a path integral over a Schwinger-Keldysh two-fold, where the perturbation is inserted on both branches with opposite signs. The response $\<\phi_0\>_{H+\delta H}$ at the $(n-1)^\text{th}$ power in $\delta H$ is then determined by the $n$-point correlators of field differences, namely
\be\label{eq:npointRP}
\begin{split}
    \cR_0\{\phi_0 \phi_1\cdots \phi_{n-1} \} &\equiv
 \cC\{ \phi_0^{\rone} \phi_1^{\rdiff}\cdots \phi_{n-1}^{\rdiff}\} \\&= \adjustbox{valign=c}{\input{tikz/rProdSK}}\,,
\end{split}
\ee
where $\phi_k^{\rdiff}(x)\equiv \phi_k^\rone(x)-\phi_k^\rtwo(x)$ is inserted on the two folds.
The same perturbation is inserted on the two branches with a relative minus sign because the same $\delta H$ is used to evolve both the bra and the ket.  Expanding the $\cC$ products, one finds, e.g., for $n=2,3$:
\begin{subequations}\begin{align}
\cR_2 \{\phi_1 \phi_2\} &= [\phi_2,\phi_1]\theta(x_2^0{-}x_1^0)\,,
\\
\cR_3 \{\phi_1\phi_2\phi_3\} &= [[\phi_3,\phi_2],\phi_1] \theta(x_3^0{-}x_2^0)\theta(x_2^0{-}x_1^0) + (1{\leftrightarrow}2)\,. \label{ret3pt}
\end{align}\end{subequations}
The explicit formula for the $n$-point \emph{retarded product} defined in \eqref{eq:npointRP} involves a similar nested commutator summed over $(n{-}1)!$ permutations of the sources.
The $\mathcal{R}$-product is Lorentz-invariant and vanishes
unless all sources $\phi_1\cdots \phi_n$ are in the past lightcone of the measurement $\phi_0$.%
\footnote{The $n$-point retarded products seem to have been first defined in a high-energy context in \cite{Lehmann:1957zz} following earlier work by Polkinghorne \cite{PolkinhorneDisp}. Note that the ``generalized'' retarded products of \cite{ruelle1961,araki1960properties} are more general: they include products of retarded products and span the same set as \eqref{many T products}.}

\begin{figure}
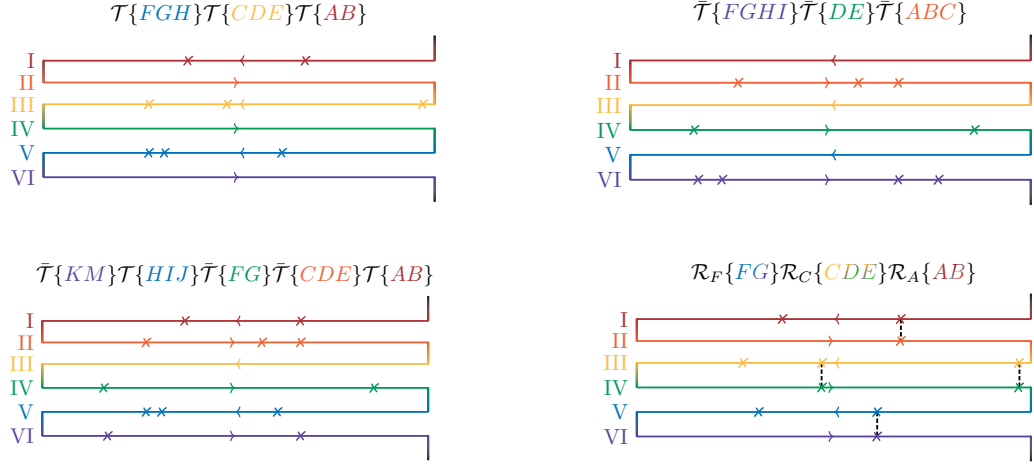

\begin{minipage}{0.49\textwidth}
\begin{tcolorbox}[colback=white, colframe=white, rounded corners]
\begin{equation*}
     \hspace{-0.7cm}\adjustbox{valign=c, width=\textwidth}{\input{tikz/3foldT}}
\end{equation*}
\end{tcolorbox}
\end{minipage}\hfill
\begin{minipage}{0.50\textwidth}
\begin{tcolorbox}[colback=white, colframe=white, rounded corners]
\begin{equation*}
     \hspace{-0.7cm}\adjustbox{valign=c, width=\textwidth}{\input{tikz/3foldTb}}
\end{equation*}
\end{tcolorbox}
\end{minipage}
\vspace{10pt}
\begin{minipage}{0.49\textwidth}
\begin{tcolorbox}[colback=white, colframe=white, rounded corners]
\begin{equation*}
     \hspace{-0.7cm}\adjustbox{valign=c, width=\textwidth}{\input{tikz/3foldTandTb}}
\end{equation*}
\end{tcolorbox}
\end{minipage}\hfill
\begin{minipage}{0.50\textwidth}
\begin{tcolorbox}[colback=white, colframe=white, rounded corners]
\begin{equation*}
\hspace{-0.7cm}
\adjustbox{valign=c, width=\textwidth}{\input{tikz/3foldR}}
\end{equation*}
\end{tcolorbox}
\end{minipage}
\caption{Various types of time-ordering products discussed in the literature and their relations to Schwinger-Keldysh time-folds. For illustrative purposes, each panel only shows a three time-fold contour, with the generalization to an arbitrary number of time-folds following the same obvious patterns. \textbf{Top left:} Product of time-ordered products $\mathcal{T}$. \textbf{Top right:} Product of anti-time-ordered products $\bar{\mathcal{T}}$. \textbf{Bottom left:} Mixed product of time- and anti-time ordered products. \textbf{Bottom right:} Product of retarded products $\mathcal{R}$. As before, time is going from right to left in each panel.
}
\label{fig:bases}
\end{figure}

In the definition of the retarded product, the measured field
could be equivalently taken to be on the second fold, $\phi_0^{\rtwo}$, thanks to the boundary condition mentioned below \eqref{example 3fold}:
\be \label{largest time equation}
  \<0|\, \cC\{ \phi_1^{\rdiff}\phi_2^{\rdiff}\cdots \phi_n^{\rdiff}\}\,|0\>=0\,.
\ee
This is known as the \emph{largest time equation} and is closely related to unitarity \cite{tHooft:1973wag}.
In fact, repeated use of this equation gives a simple way of deriving the relations between the different bases mentioned below \eqref{ret versus T}
(i.e., how to rewrite products of $\cT$ products as 
products of $\cR$ products, or of $\cTbar$ products etc.) with different number of time-folds, whose Schwinger-Keldysh representations are summarized in Fig.~\ref{fig:bases}.

As hinted above, the expectation value of a classical field $\phi(k)$ or on-shell current $j(k)$ in a scattering state $12$ can be formulated in a similar manner in terms  of a five-point process $1'2'k\ot 12$.
Starting with a vacuum state, we turn on a perturbation capable of creating or absorbing particles to, later on, measure $j(k)$.  In light of this, we wish to verify if the following is correct:
\be\begin{aligned}
{\rm Exp}_k &\stackrel{?}{=}
\<0|\,\cR_k\{j_k j_{1'}j_{2'}j_1j_2\}\,|0\>
\\ &= \<0|\, \cC\{\,j^\rone_k\,
j^{\rdiff}_{1'}\,j^{\rdiff}_{2'}\,
j^{\rdiff}_1\,j^{\rdiff}_2\,\}\,|0\>\,. \label{tentative Expk}
\end{aligned}\ee
To perform the on-shell reduction of the right-hand side,
we apply \eqref{j from a and b} to each branch.
For example, when $p^0<0$ and we go on-shell, it implies that
\be
\lim_{p^2\to-m^2} j^{\rdiff}_p \equiv (\adag_p{}^\rone-\bdag_p{}^\rone)-(\adag_p{}^\rtwo-\bdag_p{}^\rtwo)
= \adag_p{}^\rone -\adag_p{}^\rtwo \qquad (p^0<0)\,.
\ee
Here, we used the boundary condition $\lim_{t\to+\infty}(\phi^\rone-\phi^\rtwo)=0$ to identify  $\bdag_p{}^\rone=\bdag_p{}^\rtwo$.
Consequently, only surface terms at past infinity contribute.
Similarly, for $p^0>0$, $j^{\rdiff}(p)$ reduces to $(a_p^\rtwo-a_p^\rone)$.
Thus,
\be\mbox{RHS of ~\eqref{tentative Expk}}
= \<0| \cC\{
\,(b_k^\rone{-}a_k^\rone)\,
(a_{1'}^\rtwo{-}a_{1'}^\rone)\,(a_{2'}^\rtwo{-}a_{2'}^\rone)\,
(\adag_{1}{}^{\rone}{-}\adag_{1}{}^{\rtwo})\,
(\adag_{2}{}^{\rone}{-}\adag_{2}{}^{\rtwo})\,
\}|0\>\,,\label{eq:rhsExp}
\ee
where the $\cC$ operation simply orders type-$\rtwo$ operators to the left of type-$\rone$ ones.
As always, we work under the assumption that external momenta are not exactly forward, so we can discard commutators between $a$'s and $\adag$'s.
Hence, all but one term in \eqref{eq:rhsExp} are trivial: all terms with $a^{\rone}$ vanish since they annihilate the vacuum on the right, and terms with $\adag{}^{\rtwo}$ vanish since they annihilate the vacuum on the left. In summary, we find, on-shell,
\be \label{in-in reduction}
\begin{split}
\<0|\,\cR_k\{j_k j_{1'}j_{2'}j_1j_2\}\,|0\>
&= \<0| a_{1'}a_{2'} b_k \adag_1 \adag_2|0\> = {}_{\rm in}\<1'2'| b_k|12\>_{\rm in}\,,
\end{split}
\ee
thereby confirming \eqref{tentative Expk}. The right-hand side of this equation precisely corresponds to the momentum space waveform \cite{Cristofoli:2021vyo} mentioned in the introduction,
sometimes also written as $\<1'2'|S^\dag a_k S|12\>$.
Consequently, we have derived an LSZ-like reduction formula which relates the waveform to a fully retarded product.

Before moving on, we stress that, like in previous examples, \eqref{tentative Expk} is just \emph{one of many} possible reduction formulas for the same quantity. In view of the causal properties of the retarded product, this particular representation manifests the fact that the waveform depends only on physics within its past light cone.

\subsection{More reduction formulas: impulse and asymptotic fields}
\label{sec:impulse}

The above generalizes naturally to the expectation value of any Heisenberg operator $\mathcal{O}(x)$ in two-particle states.
As long as momenta are non-forward ($p_i\neq p_{i'}$),
the argument leading to \eqref{in-in reduction} yields a similar reduction formula:
\be
\underbracket[0.4pt]{{}_{\rm in}\<1'2'|\,\mathcal{O}(x)\,|12\>_{\rm in}
 = \<0|\,\cR_{\mathcal{O}}\{
 \mathcal{O}(x)j_{1'}j_{2'}j_1j_2\}\,|0\>}_{\mbox{\small (for non-forward momenta)}}\,. \label{reduction general O}
\ee
Here we discuss more applications of this formula and elaborate on the role of forward contributions in different physical settings. This subsection can safely be skipped on a first reading.

In the above retarded product (see \eqref{eq:npointRP}), it
is important physically that all four currents be inserted on both Schwinger-Keldysh time-folds. Otherwise, for example if $j_{1'}$ and $j_{2'}$ were inserted only on the second fold, particles $1'$ and $2'$ would be forced to interact directly with each other (as in \eqref{jj reduction2}), which is generally not required by the expectation value on the left-hand side of \eqref{reduction general O}. 

The non-forward condition in \eqref{reduction general O} is required because the operator ordering of terms like $a_{1'}^\rone \adag_{1}{}^{\rone}$ in \eqref{eq:rhsExp} cannot be easily predicted by \eqref{j from a and b}:
there are possible ambiguities supported on forward momenta. On the one hand, 
from the definition of the retarded product, we can see that disconnected diagrams never contribute to the right-hand side of \eqref{reduction general O}. This is because two-point functions like
$\<0| \cC\{ j_{1}^{\rdiff}j_{1'}^{\rdiff}\}|0\>$ vanish by the largest time equation \eqref{largest time equation}.
On the other hand, disconnected diagrams are potentially important on the left-hand side, especially when we explicitly integrate the in-states against a wavefunction $\psi(p_1,p_2)$ and its complex conjugate:
\be \label{in average}
 \< \mathcal{O}(x) \>_\psi \equiv
\int\Bigg[|\Psi|^2\prod_{a} \frac{\d^\D p_{a}}{(2\pi)^{\D-1}}\delta_+(p_a^2+m_a^2)\Bigg]\,
{}_{\rm in}\<1'2'|\,\mathcal{O}(x)\,|12\>_{\rm in}\,,
\ee
where the product runs over $a=1,2,1',2'$ and where $|\Psi|^2\equiv \psi(p_{1'},p_{2'})^*\psi(p_1,p_2)$. The integrals force us to think carefully about terms supported on forward kinematics.
A simple criterion for this expectation value to be expressible in terms of 
the retarded correlator \eqref{reduction general O} is that disconnected terms do not contribute, meaning that the operator is such that averages in one-particle states vanish:
\be
\mbox{\eqref{reduction general O} can be used in \eqref{in average} when and only when}\;
{}_{\rm in}\<i'|\,\mathcal{O}(x)\,|i\>_{\rm in}
=0 \ \forall~i,i'\,.\label{in reduction criterion}
\ee
This is certainly the case for the expectation value of radiation field
with real momenta, i.e., $\cO=b_p$ discussed above in \eqref{in-in reduction},
since an isolated stable particle does not radiate.
In contrast, as we will discuss shortly, position-space potentials generally receive additional ``Coulomb'' contributions.

Now consider the \emph{impulse} or \emph{momentum kick} $\Delta P^\mu(t)$, which measures the change in the momentum of a given particle after a scattering event. To do so, one first imagines that particles are distinguishable such that there exists some
operator $P^\mu_i(t)$ measuring asymptotically the momentum carried by particles of type $i$.  In the far past, we have
\be \label{P inf}
 \lim_{t\to-\infty} P_i^\mu(t)= \int\Bigg[ \frac{\d^\D p}{(2\pi)^{\D-1}}\delta_+(p^2+m_i^2)\, p^\mu\Bigg]\, \adag_{i,p} a_{i,p}\,,
\ee
with a similar formula in terms of $b$'s in the far future. Conceptually, we can think of $P_i^\mu$ as the asymptotic version of the energy-momentum operator, which is quadratic in the fields.
From our perspective, the late-time expectation value of $P_i^\mu$ is an example of a generalized 6-particle amplitude similar to the inclusive cross-section defined in \eqref{6pt N}:
\begin{subequations}
    \begin{align}
         \< P_i^\mu(t{\to}\infty)\>_{\rm \psi} &= \int_{1',2',1,2,p}\text{d}\mu \,
 \<0|\, a_{1'}a_{2'}\,\bdag_{i,p} b_{i,p}\,\adag_1\adag_2\,|0\>\\&=\int_{1',2',1,2,p}\text{d}\mu \, \<1'2'|\,S^\dag \adag_{i,p}a_{i,p}S\,|12\>\,,
 \label{KMOC 6pt}
    \end{align}
\end{subequations}
where the measure $\text{d}\mu$ collects the square-bracket factors in \eqref{in average} and \eqref{P inf}.
Here, however, each $S$ and $S^\dag$ factor can be disconnected when the bra-ket in \eqref{KMOC 6pt} is expanded into disconnected and connected parts as in Sec.~\ref{sec:blobology}.
The case where both factors are disconnected, which represents the forward contribution, can be canceled physically by focusing on the impulse, i.e., taking the expectation value of
\be
\Delta P^\mu_i(t)=P^\mu_i(t) - N_i(t) \<P^\mu_i\>\, ,
\ee
where $N_i$ asymptotically measures
the density of particles of type $i$.  At $t\to+\infty$, this replaces the bra-ket in \eqref{KMOC 6pt} by
\be \label{Delta P}
 \<1'2'| \Delta P_i^\mu|12\>= \< 1'2'| S^\dag P_i^\mu S - P_i^\mu |12\> =
 \<1'2'| S^\dag [P_i^\mu,S]\,|12\>  = \<1'2'|\, [S^\dag, P_i^\mu]S\,|12\>\,,
\ee
with $P_i^\mu$ given in the right-hand side of \eqref{P inf}.
The operator $\Delta P^\mu_i$ satisfies the criterion \eqref{in reduction criterion} and therefore its expectation value can be related to a five-point correlator:
\be \label{impulse reduction}
\< \Delta P^\mu_i \>_\psi =
\int\Bigg[|\Psi|^2\prod_{a} \frac{\d^\D p_{a}}{(2\pi)^{\D-1}}\delta_+(p_a^2+m_a^2)\Bigg] \, {}_{\rm in}\<1'2'|\,\Delta P^\mu_i(t\to\infty)\,|12\>_{\rm in}\,.
\ee
This is a reduction formula for the KMOC impulse in terms of a connected $5$-point retarded correlator, where one operator is quadratic in fields. Note that the integrand is not necessarily connected when viewed as a 6-particle amplitude as in \eqref{KMOC 6pt}, since either $S^\dag$ or $S$ could be disconnected (but not both simultaneously).

The relation between KMOC expectation values and Schwinger-Keldysh path integrals has also been discussed recently in the context of the \emph{worldline formalism} in \cite{Damgaard:2023vnx}.
One potential application of \eqref{impulse reduction} could be to help manifest
simplifications in the classical limit $\hbar\to 0$. 
On one hand, the Schwinger-Keldysh formalism offers a natural way to approach this limit: as $\hbar\to 0$,
the fields on the two time-folds become approximately equal and the difference field becomes a Lagrange multiplier that enforces the classical equations of motion (see for example \cite{Mueller:2002gd,Caron-Huot:2010fvq}).
On the other hand, very efficient ways of taking the classical limit of \eqref{impulse reduction} and related observables have been developed based on the worldline formalism and also the so-called exponential representation, which is found to remove super-classical contributions \cite{Damgaard:2021ipf}:
\be
S=\e^{\frac{i}{\hbar}\hat{N}}\,,
\ee
where $\hat{N}$ is defined by this equation.
When the $\hat{N}$ operator is approximated by its $2\ot 2$ component, substituting it into \eqref{KMOC 6pt} reproduces familiar quantum-mechanical relations between the deflection angle and the derivatives of the phase shift. These relations have been
generalized to include radiative corrections up to fourth Post-Minkowski order \cite{Damgaard:2023ttc} (building on  \cite{Bern:2019nnu,Damour:2020tta,DiVecchia:2021bdo,Bini:2021gat,Bern:2021yeh} and references therein). It could be worthwhile to better understand the $\hat{N}$ operator from the path integral perspective.

\paragraph{Relation between on-shell and asymptotic fields}

Finally, let us comment on the relation between on-shell generalized amplitudes such as \eqref{in-in reduction} and the large-distance limit of dynamical fields, such as the electromagnetic potential or the gravitational metric.  
Let us take for concreteness $\mathcal{O}(x)$ to be a scalar field $\phi(x)$, which couples to massless scalar particles, and focus on the limit where $x$ approaches future null infinity ${\cal I}^+$: $|\vec{x}|\to\infty$ with constant retarded time $x^0-|\vec{x}|$.

Since we start (by assumption) from a state that differs from the vacuum only by the addition of two particles 1 and 2
(let's take them to be massive), the expectation value
$\<\phi(x)\>_\psi$ of the massless field vanishes at past null infinity.
Defining a current $j(x)=-i\partial_x^2\phi(x)$ as in \eqref{eq:A1},
the relation between $\phi$ and the current can thus be inverted to give
\be \< \phi(x) \>_\psi
= \int \d^\D y\, G_R(x{-}y)\, \<j(y)\>_\psi
= \int \frac{\d^\D p}{(2\pi)^\D} \e^{ip\cdot x} G_R(p)\,\<j(p)\>_\psi\,, \label{avg from GR}
\ee
with $G_R(p)=\frac{-i}{p^2-i\varepsilon p^0}$ being the retarded Green's function for a free massless field.
This is a general relation between $\phi(x)$ and $j(x)$ which does not rely on equations of motion for $\phi$ but only on the boundary condition satisfied by the prescribed initial state. At this stage, the momentum in $j(p)$ is still generally off-shell. As $x$ approaches ${\cal I}^+$, we expect on-shell radiation modes to dominate. This can be seen by a complex deformation of energy integrals, which is equivalent to the following identity:
\begin{align}
\<\phi(x)\>_\psi&=
\int \frac{\d^\D p}{(2\pi)^\D}  \e^{ip\cdot x} \left[\big(G_R(p) - G_A(p)\big)+G_A(p)\right]\,\<j(p)\>_\psi
\\ &= \int \frac{\d^\D p}{(2\pi)^{\D-1}}\delta(p^2)\e^{ip{\cdot}x}\,j(p)+
\int \d^\D y\, G_A(x{-}y)\,\<j(y)\>_\psi\,. \label{avg from GR GA}
\end{align}
Here, $G_A(p)$ is the advanced Green's function, which differs from $G_R$ by the sign of $i\varepsilon$, and we have used that $G_R-G_A$ is proportional to an on-shell $\delta$-function.\footnote{Recall that, physically, the retarded Green’s function can be viewed as propagating forward in time, while the advanced Green’s function travels backward in time. Thus, the retarded one originates from a source (retarded) in past time, and the advanced one originates from a source (advanced) in the future.} The first term is the expected on-shell contribution: it features the creation/annihilation operators defined by the reduction formula \eqref{j from a and b}. Thus, \eqref{avg from GR GA} can be written as the familiar mode expansion for a free field, plus a correction term:
\be \label{reduction including Coulomb}
\begin{split}
     \lim_{x\to \cal{I}^+} \<\phi(x)\>_\psi &=
\int \frac{\d^\D p}{(2\pi)^{\D-1}}\delta_+(p^2)\,
\left[ \e^{ip{\cdot}x}\< b_p\>_\psi + \e^{-ip{\cdot}x}\<\bdag_p\>_\psi \right]\\& \qquad\qquad\qquad\qquad   +
\int \d^\D y\, G_A(x{-}y)\,\<j(y)\>_\psi\,.
\end{split}
\ee
Note that we did not assume here that $\phi$ is close to a free field, we only assumed that its correlators feature a massless pole,
and the boundary condition satisfies by the expectation value (which set $\< a_p\>_\psi=0$).
The advanced term accounts by construction for any possible effect due to interactions and will be interpreted shortly. While we discuss only scalar fields here for notational simplicity, a similar decomposition holds for gauge fields or gravity, with the first term then including only on-shell, transverse-traceless polarizations.

\begin{figure}
\centering
\hspace{-1.5cm}
\adjustbox{valign=c}{\input{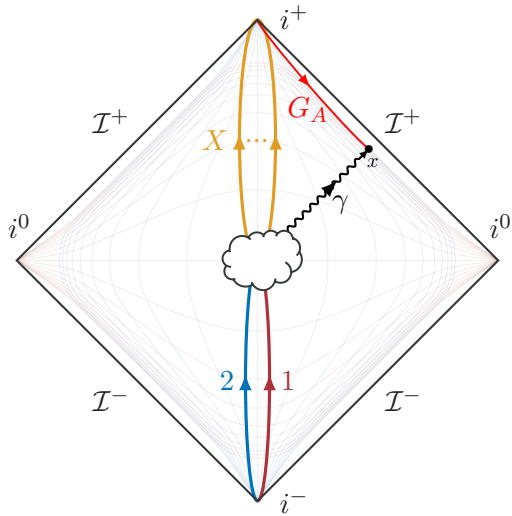}}
    \caption{Penrose diagram describing inclusive field measurements near future null infinity ${\cal I}^+$ after a collision at the origin (white cloud).
    The retarded field at a point $x$ is the sum \eqref{reduction including Coulomb} of the on-shell radiation field plus an \emph{advanced} contribution sensitive to sources in the future lightcone of $x$ and thus also near ${\cal I}^+$.  This includes the ``Coulomb'' field from sources (denoted by $X$ above) that reach timelike infinity $i^+$, which is approximately independent of the retarded time of $x$.}
    \label{fig:penrose}
\end{figure}

Let us interpret the advanced term from~\eqref{reduction including Coulomb}. It is often the case that all sources $j(x)$ are massive particles, which cannot reach future null infinity in finite retarded time.  Then the advanced contribution is independent of retarded time and is simply the sum of the static Coulomb fields $\propto 1/(x^2v_i^2-(v_i{\cdot}x)^2)^{\D-3}$ of each outgoing particle, see Fig.~\ref{fig:penrose}.  In massive QED, this situation is always realized.  More generally, the source $j(x)$ could have support along the trajectory of other radiated particles, leading to nontrivial dependence on retarded time (sourced by the charge of other radiated quanta in massless QED or Yang-Mills theory or the gravitational field surrounding outgoing radiation in general relativity).

The first term of \eqref{reduction including Coulomb} only includes fully connected diagrams (where both particles interact), since an isolated particle cannot radiate. Thus, it is precisely what is computed by the KMOC expectation value and reduction formula \eqref{in-in reduction}, namely:
\be
\< b_p \>_\psi =
\int\Bigg[|\Psi|^2\prod_{a} \frac{\d^\D p_{a}}{(2\pi)^{\D-1}}\delta_+(p_a^2+m_a^2)\Bigg]\,\<0|\cR_p\{j_pj_{1'}j_{2'}j_1j_2\}|0\>\,. \label{reduction bp}
\ee

There are certainly many situations in which the Coulomb/advanced contribution in \eqref{reduction including Coulomb} can be neglected.  For example, when computing local gauge-invariant quantities like the electromagnetic field strength or the linearized gravitational curvature (as relevant for LIGO-type waveform measurements), one finds that the Coulomb field decays faster than radiation by an additional power of $1/|\vec{x}|$, which makes it completely negligible.

Another such situation arises when computing the energy flux of the scalar field along ${\cal I}^+$, $P^\mu_{\rm rad} \propto \int \d u~T^{u\mu}$,
with $u$ being the retarded time and $T^{u\mu}\simeq\partial^u\phi\partial^\mu\phi$ for a scalar.
The derivatives acting on the field then suppress the Coulomb part.  The energy flux is quadratic in the scalar field, but if the source is sufficiently classical we expect a factorization, in which case we can apply \eqref{reduction including Coulomb} and find:
\be
\< j(p_1)j(p_2)\>_\psi = \<j(p_1)\>_\psi \<j(p_2)\>_\psi \quad\Rightarrow\quad
\< P^\mu_{\rm rad} \> = \int \frac{\d^\D p}{(2\pi)^{\D-1}} \delta_+(p^2)\,p^\mu \left|\<j(p)\>_\psi\right|^2\,,
\ee
which is the particle physicist's  formula for radiated momentum.  For electromagnetism (or gravity) one finds similar formulas with sums over transverse polarizations.
For nonrelativistic particles for example this reduces to the familiar Larmor formula:
$P^0_{\rm rad}\propto \int \d t~\ddot{\vec{x}}^2$.
Nonetheless,  one could imagine situations where the Coulomb field near infinity is important (see for example the $\delta(\omega)$ term in \cite[Eq.~(8)]{Manohar:2022dea} in the analysis of angular momentum radiation)
and thus we include it in \eqref{reduction including Coulomb} for completeness.\footnote{
The Coulomb contribution can, of course, also be understood taking Weinberg's soft factor slightly off-shell in \eqref{avg from GR}, which we believe is effectively what is done in \cite{Manohar:2022dea}.  However, in formulas involving on-shell real momenta such as \eqref{reduction bp}, matrix elements $\<1'|a_p|1\>$ are strictly zero and we believe that singularities near $p^0=0$ in on-shell integrals \eqref{reduction including Coulomb} are to be regulated by going to $\D=4-2\epsilon$ spacetime dimensions rather than by an $i\varepsilon$ prescription.  The end result for potentials is the same, of course, due to the identity \eqref{avg from GR GA}.
}

In summary, the general relation between potentials and on-shell fields is given by \eqref{reduction including Coulomb}
where the first term includes only connected contributions to the KMOC expectation value, which can be computed by the reduction formula \eqref{reduction bp}.  The complete formula, however, includes an extra ``Coulomb'' term, which may or may not be relevant depending on the physical situation of interest.

\subsection{Perturbation theory: dressing time-ordered diagrams}\label{ssec:SK pert}

The reduction formulas discussed above can be used to analyze perturbative corrections to various observables in field theory. Below, we will find perfect agreement with the graphical manipulations of blobs from the preceding section.

The Feynman rules for non-time-ordered correlators of currents can be directly derived from the (Schwinger-Keldysh) path integral \eqref{eq:phiExp}, for which only amputated diagrams contribute non-trivially.
Feynman rules in this formalism have been described in many references, see for example \cite{Haehl:2017qfl,Meltzer:2020qbr}.
Beside the traditional rules of time-ordered perturbation theory, we are to include the following extra decorations:
\begin{enumerate}[label=(\roman*)]
    \item For each diagram, sum over all possible assignments of a time-fold label (type-$\rone$ or type-$\rtwo$) to each vertex. Every external leg must connect to a vertex on the same time-fold. \label{rule i}
    \item Add an overall minus sign for each type-$\rtwo$ vertex
    and for each propagator connecting two type-$\rtwo$ vertices.
    \item Draw a positive-energy cut on each propagator going from a type-$\rone$ vertex to a type-$\rtwo$ vertex. Additionally, flip the $i\varepsilon$ in each propagator connecting two type-$\rtwo$ vertices.  More explicitly, apply the replacement rule
\be \label{SK replacement}
 \frac{1}{p^2+m^2-i\varepsilon}\mapsto \left\{\begin{array}{ll}\displaystyle
2\pi i \delta_+(p^2+m^2) \quad & \mbox{for each prop. going from $\rone$ to $\rtwo$}\,,\\ \displaystyle (p^2+m^2+i\varepsilon)^{-1} & \mbox{for each prop. between $\rtwo$ vertices}\,.
 \end{array}\right.
\ee
\end{enumerate}
It can be deduced from these rules that any given diagram can be broken down into islands consisting of type-$\rtwo$ fields, which are separated from type-$\rone$ fields by cuts. In particular, the rules are just the same ones that one would use to compute these cuts in the context of Cutkosky rules.

The rules can be easily extended to observables involving more than two time-folds (see, for example, \cite{Haehl:2017qfl}). In such cases, on the one hand, every even-type vertex ($\rtwo$, ${\rm IV}$ and so on) carries a minus sign, and propagators within even-type folds are anti-time-ordered and pick up the corresponding minus sign. On the other hand, any propagator connecting different folds is cut, with positive energy flowing towards the larger fold label. For example, a propagator extending between type-${\rm I}$ and type-${\rm III}$ vertices is the same as that between type-${\rm I}$ and type-${\rm II}$.

In order to make the above rules clearer, let us pause and discuss few explicit examples.
As a first one, we consider a triangle diagram in a theory with a
$-ig$ cubic vertex and $-i\lambda$ quadric vertex. The contribution of this diagram to the conventional amplitude is
\begin{subequations}\begin{gather} \label{2to2 cut SK}
  i\cM \supset \adjustbox{valign=c}{\input{tikz/triangleSK}}
  \\
  =
-g^2\lambda\int \frac{\d^\D\ell}{(2\pi)^\D}
  \frac{1}{(\ell^2+m_1^2-i\varepsilon)
  [(\ell{-}p_3)^2+m_2^2-i\varepsilon]
    [(\ell{+}p_4)^2+m_3^2-i\varepsilon]}\,.
\end{gather}\end{subequations}
Next, let us work out the contribution of the same diagram to the cut product defined by the left-hand side of \eqref{2->2 cut example B}, namely 
$\<0|\cTbar\{j_3j_4\}\cT\{j_1j_2\}|0\>=
\<0|\,\cC\{j_3^{\rtwo}j_4^{\rtwo}j_1^{\rone}j_2^{\rone}\}\,|0\>$.
A single assignment of $\rone$ and $\rtwo$ labels is compatible with external labels:
\begin{subequations}
\begin{align}
\mbox{Eq.~\eqref{2to2 cut SK}}&\Rightarrow
\<0|\,\cC\{j_3^{\rtwo}j_4^{\rtwo}j_1^{\rone}j_2^{\rone}\}\,|0\> \supset \adjustbox{valign=c}{\input{tikz/triCutSK}}
\\&=g^2\lambda\int \frac{\d^\D\ell}{(2\pi)^{\D-2}}
  \frac{\delta^+((\ell{-}p_3)^2+m_2^2)~
  \delta^+((\ell{+}p_4)^2+m_2^2)}{(\ell^2+m_1^2+i\varepsilon)}
  \,.\label{2to2 cut SK2}
  \end{align}
\end{subequations}
Note that the overall sign flipped from \eqref{2to2 cut SK} to \eqref{2to2 cut SK2} because there are two vertices and one propagator on the left of the cut.
Here are examples of additional and non-trivial contributions to \eqref{2->2 cut example B}
\begin{equation}
    \adjustbox{valign=c}{\input{tikz/tri_tchan_SK}}~
    \adjustbox{valign=c}{\input{tikz/triboxC1_SK}}~\adjustbox{valign=c}{\input{tikz/triboxC2_SK}}~\adjustbox{valign=c}{\input{tikz/triboxC3_SK}}\,.
\end{equation}
These examples clearly demonstrate that the computation rules for this observable in perturbation theory are consistent with those leading to the $s$-channel cut. This reflects how perturbation theory trivializes asymptotic completeness, as pointed out earlier below \eqref{2->2 cut example B}. Note that isolated islands of the $\rone$- or $\rtwo$-type vanish, for example
\begin{equation}
    \adjustbox{valign=c}{\input{tikz/triboxIsland2}}=0\,.
\end{equation}
Next, let us move on to the more interesting five-point inclusive observable
\be\label{eq:exp3SK}
 {\rm Exp}_3 \equiv
 \<0| \cC\{ j_1^\rdiff j_2^\rdiff j_4^\rdiff j_5^\rdiff\, j_3^\rone\}|0\>\,.
\ee
At tree-level, for instance, the diagram considered in Sec.~\ref{sec:treelevel} contributes to the amplitude
\begin{align}
 i\cM &\supset \begin{gathered}
\begin{tikzpicture}[baseline= {($(current bounding box.base)+(10pt,-25pt)$)},line width=1, scale=0.7,yscale=-1]
\begin{scope}
\coordinate (a) at (0,0) ;
\coordinate (b) at (1,0) ;
\coordinate (c) at ($(b)+(-40:1)$);
\draw[] (a) -- (b);
\draw[] (b) -- (c);
\draw[ ] (b) -- ++ (30:1) node[right] {\footnotesize$2$};
\draw[  ] (c) -- ++ (-150:1) node[left]{\footnotesize$3$};
\draw[ ] (c) -- ++ (-30:1) node[right] {\footnotesize$1$};
\draw[ ] (a) -- (-150:1) node[left] {\footnotesize$4$};
\draw[ ] (a) -- (150:1) node[left] {\footnotesize$5$};
\fill[black,thick] (a) circle (0.07);
\fill[black,thick] (b) circle (0.07);
\fill[black,thick] (c) circle (0.07);
\end{scope}
\end{tikzpicture}
\end{gathered} =
 \frac{-ig^3}{(-s_{45}+M^2-i\eps)(-s_{13}+M^2-i\eps)}\,.
\end{align}
To compute its contributions to the expectation value we need to decorate each vertex
except the one on which leg 3 lands with either a $\rone$ or $\rtwo$, according to
rule \ref{rule i} above:
\begin{align}
{\rm Exp}_3 &\supset \begin{gathered}
\begin{tikzpicture}[baseline= {($(current bounding box.base)+(10pt,-25pt)$)},line width=1, scale=0.7,yscale=-1]
\begin{scope}
\coordinate (a) at (0,0) ;
\coordinate (b) at (1,0) ;
\coordinate (c) at ($(b)+(-40:1)$);
\draw[] (a) -- (b);
\draw[] (b) -- (c);
\draw[ ] (b) -- ++ (30:1) node[right] {\footnotesize$2$};
\draw[  ] (c) -- ++ (-150:1) node[left]{\footnotesize$3$};
\draw[ ] (c) -- ++ (-30:1) node[right] {\footnotesize$1$};
\draw[ ] (a) -- (-150:1) node[left] {\footnotesize$4$};
\draw[ ] (a) -- (150:1) node[left] {\footnotesize$5$};
\fill[black,thick] (a) circle (0.07) node[left,below, xshift=5pt][font=\scriptsize]{$$\text{I-II}$$};
\fill[black,thick] (b) circle (0.07) node[left,above, xshift=-5pt][font=\scriptsize]{$$\text{I-II}$$};
\fill[black,thick] (c) circle (0.07) node[left,below, xshift=5pt][font=\scriptsize]{$$\text{I}$$};
\end{scope}
\end{tikzpicture}
\end{gathered}
\\&= \begin{tikzpicture}[baseline= {($(current bounding box.base)+(10pt,0pt)$)},line width=1, scale=0.7,yscale=-1]
\begin{scope}[xshift=0]
\coordinate (a) at (0,0) ;
\coordinate (b) at (1,0) ;
\coordinate (c) at ($(b)+(-40:1)$);
\draw[] (a) -- (b);
\draw[] (b) -- (c);
\draw[ ] (b) -- ++ (30:1) node[right] {\footnotesize$2$};
\draw[  ] (c) -- ++ (-150:1) node[left]{\footnotesize$3$};
\draw[ ] (c) -- ++ (-30:1) node[right] {\footnotesize$1$};
\draw[ ] (a) -- (-150:1) node[left] {\footnotesize$4$};
\draw[ ] (a) -- (150:1) node[left] {\footnotesize$5$};
\fill[black,thick] (a) circle (0.07) node[below] [font=\scriptsize]{$\rone$};
\fill[black,thick] (b) circle (0.07)  node[below] [font=\scriptsize]{$\rone$} node[right, xshift=27pt]{$+$};
\fill[black,thick] (c) circle (0.07)  node[below] [font=\scriptsize]{$\rone$};
\end{scope}
\begin{scope}[xshift=130]
\coordinate (a) at (0,0) ;
\coordinate (b) at (1,0) ;
\coordinate (c) at ($(b)+(-40:1)$);
\draw[] (a) -- (b);
\draw[] (b) -- (c);
\draw[ ] (b) -- ++ (30:1) node[right] {\footnotesize$2$};
\draw[  ] (c) -- ++ (-150:1) node[left]{\footnotesize$3$};
\draw[ ] (c) -- ++ (-30:1) node[right] {\footnotesize$1$};
\draw[ ] (a) -- (-150:1) node[left] {\footnotesize$4$};
\draw[ ] (a) -- (150:1) node[left] {\footnotesize$5$};
\fill[black,thick] (a) circle (0.07) node[below] [font=\scriptsize]{ $\rtwo$};
\fill[black,thick] (b) circle (0.07)  node[below] [font=\scriptsize]{$\rone$} node[right, xshift=27pt]{$+$};
\fill[black,thick] (c) circle (0.07)  node[below] [font=\scriptsize]{$\rone$};
\draw[color =Orange , dashed] (0.5,0.5) -- (0.5,-0.5);
\end{scope}
\begin{scope}[xshift=260]
\coordinate (a) at (0,0) ;
\coordinate (b) at (1,0) ;
\coordinate (c) at ($(b)+(-40:1)$);
\draw[] (a) -- (b);
\draw[] (b) -- (c);
\draw[ ] (b) -- ++ (30:1) node[right] {\footnotesize$2$};
\draw[  ] (c) -- ++ (-150:1) node[left]{\footnotesize$3$};
\draw[ ] (c) -- ++ (-30:1) node[right] {\footnotesize$1$};
\draw[ ] (a) -- (-150:1) node[left] {\footnotesize$4$};
\draw[ ] (a) -- (150:1) node[left] {\footnotesize$5$};
\fill[black,thick] (a) circle (0.07);
\fill[black,thick] (b) circle (0.07);
\fill[black,thick] (c) circle (0.07);
\fill[black,thick] (a) circle (0.07) node[below] [font=\scriptsize]{$\rone$};
\fill[black,thick] (b) circle (0.07)  node[below] [font=\scriptsize]{$\rtwo$} node[right, xshift=27pt]{$+$};
\fill[black,thick] (c) circle (0.07)  node[below] [font=\scriptsize]{$\rone$};
\draw[color =Orange , dashed] (0.5,0.5) -- (0.5,-0.5);
\draw[color =Orange , dashed] (0.9,-0.5) -- (1.65,0);
\end{scope}
\begin{scope}[xshift=390]
\coordinate (a) at (0,0) ;
\coordinate (b) at (1,0) ;
\coordinate (c) at ($(b)+(-40:1)$);
\draw[] (a) -- (b);
\draw[] (b) -- (c);
\draw[ ] (b) -- ++ (30:1) node[right] {\footnotesize$2$};
\draw[  ] (c) -- ++ (-150:1) node[left]{\footnotesize$3$};
\draw[ ] (c) -- ++ (-30:1) node[right] {\footnotesize$1$};
\draw[ ] (a) -- (-150:1) node[left] {\footnotesize$4$};
\draw[ ] (a) -- (150:1) node[left] {\footnotesize$5$};
\fill[black,thick] (a) circle (0.07) node[below] [font=\scriptsize]{$\rtwo$};
\fill[black,thick] (b) circle (0.07)  node[below] [font=\scriptsize]{$\rtwo$};
\fill[black,thick] (c) circle (0.07)  node[below] [font=\scriptsize]{$\rone$};
\draw[color =Orange , dashed] (0.9,-0.5) -- (1.65,0);
\end{scope}
\end{tikzpicture}\notag
\\
&= \frac{-ig^3}{(-s_{45}+M^2-i\eps)(-s_{13}+M^2-i\eps)}
+ \frac{-2\pi g^3\delta(s_{45}-M^2)}{-s_{13}+M^2-i\eps} +0+0\,.
\end{align}
Due to basic kinematic considerations and stability, the last two diagrams vanish: while the third diagram contains a propagator with an inconsistent energy flow, the fourth one breaks the stability assumption. The two non-zero terms concord precisely with \eqref{eq:Exp3res} upon setting $g_i=g$ for all $i$. The bubble cut in the example of Sec.~\ref{sec:pert_blobology} is similarly reproduced.

Next, let us consider the one-loop pentagon, which was discussed earlier at the end of Sec.~\ref{sec:pert_blobology} and will be discussed further in Sec.~\ref{sec:waveforms}.
We have that
\be\label{eq:expKID}
 {\rm Exp}_k \adjustbox{valign=c,scale={0.75}{0.75}}{\input{tikz/ID}} \equiv\adjustbox{valign=c,scale={0.75}{0.75}}{\input{tikz/IDall}}\,.
\ee
One should observe that, as per \eqref{eq:exp3SK}, the external graviton comes attached to a type-I vertex.  Thus, in principle, the implicit sum in \eqref{eq:expKID} is over $2^4=16$ potential label assignments. However, assuming once again that the external particles are stable, most options are dismissed by either energy flow or kinematic considerations. Simple instances would be
\be
\adjustbox{valign=c,scale={0.75}{0.75}}{\input{tikz/IDbc1}}~\adjustbox{valign=c,scale={0.75}{0.75}}{\input{tikz/IDbc4}}~\adjustbox{valign=c,scale={0.75}{0.75}}{\input{tikz/IDbc2}}~\adjustbox{valign=c,scale={0.75}{0.75}}{\input{tikz/IDbc3}}\,\quad \mbox{(all zero)}\,.
\ee
The first diagram vanishes because the mass of the top (now on-shell) heavy particle before and after the graviton emission is the same, thereby contradicting the stability condition. The second diagram vanishes as the energy flow within the bottom (now on-shell) heavy line has the wrong sign. Moreover, the \say{L-shaped} cut in this diagram also breaks the stability condition.  In the third diagram, the graviton is produced (from vacuum) before the other external particles can create it. Hence, the energy flow across the cut is negative and consequently, it vanishes. Finally, the fourth diagram vanishes because the $t$-channels are spacelike. In the end, only the two diagrams survive, namely 
\be\label{eq:expIDfinal}
  {\rm Exp}_k \adjustbox{valign=c,scale={0.75}{0.75}}{\input{tikz/ID}} = \adjustbox{valign=c,scale={0.75}{0.75}}{\input{tikz/IDall1}}+\adjustbox{valign=c,scale={0.75}{0.75}}{\input{tikz/IDskCut}}\,.
\ee
Finally, one may consider diagrams where some propagators have been canceled. From the reasoning that got us from \eqref{eq:expKID} to \eqref{eq:expIDfinal}, we have, for example,
\begin{align}
   {\rm Exp}_k \adjustbox{valign=c,scale={0.75}{0.75}}{\input{tikz/I5D}}&\equiv \adjustbox{valign=c,scale={0.75}{0.75}}{\input{tikz/I5DskAll}}=\adjustbox{valign=c,scale={0.75}{0.75}}{\input{tikz/I5Dsk}}\,,\label{eq:expI3}\\ {\rm Exp}_k \adjustbox{valign=c,scale={0.75}{0.75}}{\input{tikz/I3D}}&\equiv \adjustbox{valign=c,scale={0.75}{0.75}}{\input{tikz/I3DskAll}}=\adjustbox{valign=c,scale={0.75}{0.75}}{\input{tikz/I3Dsk}}+\adjustbox{valign=c,scale={0.75}{0.75}}{\input{tikz/I3Dskcut}}\,.  
\end{align}
It's worth pointing out that the cut term can't appear in \eqref{eq:expI3}, given that the external graviton is required to connect to a type-$\rone$ vertex and that cutting through the vertex is simply not allowed. Thus, all the above agree precisely with what was initially found in Sec.~\ref{sec:pert_blobology} by comparing with the blob notation, namely
\begin{equation}\label{inclusive KMOC}
  \stackrel{\adjustbox{valign=t}{\begin{tikzpicture}[line width=1]
\draw[line width = 1] (0,0.3) -- (-1.2,0.3) node[above] {\small$1'$};
\draw[line width = 1] (0,-0.3) -- (-1.2,-0.3) node[below] {\small$2'$};
\draw[line width = 1] (2,0.3) -- (3.2,0.3) node[above] {\small$1$};
\draw[line width = 1] (2,-0.3) -- (3.2,-0.3) node[below] {\small$2$};
\draw[photon] (2,0) -- (1,-1) node[left] {\small$k$};
\filldraw[fill=gray!30, very thick](0,-0.3) rectangle (2,0.3);
\draw[] (1,0) node {$X$};
\filldraw[fill=gray!5, line width=1.2](0,0) circle (0.6) node {$S^\dag$};
\filldraw[fill=gray!5, line width=1.2](2,0) circle (0.6) node {$S$};
\draw[dashed,Orange] (1,0.7) -- (1,-1.1);
\end{tikzpicture}}}{{\rm Exp}_k}
  =  \stackrel{\adjustbox{valign=t}{\begin{tikzpicture}[line width=1]
\draw[line width = 1] (2,0.3) -- (0.8,0.3) node[above] {\small$1'$};
\draw[line width = 1] (2,-0.15) -- (0.8,-0.15) node[below] {\small$2'$};
\draw[line width = 1] (2,0.3) -- (3.2,0.3) node[above] {\small$1$};
\draw[line width = 1] (2,-0.3) -- (3.2,-0.3) node[below] {\small$2$};
\draw[photon] (2,0) -- (1,-1) node[left] {\small$k$};
\filldraw[fill=gray!5, line width=1.2](2,0) circle (0.6) node {$i \mathcal{M}$};
\end{tikzpicture}}}{i\mathcal{M}} + 
  \stackrel{\adjustbox{valign=t}{\begin{tikzpicture}[line width=1]
\draw[line width = 1] (0,0.3) -- (-1.2,0.3) node[above] {\small$1'$};
\draw[line width = 1] (0,-0.3) -- (-1.2,-0.3) node[below] {\small$2'$};
\draw[line width = 1] (2,0.3) -- (3.2,0.3) node[above] {\small$1$};
\draw[line width = 1] (2,-0.3) -- (3.2,-0.3) node[below] {\small$2$};
\draw[photon] (2,0) -- (1,-1) node[left] {\small$k$};
\filldraw[fill=gray!30, very thick](0,-0.3) rectangle (2,0.3);
\draw[] (1,0) node {$X$};
\filldraw[fill=gray!5, line width=1.2](0,0) circle (0.6) node {$-i\mathcal{M}^\dag$};
\filldraw[fill=gray!5, line width=1.2](2,0) circle (0.6) node {$i\mathcal{M}$};
\draw[dashed,Orange] (1,0.7) -- (1,-1.1);
\end{tikzpicture}}}{{\rm Cut}_{1'2'}\,.} 
\end{equation}
In summary, whether one uses the blob notation or the Schwinger-Keldysh formalism, the inclusive expectation value ${\rm Exp}_k$ is computed from the amplitude precisely by subtracting its $1'2'$-channel cuts. We emphasize that the cut subtraction commutes with all familiar operations, such as numerator algebra or integration-by-parts identities.

\subsection{Comments on CPT/CRT}
\label{sec:CPT}

Using the reduction formulas, we can discuss how CPT invariance manifests itself on asymptotic observables. For a correlation function of scalar bosonic fields $\phi_i(x_i)$,
CPT guarantees that simultaneously flipping time together with an \emph{odd} number of spacelike  coordinates as well as reversing the ordering of operators gives the same correlation function \cite[Sec.~4-3]{Streater:1989vi}, namely
\be \label{CPT 1}
\< 0 | \phi_1(x_1) \phi_2(x_2) \cdots \phi_n(x_n) | 0 \> \stackrel{\text{CRT}}{=} \< 0 | \phi_n(-P'x_n) \cdots \phi_2(-P'x_2) \phi_1(-P'x_1) | 0 \>\,.
\ee
Here, we denote by $P'$ the \emph{transverse parity}, which flips the sign
of $D\,{\rm mod}\,2$ spacelike coordinates.
While \cite{Streater:1989vi} only discuss the case of four spacetime dimensions,
where $P'\simeq \mathbb{1}$ and the right-hand side of \eqref{CPT 1} can be written more simply as
$\phi_i(-x_i)$, we quoted here the general statement known as CRT, see, e.g., \cite{Witten:2018zxz}.
The requirement is that $x^\mu\mapsto -P'x^\mu$ must be an element of the complexified connected Lorentz group.  Analogous equations hold for fields with spin and fermions, at a cost of extra factors coming from the Lorentz action.

The important thing to notice for our purposes is that \eqref{CPT 1} commutes with the time-ordering symbol, i.e., the time-ordered product of $\phi(x)$'s maps to the time-ordered product of $\phi(-P'x)$'s.  We can also consider products of $\cT$'s as in \eqref{many T products}: then we get the same $\cT$ products, but in the opposite order.  Fourier-transforming both sides, this yields the following convenient form of CPT/CRT:
\be\begin{aligned}
 &\<0| \cT\{\phi_{1}(p_1)\Compactcdots \phi_{i_1}(p_{i_1})\}
 \Compactcdots
 \cT\{\phi_{i_{n-1}+1}(p_{i_{n-1}+1})\Compactcdots \phi_{i_n}(p_{i_n})\}|0\>
 \\ &~\stackrel{\text{CRT}}{=}
 \<0|\cT\{\phi_{i_{n-1}+1}(-P'p_{i_{n-1}+1})\Compactcdots \phi_{i_n}(-P'p_{i_n})\}
 \Compactcdots
 \cT\{\phi_1(-P'p_1)\Compactcdots \phi_{i_1}(-P'p_{i_1})\} |0\>\,.
\end{aligned}\ee
Applying the reduction identity \eqref{j from a and b} to a general product,
we see that result amounts to the simple substitution on a vacuum expectation value of in and out
creation/annihilation operators:
\be
\mbox{CRT reverses all products and takes:}\quad
a_{m}(p)\leftrightarrow \bdag_{\bar{m}}(P'p)\,, \quad
\adag_{m}(p)\leftrightarrow b_{\bar{m}}(P'p)\,,
\ee
where $\bar{m}$ represents the antiparticle of the particle $m$ 
(which comes about because the same field $\phi_m$ that can absorb particle $m$ will create $\bar{m}$ when the sign of its energy is reversed).

For a single time-ordered product, the above gives the relation $\cM_{B\ot A}=\cM_{P'\bar{A}\ot P'\bar{B}}$ mentioned Sec.~\ref{sec:3.1} stating that initial and final states can be swapped at the cost of a transverse parity.
In particular, in four spacetime dimensions, this is the familiar statement that
initial and final states can be simply swapped, i.e., the amplitude for electron-positron pair production from photons is the same as for the reverse process of pair annihilation. Applying CRT to another example, the expectation value $\Exp_3$ from \eqref{inclusive 5}, gives
\be
{}_{\mathrm{in}}\langle 54 | b_3 | 21 \rangle_{\mathrm{in}} = \langle 0 | a_5 a_4 b_3 a_2^\dag a_1^\dag | 0 \rangle \stackrel{\text{CRT}}{=} \langle 0 | b_{\bar{1}} b_{\bar{2}} a_{\bar{3}}^\dag b_{\bar{4}}^\dag b_{\bar{5}}^\dag | 0 \rangle(P'p) = {}_{\mathrm{out}}\langle \bar{1}\bar{2} | a_{\bar{3}}^\dag | \bar{4}\bar{5} \rangle_{\mathrm{out}} (P'p)\, ,
\ee
where the observable on the right-hand side is an out-out expectation value instead of in-in.
Note that CPT/CRT invariance is \emph{not} equivalent to crossing, which instead only swaps a subset of particles (i.e., as in the relation between pair production and Compton scattering). 

\def\ItA{\tilde{\mathcal{I}}^A}
\def\IA{\mathcal{I}^A}
\tikzstyle{every node}=[font=\small,
dot/.style = {circle, fill, minimum size=4pt,
              inner sep=0pt, outer sep=0pt}]
\tikzset{photon/.style={decorate, decoration={snake, amplitude=1pt, segment length=6pt}}}

\section{\label{sec:waveforms}Master integrals for gravitational waveforms}

In the previous sections, we showed how the $S$-matrix and its conjugate can be strung together into a host of inclusive asymptotic observables, where each amplitude lives on its own branch on a Schwinger-Keldysh contour. Among these observables is the momentum-space \emph{gravitational waveform}, which according to the Kosower-Maybee-O'Connell (KMOC) formalism~\cite{Kosower:2018adc,Cristofoli:2021vyo} takes the form
\begin{equation} \label{KMOC}
{\rm Exp}_k\equiv
    {}_{\rm in}\< 2'1'| S^\dag a_k S |12\>_{\rm in} = \< 0| a_{2'} a_{1'}\, b_k\, \adag_{2}\adag_{1}|0\>\,.
\end{equation}
Here, $b_k$ is a graviton annihilation operator, and we take the in-states to consist of two heavy scalars, such that~\eqref{KMOC} is the expectation value of the gravity field after the collision of two compact objects.
The heavy scalars can be viewed as a model for non-spinning black holes. We now have two complementary ways of interpreting these observables: either as time-ordered products on a Schwinger-Keldysh contour (see Sec.~\ref{sec:localFields}), or as the sum of amplitudes and cuts fitting the blob pattern in \eqref{inclusive KMOC}.

\begin{figure}[t]
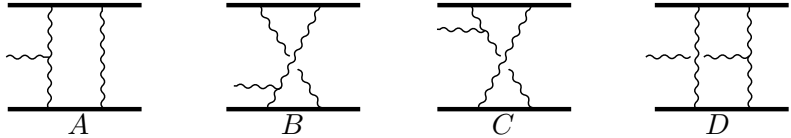
  \centering
\begin{equation*}
\stackrel{\adjustbox{valign=c,scale={0.75}{0.75}}{\input{tikz/I6}}}{A}
\qquad
\stackrel{\adjustbox{valign=c,scale={0.75}{0.75}}{\input{tikz/IB}}}{B}
\qquad
\stackrel{\adjustbox{valign=c,scale={0.75}{0.75}}{\input{tikz/IC}}}{C}
\qquad
\stackrel{\adjustbox{valign=c,scale={0.75}{0.75}}{\input{tikz/IDsource}}}{D}
\end{equation*}
\caption{\label{fig:ABCD topologies}
The one-loop topologies for the inclusive amplitude for a graviton in the background of scattering of two heavy scalars considered in this section. 
In the eikonal limit these become a single topology (up to $i\varepsilon$'s).
The cuts of these topologies were studied around \eqref{ABCD topos}.
}
\end{figure}

The aim of this section is to exemplify the computation of~\eqref{KMOC} on the one-loop topologies depicted in Fig.~\ref{fig:ABCD topologies}. 
These diagrams (and their subtopologies) account for the
emission of a massless particle (the graviton) induced by
the scattering of two compact objects via a long-range force.%
\footnote{Note that there is no need to explicitly consider other one-loop diagrams here as they can be obtained from the ones (and their subtopologies) we just mentioned, either from IBPs or partial fraction identities (see \cite[Eq. (4.20)]{Herderschee:2023fxh} for instance).} They are crucial ingredients in full waveform computations at subleading post-Minkowski order, and were recently studied by several authors~\cite{Brandhuber:2023hhy,Herderschee:2023fxh,Elkhidir:2023dco} and later in \cite{Georgoudis:2023lgf}. 

Our strategy is to first consider the contribution of diagrams to conventional five-point scattering amplitudes (i.e., the $i\cM$ term in \eqref{inclusive KMOC}),
and then add the cut as a decoration at the last possible step.
We will thus assume that the amplitude for the $2\to 3$ scattering process $i \mathcal{M}_{1'2'k \ot 12}$ is already expressed in terms of a basis of master integrals, and focus here on how to incorporate the $\cut_{1'2'}$
term at the level of these master integrals.  It should be clear from the preceding section that the cut decorations commute with integration-by-parts identities such that \eqref{KMOC} can be computed \say{master-by-master.}

Moreover, we will discuss simplifications in the eikonal limit (defined below). An important feature of this limit is that several topologies that are distinct in general kinematics become similar and differ only by $i\varepsilon$ prescriptions. Even so, we find it helpful structure the computations below to calculate each topology separately. 

\subsection{Setup}
\begin{figure}
    \centering
\begin{equation}
\begin{tikzpicture}[scale=0.6,thick,
baseline={([yshift=-0.4ex]current bounding box.center)}]
    \draw[line width=2] (-1,1) -- (0,0) -- (1,1)  (-1,-1) -- (0,0) -- (1,-1);
    \draw[photon] (-0.5,0) node[dot]{} -- +(-1.2,0);
    \filldraw[fill=gray!5, line width=1.2](0,0) circle (0.8);
    \node[] at (2,1.5) {$p_1=\bar{p}_1 + \frac{q_1}{2}$};
    \node[] at (-2,1.5) {$p_1'=\bar{p}_1 - \frac{q_1}{2}$};
    \node[] at (2,-1.5) {$p_2=\bar{p}_2 + \frac{q_2}{2}$};
    \node[] at (-2,-1.5) {$p_2'=\bar{p}_2 - \frac{q_2}{2}$};
    \node[] at (-3.5,0) {$k=q_1+q_2$};
\end{tikzpicture}
\hspace{2cm}
\adjustbox{valign=c,scale={1}{1}}{\input{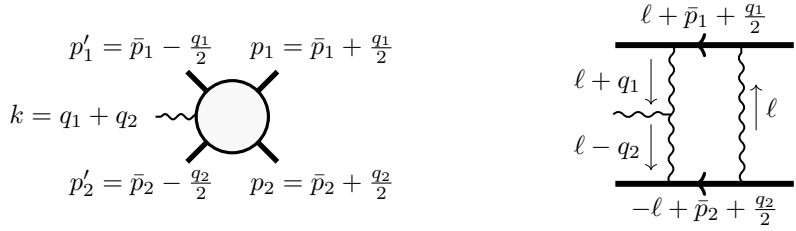}}
\nonumber
\end{equation}
\caption{Labeling of the kinematics for a five-point scattering process (left) and the internal momentum labeling of the topology $G^A$ (right).}
\label{fig:momlabel}
\end{figure}

We use the same momentum labeling for the topologies in Fig.~\ref{fig:ABCD topologies} as in \cite{Brandhuber:2023hhy}, which is summarized in Fig.~\ref{fig:momlabel}. The diagrams in Fig.~\ref{fig:ABCD topologies} correspond to the following family of integrals:
\be\begin{aligned} \label{5pt family}
    G^A_{a_1,a_2,a_3,a_4,a_5} & = \e^{\epsilon\gammaE}\int \frac{\d^\D\ell}{i\pi^{\D/2}} \frac{1}{[\ell^2]^{a_1} [(\ell+\bar{p}_1+\frac{1}{2}q_1)^2+m_1^2]^{a_2} [(\ell+q_1)^2]^{a_3}}
    \\ & \qquad \qquad \qquad \qquad \qquad \qquad \times \frac{1}{[(\ell-q_2)^2]^{a_4} [(\ell-\bar{p}_2-\frac{1}{2}q_2)^2+m_2^2]^{a_5}} \,,
\end{aligned}\ee
where the expression is written for the $A$ topology as indicated by the superscript,
and $-i\varepsilon$ should be added to each square bracket to compute the conventional time-ordered amplitude.

We will sometimes leave the dependence on the $a_i$ implicit in cases where they do not matter.

Master integrals for the other topologies are defined by simple permutations of $\bar{p}_1$ and  $\bar{p}_2$:
$B$ is obtained by $\bar{p}_1\to -\bar{p}_1$, $C$ by $\bar{p}_2 \to -\bar{p}_2$,
and $D$ by $\bar{p}_i \to -\bar{p}_i$ for both $i=1,2$.  As emphasized in \cite{Brandhuber:2023hhy,Herderschee:2023fxh}, the fact that permutations act simply on $\bar{p}_i^\mu=\frac12(p_i+p_i')^\mu$
makes these variables particularly convenient for eikonal expansions.
In this notation, the total scattering amplitude can be written as
\be\begin{aligned} \label{amp 4 terms}
    i\mathcal{M} & = \sum_i \Big[c_i(\barp_1,\barp_2) G^A_i + c_i(-\barp_1,\barp_2) G^B_i + c_i(\barp_1,-\barp_2) G^C_i + c_i(-\barp_1,- \barp_2) G^D_i
    \Big] \,,
\end{aligned}\ee
where the sum in $i$ runs over some basis of master integrals given by the $G_i^X$, and $c_i$ are their coefficients.%
\footnote{
Integrals with $a_2\leq 0$ or $a_5\leq 0$ can be viewed as members of several topologies. To keep the notation uniform, we will assume with no loss of generality that the corresponding $c_i$'s have been chosen such that~\eqref{amp 4 terms} holds.
Note also that the $G$'s defined in \eqref{5pt family} are real in Euclidean kinematics and
so the coefficients $c_i$ in \eqref{amp 4 terms} that arise from Feynman rules would be purely imaginary.}

When computing the in-in expectation value $\Exp_k$, we need to add to the scattering amplitude the contribution from the $\cut_{1'2'}$ term in \eqref{inclusive KMOC}.
In fact, out of the one-loop diagrams shown in Fig.~\ref{fig:ABCD topologies}, only the $D$ topology admits a non-zero cut in this channel.  This was previously exemplified in Secs.~\ref{sec:pert_blobology} and \ref{ssec:SK pert}.
Thus, for the $A$, $B$ and $C$ topologies, there is actually no difference between the time-ordered amplitude $i \mathcal{M}$ and ${\rm Exp}_k$.
The only term to add is the cut of $D$ obtained by taking the discontinuity of two propagators and including appropriate signs
as determined by the rules \eqref{SK replacement}: 

\begin{align}
    \cut \, G_{a_1,a_2,a_3,a_4,a_5}^D = 4 \pi^2
    \e^{\epsilon\gammaE}\int \frac{\d^\D\ell}{i\pi^{\D/2}} \frac{1}{(\ell^2+i\varepsilon)^{a_1} \left( (\ell{+}q_1)^2-i\varepsilon\right)^{a_3} \left( (\ell{-}q_2)^2 - i \varepsilon\right)^{a_4}}
        \label{eq:cutD}
    \\  \times
     \delta_+^{[a_2]} \Big[ (\ell{-}\bar{p}_1{+}\tfrac{1}{2}q_1)^2+m_1^2 \Big]
    \,\delta_+^{[a_5]} \Big[ (\ell{+}\bar{p}_2{-}\tfrac{1}{2}q_2)^2+m_2^2 \Big]\nonumber\,,
\end{align}
where we introduced the shorthand notation 
\be
\delta^{[a]} \equiv \frac{(-1)^{a-1}}{(a-1)!} \delta^{(a-1)}(x)\, .
\ee
Here, $\delta^{(n)}(x)$ denotes the $n$-th derivative of the Dirac $\delta$-function $\delta(x)$, and the subscript \say{$+$} denotes positive-energy flow as before. 
Recall that the derivative can be evaluated using integration by parts 
\begin{equation}
    \int \d x\, \delta^{(n)}(x) f(x) = (-1)^{n} \int \d x\, \delta(x) f^{(n)}(x)\,,
\end{equation}
for any well-behaved test function $f(x)$.
The observable $\Exp_k$ then becomes
\be\begin{aligned}
    \Exp_k & = \sum_i \Big[c_i(\barp_1,\barp_2) G^A_i + c_i(-\barp_1,\barp_2) G^B_i  \\ & \hspace{2cm}
    + c_i(\barp_1,-\barp_2) G^C_i + c_i(-\barp_1,- \barp_2) \left(  G^D_i + \cut \, G^D_i \right)
    \Big] \,, \label{Expk 5 terms}
\end{aligned}\ee
where the coefficients $c_i(\barp_1,\barp_2)$ are \emph{the same} as in \eqref{amp 4 terms}.
This formula highlights that the ingredients used to compute $i\cM$ can be recycled to compute $\Exp_k$.

\subsection{Kinematics and tensor reduction in the eikonal limit}\label{ssec:eik}

As shown in \cite{Brandhuber:2023hhy,Herderschee:2023fxh,Elkhidir:2023dco}, the classical limit of the waveform is obtained from the \emph{heavy-mass effective field theory} (HEFT) \emph{expansion}, often also referred to as the \emph{eikonal expansion}. This can be understood as an expansion in small momentum transfer $q_i\sim \hbar$ but can also be organized
in terms of inverse powers of masses when dealing with heavy objects.  Defining the barred heavy masses as $\bar{m}_i^2=-\bar{p}_i^2$, this expansion is organized in powers of $\bar{m}_i^{-1}$.
For convenience, we also define the barred velocities $\bar{v}_i$ using $\bar{p}_i = \bar{m}_i \barv_i$, as well as the scalar products
\begin{align}
    \bary =  - \barv_1 \cdot \barv_2\,, \qquad q_i^2, \quad \text{and} \quad \barw_i = - \barv_i \cdot (q_1+q_2) \,, \label{5pt variables}
\end{align}
for $i=1,2$.
By momentum conservation, the barred velocities
are orthogonal to the respective momentum transfer (see \cite{Brandhuber:2023hhy,Herderschee:2023fxh,Elkhidir:2023dco}):
\begin{equation}
    \barv_i^2=-1, \qquad \barv_i{\cdot}q_i=0 \qquad (i=1,2)\,.
\end{equation}

Before we embark on an expansion for the scalar integrals $G_i^X$, let us briefly
discuss the role of numerators. A three-point vertex between the graviton and the heavy scalar with momentum $p_i$ is given by 
\begin{equation}
    \begin{tikzpicture}[scale=0.6,thick,
    baseline={([yshift=-0.4ex]current bounding box.center)}]
    \draw[line width=2] (-1,-1)--(1,-1);
    \draw[photon] (0,-1)--(0,0);
    \end{tikzpicture}
    \propto
    \sqrt{\GN} \bar{m}_i^2 \,,
    \label{eq:threeptgrav}
\end{equation}
where $\GN$ is Newton's constant.
Thus, in the mass power counting, a three-point vertex contributes a factor of $\bar{m}_i^2$ and hence the numerators of the pentagon topology (before integral reduction) go like $\sim \bar{m}_1^4\bar{m}_2^4$.
We will not go into the details of numerator computations here (see the recent computations in \cite{Brandhuber:2023hhy,Herderschee:2023fxh,Elkhidir:2023dco} for details), but we stress one important feature:
in applications, one will be interested in contributions that are up to $\frac{1}{\bar{m}}$-suppressed relative to it, that is, corresponding to $\bar{m}_1^4 \bar{m}_2^3$ or $\bar{m}_1^3 \bar{m}_2^4$ in the pentagon numerator.
It will thus be important to retain subleading terms in all steps. The leading contribution in the expansion of the pentagon topologies is of order $\bar{m}_1^3 \bar{m}_2^3$, and is known as the ``superclassical'' or ``hyperclassical'' term.

The large-mass expansion of the integrals $G_i$ defined in \eqref{5pt family} and their cuts can be analyzed at the level of integrands using the method of regions~\cite{Beneke:1997zp,Jantzen:2011nz,Jantzen:2012mw,Heinrich:2021dbf}.
The most important region is that of momenta $|\ell^\mu|\ll \bar{m}_i$, where one can directly expand the definition \eqref{5pt family} in powers of $1/\bar{m}_i$ after plugging in that
$\bar{p}_i=\bar{m}_i\barv_i$ (and using $\bar{p}_i^2+\bar{m}_i^2=0$).
We refer to it as the \emph{eikonal region} (this region is also called \emph{soft} in the literature on post-Minkowskian expansions \cite{Parra-Martinez:2020dzs}).
The contribution from this region is expressed in terms of integrals in which the heavy propagators become linear in $\ell$:
\be \label{series one propagator}
 \frac{1}{(\ell+\bar{p}_1+\frac{1}{2}q_1)^2+m_1^2}  =
 \frac{1}{\bar{m}_1} \frac{1}{[2\ell{\cdot}\barv_1]}
- \frac{1}{\bar{m}_1^2} \frac{\ell{\cdot}(\ell+q_1)}
{[2\ell{\cdot}\barv_1]^2}+\ldots
\ee
The result can be expressed in terms of a family of eikonal integrals:
\begin{equation} \label{eik topology}
 \GeikA_{a_1,a_2,a_3,a_4,a_5} = \e^{\epsilon\gammaE}\int \frac{\d^\D\ell}{i\pi^{\D/2}}
 \frac{1}{[\ell^2]^{a_1}[2\ell{\cdot}\barv_1]^{a_2} [(\ell{+}q_1)^2]^{a_3}[(\ell{-}q_2)^2]^{a_4}
 [-2\ell{\cdot}\barv_2]^{a_5}}\,,
\end{equation}
where, as in \eqref{5pt family}, each square bracket carries a $-i\varepsilon$.
In the $G$ notation, multiplication by $\ell$-dependent factors is visually awkward but straightforward using the identity
$-2\ell{\cdot}(\ell+q_1)=q_1^2-\ell^2-(\ell{+}q_1)^2$,
which gives the following expansion for example for the eikonal region contribution to a scalar box:
\begin{align} \label{scalar box expansion}
G^A_{1,1,0,1,1}\Big|_{\rm eik}
=&\phantom{+} \frac{1}{\bar{m}_1\bar{m}_2}\GeikA_{1,1,0,1,1}
\nonumber\\&+ \frac{1}{2\bar{m}_1^2\bar{m}_2}
\left(
q_1^2\GeikA_{1,2,0,1,1}-\GeikA_{0,2,0,1,1}-\GeikA_{1,2,-1,1,1}\right)
\nonumber\\&+\frac{1}{2\bar{m}_1\bar{m}_2^2}
\left(
q_2^2\GeikA_{1,1,0,1,2}-\GeikA_{0,1,0,1,2}-\GeikA_{1,1,0,0,2}\right)+ \ldots,
\end{align}
where the omitted terms are further suppressed in large masses.  All mass dependence is now explicit since the eikonal integrals \eqref{eik topology} do not depend on $\bar{m}_i$.
Of course, at this stage the right-hand side could be simplified using integration-by-parts identities appropriate to the eikonal integrals.
Focusing on the third line contribution, which will also be discussed below, we find for example (choosing the same basis of masters as in \eqref{remaining ints} below):
\begin{align} \label{reduced triangle}
G^A_{1,1,0,1,1}\Big|_{{\rm eik},1/\bar{m}_1\bar{m}_2^2}
=&\phantom{+}
\frac{(\D-5)\bar{w}_1\bar{y}}{2(\bar{y}^2-1)}\GeikA_{1,1,0,1,1}
+\frac{\D-4}{2(\bar{y}^2-1)}
\left(
 \bar{y}\,\GeikA_{1,0,0,1,1}+\frac{q_2^2+\bar{w}_1^2}{q_2^2}\GeikA_{1,1,0,1,0}
\right)
\nonumber\\&
-\frac{\barw_1}{2(\bar{y}^2-1)}
\left(
\GeikA_{1,0,0,2,0}+\frac{\barw_1}{q_2^2}\GeikA_{0,1,0,2,0}
\right)\,.
\end{align}

It is worth emphasizing that the eikonal region does \emph{not} capture the full large-mass limit of the integrals $G_i$ in \eqref{5pt family},
i.e.,~the large-mass expansion of the integrand does \emph{not} commute with integration.
This happens because the integration includes momenta that extend far outside the eikonal region where the expansion \eqref{series one propagator} is valid.   The actual limit includes a further contribution from the \emph{hard region}
where the exchanged gravitons carry large off-shell momenta $|\ell^\mu|\sim \bar{m}_i$:
\be \label{G from eik plus hard}
 \lim_{\bar{m}_i\to\infty} G_i =
 G_i\big|_{\rm eik} + G_i\big|_{\rm hard}\,.
\ee
The main claim to fame of the method of regions is that such a sum holds to all orders in the mass expansion for dimensionally-regulated integrals.
The hard contribution can be computed by Taylor-expanding in small $q_i^\mu$ and $k^\mu$, which results in integrals of the type
\be G^{\rm hard}_{b_1,b_2,b_3} = 
\e^{\epsilon\gammaE}\int \frac{\d^\D\ell}{i\pi^{\D/2}}
 \frac{1}{[\ell^2]^{b_1}[(\ell{+}\bar{p}_1)^2+\bar{m}_1^2]^{b_2}
 [(\ell{-}\bar{p}_2)^2+\bar{m}_2^2]^{b_3}}\,.
\ee
These integrals depend only on $\bar{y}$, $\bar{m}_1$ and $\bar{m}_2$ and have coefficients
that are polynomial in $q_i^\mu$.  Therefore, even though it is possible that the hard contribution to \eqref{G from eik plus hard} might be leading in the $\bar{m}_i \to \infty$ limit, its Fourier transform vanishes at nonzero impact parameter.
For this reason, we will not discuss the hard region further and we focus below on the eikonal region contribution,
as also done in \cite{Brandhuber:2023hhy,Herderschee:2023fxh,Elkhidir:2023dco}.

The above manipulations can be carried out for each topology independently. We define the eikonal integrals
for the $B$, $C$ and $D$ topologies by relabelling appropriate signs $\bar{v}_i\mapsto \pm \bar{v}_i$ in \eqref{eik topology}, similarly to what was
described below \eqref{5pt family}.
Note that while the permuted eikonal integrals all formally have the same integrand up to overall signs, they actually have distinct $i\varepsilon$'s and for this reason we treat the $A$, $B$, $C$ and $D$ eikonal integrals as distinct topologies.
Their coefficients, of course, end up related by simple sign changes so in practice the algebra only needs to be carried out for one topology.
Specifically, writing out the coefficients
in terms of $\bar{m}_i$, $\bar{w}_i$, $\bar{y}$ and $q_i$ variables of \eqref{5pt variables}, the eikonal region contribution to the in-in expectation value \eqref{Expk 5 terms} becomes
\be\begin{aligned}
    \Exp_k\Big|_{\rm eik} & = \sum_{i}\Big[
    c_i'(\bar{w}_1,\bar{w}_2,\bar{y}) \GeikA_i+
    c_i'(-\bar{w}_1,\bar{w}_2,-\bar{y}) G^{{\rm eik},B}_i
\\ & \hspace{11mm}
    +c_i'(\bar{w}_1,-\bar{w}_2,-\bar{y}) G^{{\rm eik},C}_i+
    c_i'(-\bar{w}_1,-\bar{w}_2,\bar{y})\left( 
G^{{\rm eik},D}+ \cut\, G^{{\rm eik},D}\right)\Big]\,,
\label{Expk 5 terms eik}
\end{aligned}\ee
where the coefficients $c'_i$ are obtained from those in \eqref{Expk 5 terms} through series expansion and possible IBP reduction as in (\ref{series one propagator}--\ref{reduced triangle}).
Alternatively, the $c_i'$ numerators can be obtained directly 
using heavy-mass effective theory (HEFT), see for example \cite{Brandhuber:2021bsf}. The cut eikonal integrals are defined by taking discontinuity of two propagators as in \eqref{eq:cutD}.

Equation \eqref{Expk 5 terms eik} provides a robust way to organize the large-mass expansion of the in-in expectation value in terms of eikonal integrals.
Note that a priori it is important to know the coefficients in the $A$, $B$, $C$, $D$ topologies separately. However, because the original coefficients in \eqref{amp 4 terms}
depend only on $\bar{m}_i$ and $\bar{v}_i$ through their product $\bar{p}_i$, they differ by relative signs that are completely determined by the considered order in $\bar{m}_i$.
Thus, it is useful to combine the four topologies into
combinations with definite parity in each $\bar{m}_i$:
\be\begin{aligned}\label{eq:Gsum}
\Exp_k[G_{a_1\ldots\, a_5}^{{\rm eik},\sigma_1\sigma_2}] &\equiv
G_{a_1\ldots\, a_5}^{{\rm eik},A} + (-1)^{a_2}\sigma_1 G_{a_1\ldots\, a_5}^{{\rm eik},B} 
\\&\qquad
+\sigma_2(-1)^{a_5} G_{a_1\ldots\, a_5}^{{\rm eik},C}+\sigma_1 \sigma_2(-1)^{a_2+a_5} \left(G_{a_1\ldots\, a_5}^{{\rm eik},D}+\cut \, G_{a_1\ldots\, a_5}^{{\rm eik},D}\right),
\end{aligned}\ee
with the sign choice $\sigma_i=(-1)^{j_i}$
corresponding to a contribution of order $\bar{m}_1^{j_1}\bar{m}_2^{j_2}$.

The same logic applied to the time-ordered amplitude $i\cM$
give expressions identical to \eqref{Expk 5 terms eik} and \eqref{eq:Gsum}, simply without the $\cut$ term.
Because the cut only affects integrals with the $D$ topology, knowing the total \emph{integrated} amplitude $i\cM$ would generally be insufficient to recover the expression for $\Exp_k$.

\subsection{Aside: \texorpdfstring{$i\varepsilon$}{iε} and relation to principal value integrals}

The above formalism is based on conventional time-ordered integrals and their cuts.
In order to make it less abstract, we elaborate here on the contribution of the eikonal region
to \eqref{Expk 5 terms}, accounting for a single integral $G_i^A$ and its $B$, $C$ and $D$ permutations.
In an effort to make closer contact with the recent works \cite{Brandhuber:2023hhy,Herderschee:2023fxh,Elkhidir:2023dco,Georgoudis:2023lgf},
we will at the same time make explicit all relevant $i\varepsilon$ prescriptions.  This step is not required to understand the next subsections.

Cut contributions can be expanded in the eikonal region using a formula similar to \eqref{series one propagator}:
\begin{align}
    \delta\Big[(\ell+\bar{p}_2-\tfrac{1}{2}q_2)^2+m_2^2\Big] & = 
    \frac{ \delta \left(2 \ell \cdot \barv_2 \right) }{\bar{m}_2} + \frac{\ell{\cdot}(\ell-q_2)}{\bar{m}_2^2} \delta' \left(2 \ell \cdot \barv_2 \right) + \mathcal{O}\left(\bar{m}_2^{-3}\right)\,.
    \label{eq:delta-expansion}
\end{align}
Introducing a shorthand for passive factors,
\begin{equation}
    \int \frac{\dbar^\D \ell}{\Dden_1 \Dden_3 \Dden_4} \equiv \e^{\epsilon\gammaE} \int \frac{\rd^\D\ell}{i\pi^{\D/2} [\ell^2]^{a_1} [(\ell{+}q_1)^2]^{a_3} [(\ell{-}q_2)^2]^{a_4}} \,,
    \label{eq:elldenominator}
\end{equation}
the individual contributions
to first subleading order in $1/\bar{m}_2$
can be compactly written as:
\begin{subequations}\begin{align}
{G_i^A \Big\vert_{\text{eik}} = }
\adjustbox{valign=c,scale={0.65}{0.65}}{\input{tikz/I6}}
&  {\approx  \int \frac{\dbar^\D \ell}{\Dden_1 \Dden_3 \Dden_4}
\frac{(\bar{m}_1 \bar{m}_2)^{-1}}{(2\ell{\cdot}\barv_1 {-}i\varepsilon)({-}2\ell{\cdot}\barv_2 {-} i\varepsilon)}
\left[1 {-} \medmath{\frac{1}{\bar{m}_2}\frac{\ell{\cdot}(\ell{-}q_2)}{({-}2 \ell \cdot \barv_2 {-} i\varepsilon)}}  \right],}
\\
{G_i^B \Big\vert_{\text{eik}} = }
\adjustbox{valign=c,scale={0.65}{0.65}}{\input{tikz/IB}}
&  {\approx \int \frac{\dbar^\D \ell}{\Dden_1 \Dden_3 \Dden_4}
\frac{(\bar{m}_1 \bar{m}_2)^{-1}}{({-}2\ell{\cdot}\barv_1 {-}i\varepsilon)({-}2\ell{\cdot}\barv_2 {-} i\varepsilon)}\!
\left[1 {-} \medmath{\frac{1}{\bar{m}_2}\frac{\ell{\cdot}(\ell{-}q_2)}{({-}2 \ell \cdot \barv_2 {-} i\varepsilon)}}  \right]\!,}
\\
{G_i^C \Big\vert_{\text{eik}} =} \adjustbox{valign=c,scale={0.65}{0.65}}{\input{tikz/IC}}
& {\approx  \int \frac{\dbar^\D \ell}{\Dden_1 \Dden_3 \Dden_4}
\frac{(\bar{m}_1 \bar{m}_2)^{-1}}{(2\ell{\cdot}\barv_1 {-}i\varepsilon)(2\ell{\cdot}\barv_2 {-} i\varepsilon)}
\left[1 {-} \medmath{\frac{1}{\bar{m}_2}\frac{\ell{\cdot}(\ell-q_2)}{(2 \ell \cdot \barv_2- i\varepsilon)}}  \right],}
\\
{G_i^D\Big\vert_{\text{eik}} = }
\adjustbox{valign=c,scale={0.65}{0.65}}{\input{tikz/IDsource}}
&  {\approx  \int \frac{\dbar^\D \ell}{\Dden_1 \Dden_3 \Dden_4}
\frac{(\bar{m}_1 \bar{m}_2)^{-1}}{({-}2\ell{\cdot}\barv_1 {-}i\varepsilon)(2\ell{\cdot}\barv_2 {-} i\varepsilon)}
\left[1 {-} \medmath{\frac{1}{\bar{m}_2}\frac{\ell{\cdot}(\ell-q_2)}{(2 \ell \cdot \barv_2- i\varepsilon)}} \right],} \\
{\cut \,G_i^D \Big\vert_{\text{eik}} =} \adjustbox{valign=c,scale={0.65}{0.65}}{\input{tikz/nontrivial1}}
& \approx \int \frac{ 4 \pi^2 \dbar^\D \ell}{\Dden_1 \Dden_3 \Dden_4} \, 
    \frac{\delta(2 \ell \cdot \barv_1)}{\bar{m}_1 \bar{m}_2} \Big[ \delta (2 \ell \cdot \barv_2 ) + \medmath{\frac{\ell{\cdot}(\ell-q_2)}{\bar{m}_2}} \delta' (2 \ell \cdot \barv_2) \Big].
    \label{eq:CutGDeik}
\end{align} \label{eq:GvsGeik}\end{subequations}
For simplicity we display integrals with single powers $a_2=a_5=1$ of eikonal propagators, and $\approx$ means we kept only the terms up to the first subleading order in $1/\bar{m}_2$. In the last line, we ignore the $i\varepsilon$ in the denominator $\Dden_1$, since it cannot vanish.

When summing up these contributions, it matters whether the coefficient $c_i$ of the considered integral $G_i^A$ is odd or even in the $\bar{p}_i$.
As a first illustration, let us consider a numerator which is even and scales like the superclassical pentagon numerator: $c_{i}=\bar{p}_1^4\bar{p}_2^4=\bar{m}_1^4\bar{m}_2^4$.
The contribution of $c_i G_i$ to the time-ordered amplitude \eqref{amp 4 terms} is then the sum of the first four lines in \eqref{eq:GvsGeik}, which simplifies at leading power to
\be \label{Geik: iM}
i\cM[\bar{p}_1^4\bar{p}_2^4G_i]\,\Big\vert_{{\rm eik},\bar{m}_1^3\bar{m}_2^3}
= -4\pi^2\int \frac{\dbar^\D \ell}{\Dden_1 \Dden_3 \Dden_4}
\delta (2 \ell \cdot \barv_1)
\delta (2 \ell \cdot \barv_2)\,.
\ee
For the in-in expectation value, we similarly sum up all five lines and now find a simple zero:
\be \label{Geik: super 0}
\Exp_k[\bar{p}_1^4\bar{p}_2^4 G_i]\,\Big\vert_{{\rm eik},\bar{m}_1^3\bar{m}_2^3} = 0\,.
\ee
This shows that the superclassical contribution to the in-in expectation value neatly cancels, as was previously noted in \cite{Brandhuber:2023hhy,Herderschee:2023fxh,Elkhidir:2023dco,Georgoudis:2023lgf}.

When we look at the $1/\bar{m}_2$-suppressed pieces
in \eqref{eq:GvsGeik},
the relative signs between the five lines play out differently. By summing the first four and five lines, for the amplitude and in-in expectation value we find, respectively,
\begin{equation}
    \begin{split}
        \label{Geik: sub}
\left\{i\cM,\Exp_k\right\}&[\bar{p}_1^4\bar{p}_2^4 G_i]\Big\vert_{{\rm eik},\bar{m}_1^3 \bar{m}_2^2}
\\&= \int \frac{\dbar^\D \ell\,\ell{\cdot}(\ell-q_2)}{\Dden_1 \Dden_3 \Dden_4} \delta (2 \ell \cdot \barv_1) 
\left\{{\rm PV}\frac{-4\pi i}{(-2 \ell \cdot \barv_2)^2},
\frac{-4\pi i}{(-2 \ell \cdot \barv_2-i\eps)^2} \right\},
    \end{split}
\end{equation}
with principal value ${\rm PV}\frac{1}{x^2}=\frac12\left[\frac{1}{(x{+}i\varepsilon)^2}+\frac{1}{(x{-}i\varepsilon)^2}\right]$. The difference between the two cases is the $\delta'$ contribution in \eqref{eq:CutGDeik}.

Finally, let us consider the example of a numerator with classical scaling, and which is odd under $\bar{p}_2$: $c_{i}=\bar{p}_1^4\bar{p}_2^2 \bar{p}_2{\cdot}k\sim \bar{m}_1^4\bar{m}_2^3$.
Now, we only keep the leading term in each of \eqref{eq:GvsGeik}, but the relative signs again play out as in the preceding example:
\begin{equation}
    \begin{split}
        \label{Geik: odd}
\left\{i\cM,\Exp_k\right\}&[\bar{p}_1^4\bar{p}_2^2 \bar{p}_2{\cdot}k G_i]\,\Big\vert_{{\rm eik},\bar{m}_1^3 \bar{m}_2^2}\\&=
\bar{w}_2\int \frac{\dbar^\D \ell}{\Dden_1 \Dden_3 \Dden_4} \delta (2 \ell \cdot \barv_1)
\left\{
{\rm PV}\frac{2\pi i}{-2 \ell \cdot \barv_2},
\frac{2\pi i}{-2\ell\cdot\barv_2-i\varepsilon}\right\}\,.
    \end{split}
\end{equation}

There are a few things to note about the above expressions.  First, notice that pentagon integrals with classical-scaling numerators $\sim \bar{m}_1^4\bar{m}_2^3$ (or $\bar{m}_1^3\bar{m}_2^4$, which are similar, but not exemplified above) can come from either of two sources: from expansion of the original non-eikonal propagators as in \eqref{Geik: sub}, or from subleading numerators as in \eqref{Geik: odd}.
In both cases, the contribution to the scattering amplitude $i\cM$ has one heavy propagator on-shell and another with principal value.  These are precisely the integrals found in \cite{Brandhuber:2023hhy,Herderschee:2023fxh}.
Of course, these two sources are not fundamentally distinct since they both arise together if one expands the full expression for the integrand in $1/\bar{m}_i$, and more generally they can be related by integration-by-parts identities.

A second comment is that the in-in expectation value
from both sources \eqref{Geik: sub} and \eqref{Geik: odd} is \emph{not} given by a principal value integral. The difference comes respectively from the $\delta'$ and $\delta$ terms in the cut contribution \eqref{eq:CutGDeik}.
This difference is generally nonzero
(as is manifestly the case in \eqref{Geik: odd}, where it is a sign-definite integral). 
We conclude that the role of the $\Cut_{1'2'}$ term in \eqref{inclusive KMOC} is not \emph{solely} to cancel superclassical contributions as in \eqref{Geik: super 0}: it can also contribute to classical terms. This
is potentially in tension with conclusions reached in
\cite[Eqs.~(5.41) and (8.18)]{Brandhuber:2023hhy} and \cite[Sec.~2 and App.~A]{Herderschee:2023fxh}
regarding the cancellation of two-massive-particle-reducible (2MPR) graphs
(previously tested for $2\ot 2$ processes \cite{Brandhuber:2021eyq}).
Of course, this analysis does not exclude the possibility of nontrivial cancellations among diagrams: whether
these cut contributions
affect
the inferred gravitational waveform is a separate question which should be further analyzed.  From the discussion of infrared divergences in Sec.~\ref{ssec:IR} below, we however find this unlikely.

The above examples can be simply understood using the expansion in eikonal integrals introduced in \eqref{Expk 5 terms}-\eqref{eq:Gsum}. Completing $\ell$-dependent factors into inverse propagators similarly to \eqref{scalar box expansion}, the two examples correspond to the following expansions:
\begin{align}
 \bar{p}_1^4\bar{p}_2^4G^A_{a_1,1,a_3,a_4,1} &=
 \bar{m}_1^3\bar{m}_2^3\GeikA_{a_1,1,a_3,a_4,1}
+\frac12\bar{m}_1^3\bar{m}_2^2
 \Big(
 q_2^2\GeikA_{a_1,1,a_3,a_4,2}-\GeikA_{a_1-1,1,a_3,a_4,2}
  \nonumber\\&\hspace{55mm}
 -\GeikA_{a_1,1,a_3,a_4-1,2}\Big) +\ldots
\\
\bar{p}_1^4\bar{p}_2^2 \bar{p}_2{\cdot}k
 G^A_{a_1,1,a_3,a_4,1} &=
\bar{m}_1^3\bar{m}_2^2\bar{w}_2\GeikA_{a_1,1,a_3,a_4,1} + \ldots,
\end{align}
where the terms omitted do not contain the $\bar{m}_1^3\bar{m}_2^3$ or $\bar{m}_1^3\bar{m}_2^2$ powers exemplified above.
Thus, equations (\ref{Geik: iM}--\ref{Geik: odd}) are simply special cases of the general formulas for the even/odd combinations of eikonal integrals defined in \eqref{eq:Gsum}, namely
\begin{equation}
    \begin{split}
        \label{M and Exp from Geff}
&\begin{pmatrix}
i\cM [G_{a_1\ldots\, a_5}^{{\rm eik},--}]
&\Exp_k[G_{a_1\ldots\, a_5}^{{\rm eik},--}]
\\[3pt]
i\cM [G_{a_1\ldots\, a_5}^{{\rm eik},-+}]
&\Exp_k[G_{a_1\ldots\, a_5}^{{\rm eik},-+}]
\end{pmatrix}
 \\& \qquad\qquad\qquad\qquad\qquad =
\int \frac{\dbar^\D \ell\,\delta^{[a_2]}(2 \ell \cdot \barv_1)}{\Dden_1 \Dden_3 \Dden_4} 
\begin{pmatrix}
4\pi^2 \delta^{[a_5]}[-2\ell{\cdot}\bar{p}_2]&0\\[3pt]
{\rm PV}\frac{2\pi i}{(-2 \ell \cdot \barv_2)^{a_5}} &
\frac{2\pi i}{(-2 \ell \cdot \barv_2-i\varepsilon)^{a_5}}
\end{pmatrix}\,,
    \end{split}
\end{equation}
where the measure $\int \frac{\dbar^\D \ell}{\Dden_1 \Dden_3 \Dden_4}$ was defined in~\eqref{eq:elldenominator}. To summarize, we have seen that the $\cut_{1'2'}$ term in \eqref{inclusive KMOC} affects the classical part of waveform integrals.
Mathematically, it changes all PV integrals to the $-i\varepsilon$ denominators shown in the second line. 

The second line of the \eqref{M and Exp from Geff} expansion makes it clear that exclusive and inclusive amplitudes are distinct classical observables.  They are both computed using exactly the same classical numerators. 
They differ in $i\varepsilon$ and thus answer different questions, in much the same way that solving the same classical equations of motion with different boundary conditions answers different questions.

We can give the $\cut_{1'2'}$ term a simple physical interpretation.  The exclusive amplitude $i\cM$ is sensitive to interactions between the heavy particles after the graviton is emitted.  However, such interactions cannot physically contribute to the waveform, which can only depend on interactions in its past lightcone, as indeed made manifested by the in-in reduction formula \eqref{in-in reduction}.  The role of the $\cut_{1'2'}$ term is precisely to remove those interactions and restore causality.  This leads to the $i\varepsilon$ in the lower-right entry of \eqref{M and Exp from Geff}, which
corresponds to a retarded propagator enforcing the correct causal ordering of vertices along one of the heavy trajectories and therefore the correct boundary conditions.  In contrast, time-ordered integrals do not compute a causal waveform.

\paragraph{Check that classical cut contribution is nonzero}

This contribution from the cut does not vanish in general. To show explicitly that it contributes, let us compute it for the following topology,
\begin{equation}
G_{01011}^D \equiv 
\adjustbox{valign=c,scale={0.75}{0.75}}{\input{tikz/GDcut}}\,.
\end{equation}
This cut is an example of a ``two-matter-particle-reducible'' (2MPR) one in \cite{Brandhuber:2023hhy,Herderschee:2023fxh}.
The cut that mixes with the scalar integrals is given by
\begin{equation}
    \cut \, G_{11011}^{D} \Big|_{{\rm eik},1/\bar{m}_1\bar{m}_2} = 4 \pi^2
    \e^{\epsilon\gammaE} \int \frac{\d^\D\ell}{i\pi^{\D/2}} \frac{\delta_+ ({-}2 \ell \cdot \barv_1)
    \,\delta_+ ( 2 \ell \cdot \barv_2 )}{(\ell^2+i\varepsilon) \left[ (\ell{-}q_2)^2 - i \varepsilon \right]}\,.
\end{equation}
To compute the cut, we go to the center-of-mass frame of $\barv_2$ (as in \cite{Herderschee:2023fxh}). Recalling that $\barv_1 \cdot \barv_2 = - \bary$, $q_2 \cdot \barv_1 = - \barw_1$ and $q_2 \cdot \barv_2 = 0$, the momentum components become
\begin{equation}
    \barv_2 = \big(1,0,\vec{0}_T\big) \,, \qquad
    \barv_1 = \big(\bary,\sqrt{\bary^2-1},\vec{0}_T\big) \,, \qquad
    q_2 = \big(0,\frac{\barw_1}{\sqrt{\bary^2-1}}, \vec{q}_{2,T} \big) \,.
\end{equation}
The $\delta$-functions set $\ell^0=0$ and $\ell^1=0$, with a Jacobian of $\frac{1}{4 \sqrt{\bary^2-1}}$, which gives
\begin{equation}
    \cut \, G_{11011}^D \Big|_{{\rm eik},1/\bar{m}_1\bar{m}_2} = \frac{\pi^2 \e^{\epsilon\gammaE}}{\sqrt{\bary^2-1}}
 \int \frac{\d^{\D-2}\vec{\ell}_T}{i\pi^{\D/2}} \frac{1}{\vec{\ell}_T^2 \left[ (\vec{\ell}_T{-}\vec{q}_{2,T})^2 - (q_2^0)^2+(q_2^1)^2 \right]} \,,
\end{equation}
where we have denoted the transverse part of the momenta with $\vec{p}_T=(p^2,\cdots,p^\D)$. This last integral is now equivalent to a Euclidean bubble integral in $(\D-2)$ spacetime dimensions, and can easily be performed (e.g.,\ using Schwinger parameters) to give
\begin{subequations}
\begin{align}
    \cut \, G_{11011}^{D} \Big|_{{\rm eik},1/\bar{m}_1\bar{m}_2} & =
    \frac{i \pi \e^{\epsilon \gamma_E}}{\sqrt{\bary^2-1}} (q_2^2)^{-1-\epsilon} \Gamma(\epsilon) \, _2 F_1 \Big[-\epsilon,1+\epsilon,1-\epsilon,1-\frac{\barw_1^2}{q_2^2(\bary^2-1)} \Big]
    \\
    & = 
    \frac{i \pi}{\epsilon q_2^2 \sqrt{\bary^2-1}}
    \Bigg[ 1
    -
    \epsilon \log \frac{q_2^4 (\bary^2-1)}{\barw_1^2} + \mathcal{O} (\epsilon^2)\Bigg]
    \,.
    \label{eq:CutEvenDexp}
\end{align}
\end{subequations}
We will also compute this cut in the next subsection, since it is a contribution to $\{ \mathcal{I}_3^{-+} \}$ in~\eqref{KMOC integrals}. When accounting for the normalization factor of $2 \epsilon^2 q_2^2 \bary \sqrt{1-\bary^{-2}}$ given in~\eqref{I3remainingint}, the expression in the square brackets for $\{\mathcal{I}_3^{-+}\}$ in ~\eqref{KMOC integrals} is precisely $\Cut\, G^{\rm eik,D}_{11011}=\cut \, G_{11011}^{D} \Big|_{{\rm eik},1/\bar{m}_1\bar{m}_2}$, and agrees with~\eqref{eq:CutEvenDexp} above. 

We can also compute the subleading $\frac{1}{\bar{m}_1\bar{m}_2^2}$ term of the cut, as given by the second term of Eq.~\eqref{eq:CutGDeik},
\begin{equation}
    \cut \, G_{11011}^{D} \Big|_{{\rm eik},1/\bar{m}_1\bar{m}_2^2} = 4 \pi^2
    \e^{\epsilon\gammaE} \int \frac{\d^\D\ell}{i\pi^{\D/2}} \frac{\ell \cdot (\ell{-}q_2)}{(\ell^2+i\varepsilon) \left[ (\ell{-}q_2)^2 - i \varepsilon \right]}
    \,\delta_+ (2 \ell \cdot \barv_1)
    \,\delta_+' ( 2 \ell \cdot \barv_2 ) \,.
\end{equation}
As in~\eqref{scalar box expansion}, we can rewrite the numerator as $\ell \cdot (\ell{-}q_2)= -\frac{1}{2} \left(q_2^2-\ell^2-(\ell{-}q_2)^2 \right)$.
Recall that when integrating over the derivative of the delta function, we must differentiate the integrand in $\ell^0$ before setting $\ell^0=0$. Putting everything together results in
\begin{equation}
    \cut \, G_{11011}^D \Big|_{{\rm eik},1/\bar{m}_1\bar{m}_2^2} = -\frac{\bary \pi^2\e^{\epsilon\gammaE}}{2 (\bary^2-1)^{3/2}} \int \frac{\d^{\D-2}\vec{\ell}_T}{i\pi^{\D/2}} \frac{ q_2^1\big(q_2^2 - \vec{\ell}_T^2 \big)}{\vec{\ell}_T^2 \left[ (\vec{\ell}_T{-}\vec{q}_{2,T})^2 - (q_2^0)^2+(q_2^1)^2 \right]^2} \,.
    \label{eq:cutres1}
\end{equation}
This last expression is a sum of a bubble with a doubled propagator and a tadpole. The integrals are straightforward to compute, and the result is
\begin{subequations}
\begin{align}
    \cut \, G_{11011}^{D} \Big|_{{\rm eik},1/\bar{m}_1\bar{m}_2^2} & =
    - \frac{i \pi^2 \e^{\epsilon \gamma_E} \bary (q_2^2)^{-1-\epsilon} }{2 (\bary^2-1)^{3/2} \,\barw_1 \, \Gamma(1-\epsilon) \sin (\pi \epsilon)} 
    \Bigg\{ \epsilon [q_2^2(\bary^2-1)]^{1+\epsilon} (\barw_1^2)^{-\epsilon}
    \nonumber
    \\
    & \hspace{1.5cm} +
    (1+\epsilon) \barw_1^2\, _2 F_1 \Big[-\epsilon,2+\epsilon,1-\epsilon; 1-\frac{\barw_1^2}{q_2^2(\bary^2-1)} \Big] \Bigg\}
    \\
    & = - \frac{i \pi \barw_1 \bary}{2 \epsilon q_2^2 (\bary^2-1)^{3/2} } \left[ 1 - \epsilon \log \frac{q_2^4 (\bary^2-1)}{w_1^2} + 2\epsilon + \mathcal{O} (\epsilon^2) \right] 
    \label{eq:expCutD}\,.
\end{align}
\end{subequations}
We can check using identities for the hypergeometric functions that the two cuts computed here are consistent with the integration-by-part identity recorded in~\eqref{reduced triangle}, where only a single term has the cut:
\begin{equation}
    \cut \, G_{11011}^{D} \Big|_{{\rm eik},1/\bar{m}_1\bar{m}_2^2} = \frac{(\D-5)\barw_1 \bary}{2 (\bary^2-1)} \cut \, G_{11011}^{\rm eik,D} \,,
    \label{eq:cutres2}
\end{equation}
where we have used that the first term in~\eqref{reduced triangle} is the only one giving a cut contribution to $\Exp_k$.

This confirms that in practice one can always eliminate $\delta'$ eikonal integrals using integration-by-parts identities, which reduce all computations to the master eikonal integrals recorded in \eqref{KMOC integrals} below.

To obtain the contribution to the waveform, we must Fourier transform these expressions to impact-parameter space by integrating over $q_i$ according to, e.g.,~\cite{Kosower:2018adc,Cristofoli:2021vyo,Brandhuber:2023hhy,Herderschee:2023fxh}.
From the non-analytic $q_2$ dependence in~\eqref{eq:CutEvenDexp} and~\eqref{eq:expCutD}, it is clear that the Fourier transform of the above scalar-box cut integral is nonvanishing at non-zero impact parameter.

\subsection{Computation of eikonal master integrals}\label{ssec:masterEik}

Having explained in Sec.~\ref{ssec:eik}
the structure of large-mass expansions in terms of
eikonal and hard integrals,
let us finally turn to the evaluation of a basis of master
integrals for the family of eikonal pentagons
defined in \eqref{eik topology}.
We will use the differential equation method and start by providing boundary conditions in the Regge-soft defined below.

Because this is a one-loop problem, it is straightforward to come up with a canonical basis of integrals (one in which the differential equation is linear in $\epsilon$ with vanishing constant term \cite{Henn:2013pwa}): there is a single integral per topology and we only need to normalize it canonically.
We label the integrals as in \cite{Brandhuber:2023hhy}.
There are 9 topologies, making up a total of 16 integrals including permutations.
Three topologies are contact integrals, denoted with a tilde:
\begin{subequations}
\begin{alignat}{4}
 \ItA_{1} &= \epsilon \barw_1  \GeikA_{01020} &=&
 \adjustbox{valign=c,scale={0.65}{0.65}}{\input{tikz/Itilde1}} 
 ~,\quad 
  &\ 
 \ItA_{2} &= 2\epsilon^2 \bary \sqrt{1-\bary^{-2}}\, \GeikA_{01011} &=&
 \adjustbox{valign=c,scale={0.65}{0.65}}{\input{tikz/Itilde2}} 
~,
\label{eq:Itilde2}
\\
 \ItA_{4} &= 4\epsilon^2 \barw_1\barw_2\, \GeikA_{01111} &=& 
 \adjustbox{valign=c,scale={0.65}{0.65}}{\input{tikz/Itilde4}} 
 ~.&&&&
\end{alignat}
\end{subequations}
The remaining six integrals are:
\begin{subequations}
\begin{alignat}{4}
 \IA_0 &= -\epsilon q_2^2\, \GeikA_{10020} &=&
 \adjustbox{valign=c,scale={0.65}{0.65}}{\input{tikz/Itilde0}} 
,\quad &\
 \IA_1 &= 2\epsilon^2\sqrt{q_2^2+\barw_1^2}\, \GeikA_{11010} &=& \adjustbox{valign=c,scale={0.65}{0.65}}{\input{tikz/I1}} 
, \\
 \IA_2 &= 2\epsilon^2\sqrt{q_1^2}\,\GeikA_{11100} &=&
 \adjustbox{valign=c,scale={0.65}{0.65}}{\input{tikz/I2}} 
, &\ 
 \IA_3 &= 2\epsilon^2q_2^2\bary \sqrt{1-\bary^{-2}}\,\GeikA_{11011}\, &=&
 \adjustbox{valign=c,scale={0.65}{0.65}}{\input{tikz/I3}} 
 \label{I3remainingint}
, \\
 \IA_5 &= -2\epsilon^2q_1^2\barw_1\, \GeikA_{11110}\, &=&
 \adjustbox{valign=c,scale={0.65}{0.65}}{\input{tikz/I5}} 
, &\ 
 \IA_6 &= \epsilon^3\sqrt{\rm Gram}\,G^{\rm eik,A\,(6d)}_{11111} &=&
 \adjustbox{valign=c,scale={0.65}{0.65}}{\input{tikz/I6}} 
. \label{remaining ints}
\end{alignat}
\end{subequations}
Finally, seven additional integrals are permutations of the above:
\begin{equation}
    \left\{
\tilde{\mathcal{I}}_1^{A\prime},
\tilde{\mathcal{I}}_2^{A\prime},
\mathcal{I}_0^{A\prime},
\mathcal{I}_1^{A\prime},
\mathcal{I}_2^{A\prime},
\IA_4,
\mathcal{I}_5^{A\prime}\right\}
 \equiv
\left\{
    \ItA_1,
    \ItA_2,
\IA_0,\IA_1,\IA_2,\IA_3,\IA_5\right\}_{1\leftrightarrow 2}.
\label{eq:Itilde_perms}
\end{equation}
More precisely, the permutation interchanges $\{ \barw_1,q_1,a_2,a_3\} \leftrightarrow\{ \barw_2,q_2,a_5,a_4\}$. Note that $\IA_0$ and the integrals with a prime were not discussed in \cite{Brandhuber:2023hhy} since they do not contain a $1/(\ell{\cdot}\barv_1)$ propagator that can be cut; Ref. \cite{Brandhuber:2023hhy}  focused on the ``single cut'' proportional to $\delta(\ell{\cdot}\barv_1)$
and a principal-value propagator.
For the reasons explained below \eqref{M and Exp from Geff}, we organize calculations in terms of conventional time-ordered propagators and their cuts and we do not use principal-value propagators.\footnote{
We warn the reader that our precise definitions differ by various constants and
we also use the opposite metric signature.}

In \eqref{remaining ints}, the pentagon integral $\IA_6$ is dimensionally-shifted from $6-2\epsilon$ spacetime dimensions. The integrated expression for it is usually not needed in practice since it only contributes to amplitudes at $\mathcal{O}(\epsilon)$; for this reason, we do not spell out the Gram determinant in its normalization and do not discuss it further.

We define master integrals for the $B$, $C$ and $D$ topologies by the same expressions with appropriate permutations, which change the signs of $\barw_i$ and $\bary$ as in \eqref{Expk 5 terms eik}.
To avoid sign confusion, we wrote all square roots in such a way that their arguments are always positive, and we always choose the principal branch. 

A technical simplification of the eikonal limit is that particle masses have disappeared from the problem. The effect of this is reflected in the differential equation, which simplifies considerably in this limit; we record it in App.~\ref{sec:eikonalDiffEq}. The counting of master integrals is also different---if we had kept two distinct nonvanishing masses $m_1\neq m_2$, we would have found $22$ master integrals for this topology instead of $16$ in the eikonal limit \cite{Smirnov:2019qkx,Klappert:2020nbg}. 

The differential equation enables us to focus on integration constants in suitable limits.
We find convenient the \emph{Regge-soft} limit
in which we take the emitted graviton to be soft, $\barw_i^2\ll q_i^2$, and subsequently take $|\bary|$ to be large,  $|\bary|\gg 1$ (which implies that $q_1^2\approx q_2^2\to q^2$
for real external momenta).\footnote{
The order of limits is important. Other scalings could be considered, for example one could take the limit with $w_i y$ fixed, however these generally lead to more complicated formulas.
}
For the conventional time-ordered amplitudes, we then find the following limits:
\be\begin{aligned} \label{eikonal MRK}
\ItA_1, \IA_5 &\to r_\Gamma\frac{(2W_1)^{-2\epsilon}}{2\cos(\pi\epsilon)}\,,
& \hspace{0cm}
\IA_1,\IA_2&\to c_1 q^{-2\epsilon} 4^{\epsilon}\sin(\pi\epsilon)^2\,,
\\
\ItA_2 &\to \ItA_1 -\frac{c_1}{2}W_1^{-2\epsilon}Y^{2\epsilon}\,,
& \hspace{0cm}
\IA_3&\to\ItA_2+2r_\Gamma q^{-2\epsilon}\epsilon \log(2Y)\,,
\\
\ItA_4 &\to 
\begin{aligned}
    r_\Gamma &W_1^{-\epsilon}W_2^{-\epsilon}Y^\epsilon \frac{2^{1-\epsilon}\pi \epsilon}{\sin(\pi \epsilon)}\\&-\frac{c_1}{2}Y^{2\epsilon}\left[W_1^{-2\epsilon} + W_2^{-2\epsilon}\right]
\end{aligned},
& \hspace{0cm}
\IA_0&\to r_\Gamma q^{-2\epsilon}\,,
\end{aligned}\ee
where $W_i\equiv -\barw_i-i\varepsilon$,\ $Y\equiv -\bary-i\varepsilon$. Remaining integrals are given by simple permutations:
\begin{align}
    \left\{\ItA_1{'},\ItA_2{'},\IA_4,\IA_5{'}\right\}&\to
\left\{\ItA_1,\ItA_2,\IA_3,\IA_5\right\}\Big\vert_{W_1 \mapsto W_2}\,,\\
\quad
\left\{\IA_0{'},\IA_1{'},\IA_2{'}\right\}&\to
\left\{\IA_0,\IA_1,\IA_2\right\}\,.
\end{align}
The two constants are
\begin{equation}
    r_\Gamma= \frac{\e^{\epsilon\gammaE} \Gamma(1-\epsilon)^2\Gamma(1+\epsilon)}{\Gamma(1-2\epsilon)} \quad \text{and} 
\quad
c_1 = \e^{\epsilon\gammaE} \Gamma(1{+}\epsilon)\Gamma(1{-}2\epsilon)\Gamma(1{+}2\epsilon)\,.
\end{equation}
These limits were obtained using the method of regions for the Regge-soft limit; an example of how to compute $\ItA_2$ is shown in App.~\ref{sec:eikonalDiffEq}.
A useful observation is that the integrals or combinations $\IA_0$, $\IA_1$, $\IA_2$, $(\IA_3-\ItA_2)$ and $(\IA_5-\ItA_1)$ all have smooth soft limits as $\barw_i\to 0$, as can be verified independently from the differential equation.  This feature makes it possible to obtain the boundary constants of all non-contact diagrams in terms of contact integrals, or integrals in which we set $\barw_i=0$. 
In addition, we have tested the above by integrating the differential equation to the back-to-back limit $\bary\to -1$, where integrals simplify.
We do not include a boundary constant for $\mathcal{I}_6$ as it decouples in the limit $\D\to 4$.

We stress that \eqref{eikonal MRK} apply for the standard (Feynman) $i\varepsilon$, so standard methods are available to verify the integrals and it is possible to cross-check with the literature as discussed below. From now on, we will focus on the relationship between time-ordered and in-in amplitudes.

A useful feature of the Regge-soft limit is that \eqref{eikonal MRK} uniformly describes
all topologies (see Fig.~\ref{fig:ABCD topologies}),
by simply accounting for the phases dictated by the sign of $\barw_1$ after permutation, namely
\be
 W_1^{-2\epsilon} = \left\{\begin{array}{ll}
     |\barw_1|\e^{-2i\pi\epsilon}& \mbox{for $A$ and $C$}, \\
     |\barw_1|\,& \mbox{for $B$ and $D$}.
 \end{array}\right.
\ee
It is somewhat surprising to find such a ``one case fits all'' formula but here this can be understood as follows.  The contact integrals $\ItA_k$ are guaranteed to be analytic in upper-half planes ${\rm Im}\, \barw_i>0$, since the dependence on these variables enters through half-sided Fourier transforms $\int_0^{\infty}\d t\, \e^{-i t \barv_i{\cdot}k}$.
The singular terms in the soft limit $\barw_i\to 0$ of the remaining integrals are then controlled by an integration region where the third and fourth propagator are small, which is equivalent to a contact integral.  Thus, all one-loop integrals for this process can be expanded into a part which is analytic in upper-half $\barw_i$ planes, plus parts that are nonsingular as $\barw_i\to 0$.
We would not expect such a clean decomposition,
and therefore we also do not expect such uniform formulas, for generic two-loop integrals.

Having described the time-ordered integrals we can immediately move on to the in-in expectation value. The extra ingredient is the $\cut_{1'2'}$ term in \eqref{inclusive KMOC}, which only affects $D$ topologies.  We take advantage of the following simple observation: the only possible cut of the $D$ integrals is the one in the $1'2'$ channel!  Thus, the Cutkosky rules simply relate $\cut_{1'2'}$ to the imaginary part of $D$ integrals:
\be
\cut_{1'2'}\,\mathcal{I}_i^D=-2i\,{\rm Im}\, \mathcal{I}_i^D
\qquad\mbox{(for the considered one-loop integrals)}.
\ee
The contribution of the $D$ topology to the in-in expectation value in \eqref{Expk 5 terms eik} can thus be simply obtained from its conventional (time-ordered) expression:
\begin{equation} \label{one loop Expk}
\Exp_k\left[\mathcal{I}_i^D\right]
\equiv \mathcal{I}_i^D + \cut_{1'2'}
\mathcal{I}_i^D = (\mathcal{I}_i^D)^*\,.
\end{equation}
This is somewhat similar to what was observed for bubble integrals in \eqref{cut from Im Sigma}.
We stress that this simplification is specific to the one-loop integrals under consideration.  Integrals of the remaining topologies ($A$, $B$ and $C$) do not get complex-conjugated and contribute the same to $i\cM$ as to $\Exp_k$. 
Let us exemplify this by using \eqref{eikonal MRK} to predict the contribution from $\tilde{\mathcal{I}}_4$ to the in-in expectation value in the Regge-soft limit. We find
\begin{equation} \label{Expk Itilde4}
\newcommand{\bl}[1]{{\color{RoyalBlue}#1}}
\newcommand{\ph}{\phantom{\e^{2\pi i\epsilon}}}
\begin{aligned} 
{\rm Exp}_k\left[\tilde{\mathcal{I}}_4^A\right]
 &\to\bl{\e^{i\pi\epsilon}}r_\Gamma \barw_1^{-\epsilon}\barw_2^{-\epsilon}\bary^\epsilon 2^{1-\epsilon} \frac{\pi \epsilon}{\sin(\pi \epsilon)}-\frac{c_1}{2}\left[\ph \barw_1^{-2\epsilon}\bary^{2\epsilon} + \ph \barw_2^{-2\epsilon}\bary^{2\epsilon}\right],
\\
{\rm Exp}_k\left[\tilde{\mathcal{I}}_4^B\right]
 &\to \bl{\e^{i\pi\epsilon}}r_\Gamma \barw_1^{-\epsilon}\barw_2^{-\epsilon}\bary^\epsilon 2^{1-\epsilon} \frac{\pi \epsilon}{\sin(\pi \epsilon)}-\frac{c_1}{2}\left[\ph \barw_1^{-2\epsilon}\bary^{2\epsilon} + \bl{\e^{2\pi i \epsilon}} \barw_2^{-2\epsilon}\bary^{2\epsilon}\right],
\\
{\rm Exp}_k\left[\tilde{\mathcal{I}}_4^C\right]
 &\to \bl{\e^{i\pi\epsilon}}r_\Gamma \barw_1^{-\epsilon}\barw_2^{-\epsilon}\bary^\epsilon 2^{1-\epsilon} \frac{\pi \epsilon}{\sin(\pi \epsilon)}-\frac{c_1}{2}\left[\bl{\e^{2\pi i \epsilon}}\barw_1^{-2\epsilon}\bary^{2\epsilon} + \ph \barw_2^{-2\epsilon}\bary^{2\epsilon}\right],
\\
{\rm Exp}_k\left[\tilde{\mathcal{I}}_4^D\right]
 &\to \bl{\e^{i\pi\epsilon}}r_\Gamma \barw_1^{-\epsilon}\barw_2^{-\epsilon}\bary^\epsilon 2^{1-\epsilon} \frac{\pi \epsilon}{\sin(\pi \epsilon)}-\frac{c_1}{2}\left[\bl{\e^{2\pi i \epsilon}}\barw_1^{-2\epsilon}\bary^{2\epsilon} + \bl{\e^{2\pi i \epsilon}}\barw_2^{-2\epsilon}\bary^{2\epsilon}\right].
\end{aligned}\end{equation}
In each line, we have translated to the common $\barw_i>0$ and $\bary>0$ variables of the $A$ topology, such that all phases are explicit (and highlighted in blue).
As another example, we find for the Regge-soft limit of $\mathcal{I}_3$
\begin{equation}
\newcommand{\bl}[1]{{\color{RoyalBlue}#1}}
\newcommand{\ph}{\phantom{\e^{2\pi i\epsilon}}}
\begin{aligned} 
{\rm Exp}_k\left[\mathcal{I}_3^A\right]
 &\to \bl{\e^{2i\pi\epsilon}}r_\Gamma\frac{(2\barw_1)^{-2\epsilon}}{2\cos(\pi\epsilon)}\ph-\frac{c_1}{2}\barw_1^{-2\epsilon}\bary^{2\epsilon} +2r_\Gamma q^{-2\epsilon}\epsilon\left(\log(2\bary)\bl{-i\pi}\right)\, ,\\
{\rm Exp}_k\left[\mathcal{I}_3^B\right]
 &\to \ph r_\Gamma\frac{(2\barw_1)^{-2\epsilon}}{2\cos(\pi\epsilon)}\ph-\frac{c_1}{2}\barw_1^{-2\epsilon}\bary^{2\epsilon} +2r_\Gamma q^{-2\epsilon}\epsilon\log(2\bary)\, ,\\
{\rm Exp}_k\left[\mathcal{I}_3^C\right]
 &\to \bl{\e^{2i\pi\epsilon}}r_\Gamma\frac{(2\barw_1)^{-2\epsilon}}{2\cos(\pi\epsilon)}-\bl{\e^{2i\pi\epsilon}}\frac{c_1}{2}\barw_1^{-2\epsilon}\bary^{2\epsilon} +2r_\Gamma q^{-2\epsilon}\epsilon\log(2\bary)\, ,\\
{\rm Exp}_k\left[\mathcal{I}_3^D\right]
 &\to \ph r_\Gamma\frac{(2\barw_1)^{-2\epsilon}}{2\cos(\pi\epsilon)}-\bl{\e^{2i\pi\epsilon}}\frac{c_1}{2}\barw_1^{-2\epsilon}\bary^{2\epsilon} +2r_\Gamma q^{-2\epsilon}\epsilon\left(\log(2\bary)\bl{+i\pi}\right)\, .
\end{aligned}\label{Expk: I3}\end{equation}
The Regge-soft limit contributions to the in-in expectation value from the remaining masters are obtained in a similar fashion, since the formulas in \eqref{eikonal MRK} involve the same ingredients.

Finally, we can collect the four topologies into the definite-parity combinations \eqref{eq:Gsum}, which control individual orders in the $1/\bar{m}_i$ expansion.  For odd-odd terms (e.g., order $\bar{m}_1^3\bar{m}_2^3$),
it is easy to check in the examples
\eqref{Expk Itilde4} and \eqref{Expk: I3}
that the four lines with appropriate signs sum up to zero. More generally one can verify that
\be
 \Exp_k\left[\mathcal{I}_{i}^{--}\right]=0\,,
\ee
for all master integrals in accordance with the cancellation of superclassical terms noted previously
in \eqref{Geik: super 0} and \eqref{M and Exp from Geff}.
This cancellation crucially relies on including the $\cut_{1'2'}$ contribution, which led to the $D$ topology being complex conjugated in \eqref{one loop Expk}.

For odd-even terms (e.g., order $\bar{m}_1^3\bar{m}_2^2$) we generally find non-vanishing answers, for example
\begin{subequations}
\begin{align}
\Exp_k\left[\mathcal{I}_3^{-+}\right] &=
\Exp_k\left[ \mathcal{I}_3^A-\mathcal{I}_3^B+\mathcal{I}_3^C-\mathcal{I}_3^D\right]
\\
&=\mathcal{I}_3^A-\mathcal{I}_3^B+\mathcal{I}_3^C-(\mathcal{I}_3^D)^*
\\
&\to \left(\e^{2\pi i \epsilon}-1\right)r_\Gamma \frac{(2\barw_1)^{-2\epsilon}}{\cos(\pi\epsilon)}
-4\pi i r_\Gamma q^{-2\epsilon}\epsilon\,,
\end{align}
\end{subequations}
in the Regge-soft limit.
(Note that relative signs on the first line include the effect of $\bary$ in the normalization of $\mathcal{I}_3$
in addition to those from the propagators.)
By integrating these boundary conditions order-by-order in $\epsilon$ using the differential equation,
we can then obtain results for the contribution of all the master integrals to the inclusive expectation values.
Working up to transcendental weight 2 (as relevant for the four-dimensional limit of observables)---that is dropping $\mathcal{O}(\epsilon^3)$ terms in canonically normalized integrals---the integration is trivial and we finally find for all odd-even contributions (all logarithms become dimensionless after multiplying by the dimensional regularization $\mu^{2\epsilon}$):
\begin{equation}\label{KMOC integrals}
\def\expk#1{\left\{\frac{#1}{2\pi i\epsilon}\right\}}
\begin{aligned}
\expk{\tilde{\mathcal{I}}_1^{-+}}
&= 1+i\pi\epsilon+ \epsilon \log \frac{1}{4\barw_1^2}\,,
&&
\\
\expk{\mathcal{I}_1^{-+}} &= -2\pi i \epsilon +2\epsilon\log \frac{\sqrt{q_2^2+\barw_1^2}+\barw_1}{\sqrt{q_2^2+\barw_1^2}-\barw_1},
&
\expk{\mathcal{I}_2^{-+}} &= -2\pi i \epsilon\,,
\\
\expk{\tilde{\mathcal{I}}_2^{-+}} &= \epsilon (i\pi{-}2\eta) +
\left[ 1+\epsilon \log\frac{\bary^2{-}1}{\barw_1^2}\right],
&
\expk{\tilde{\mathcal{I}}_2^{\prime -+}} &= i\pi\epsilon +\left[ 1+\epsilon \log\frac{\bary^2{-}1}{\barw_2^2}\right],
\\
\expk{\mathcal{I}_3^{-+}} &=\epsilon(i\pi{-}2\eta) +\left[-1+\epsilon \log\frac{q_2^4(\bary^2{-}1)}{\barw_1^2}\right],
&
\expk{\mathcal{I}_4^{-+}} &=i\pi\epsilon + \left[ -1+\epsilon \log\frac{q_1^4(\bary^2{-}1)}{\barw_2^2}\right],
\\
\expk{\tilde{\mathcal{I}}_4^{-+}} &=1 +i\pi \epsilon + \epsilon\log \frac{1}{4\barw_2^2} + \Big[2\epsilon\eta\Big],&
\expk{\mathcal{I}_5^{-+}} &= 1 +i\pi \epsilon + \epsilon\log \frac{q_2^4}{4\barw_1^2q_1^4}\,,
\end{aligned}
\end{equation}
where $\{\ldots\}$ is a shorthand for ${\rm Exp}_k \{\cdots\}$ and $\eta=\log(\bary+\sqrt{\bary^2-1})$.
All remaining integrals vanish 
\begin{equation}
    \{\mathcal{I}_0^{-+},\mathcal{I}_0^{\prime-+},\tilde{\mathcal{I}}_1^{\prime-+},\mathcal{I}_1^{\prime-+},\mathcal{I}_2^{\prime-+},\mathcal{I}_5^{\prime-+} \}=0\,,
\end{equation}
which can be understood from \eqref{M and Exp from Geff} and their lack of an $\ell \cdot \bar{v}_1$ cut.
In \eqref{KMOC integrals}, we have separated the time-ordered amplitudes outside the square brackets, while the square brackets display the additional contributions from ${\rm Cut}_{1'2'}$ in \eqref{inclusive KMOC}.

As a concluding remark to this subsection, we note that computing the cut contribution \eqref{inclusive KMOC} at one-loop is somewhat trivial as it is proportional to the imaginary part of \emph{certain diagrams} contributing to the amplitude, by the Cutkosky rules. 
The situation becomes more complicated at higher loops, where there are diagrams for which ${\rm Cut}_{1'2'}$
is not simply related to the imaginary part of the diagram, for example 
\begin{equation}
\underbracket[0.4pt]{\adjustbox{valign=c,scale={0.75}{0.75}}{\input{tikz/twoloop2}}\,,\adjustbox{valign=c,scale={0.75}{0.75}}{\input{tikz/twoloop3}}\,,\adjustbox{valign=c,scale={0.75}{0.75}}{\input{tikz/twoloop4}}\,.}_{\text{($\cut_{1'2'}$ is not equal to imaginary part of uncut diagram)}}
\end{equation}
Calculations of waveforms at the next order may thus require explicit computations of cut Feynman integrals,
perhaps exploiting reverse unitarity \cite{Anastasiou:2002yz,Herrmann:2021lqe} or other ideas.

\subsection{Infrared divergence} \label{ssec:IR}

The infrared divergent part of the five-point one-loop exclusive amplitudes in the eikonal limit
was calculated using Weinberg's soft factors in
\cite{Brandhuber:2023hhy,Herderschee:2023fxh,Elkhidir:2023dco,Georgoudis:2023lgf} and shown to be have a classical part which is
a simple phase: 
\begin{equation}
\label{M IR classical}
    \mathcal{M}_{3\ot 2}^{(1\text{-loop})}\Big\vert_{\rm div,cl.}
    =  \mathcal{M}_{3\ot 2}^{({\rm tree})} \times -2i \GN \, k{\cdot}(\bar{p}_1+\bar{p}_2) \log\frac{\Lambda}{\mu_{\rm IR}}\,,
\end{equation}
with $\Lambda\sim b^{-1}$ being a relevant momentum scale.
It has a simple interpretation, as discussed in these references: it represents a classical phase shift whose $k$-derivative is a logarithmically divergent time delay
in the retarded time of the observed radiation.
This is simply the Shapiro time delay suffered by the massless radiation as it escapes the bodies' gravity field. If the observation is performed at a distance $r_{\rm obs}\sim \mu_{\rm IR}^{-1}$, 
its logarithmically divergent part is
\begin{align}
 \Delta t_{\rm obs}=
   -2 \GN\, \hat{k}{\cdot}(\bar{p}_1+\bar{p}_2) \log\frac{r_{\rm obs}}{b}\,,
\label{Shapiro}
\end{align}
with $\hat{k}^\mu = k^\mu/k^0$. (This $\Delta t_{\rm obs}$ is a positive number
since we work with mostly-plus signature.)

We believe that the above result, while correct for exclusive amplitudes, is incomplete in the context of inclusive radiation produced following a two-body collision,
where the initial conditions are set in the far past.  One must then also account for the propagation of the massive bodies towards their interaction point, which leads to an additional divergent time shift $\Delta t_{\rm in}$.

It can be readily calculated by considering geodesics in a Schwarzschild black hole of total mass $M$:
\begin{equation}
    \d s^2 = -\d t^2(1-2\GN M/r) + \frac{\d r^2}{1-2\GN M/r}+ r^2 \d\Omega_2^2\,.
\end{equation}
We can write the geodesic equation
in terms of the physical radial
velocity $v_r\equiv \frac{\d r/\d t}{1-2\GN M/r}$ as:
\begin{equation}
   \frac{\d}{\d t} v_r = \frac{-\GN M}{r^2} (1-v_r^2) + \mathcal{O}(r^{-3})\,. \label{geodesic large r}
\end{equation}
Integrating in a large-$r$ expansion one finds the following relation between time and radius 
\begin{equation} \label{t matter}
    t = -\frac{r}{v_\infty}
    +\frac{\GN M}{v_{\infty}^3}(3v_\infty^2-1)\log\frac{r_{\rm in}}{r} + \mathcal{O}(r^{-1})\,,
\end{equation}
where we integrated from a large distance $r_{\rm in}$ that initially separates the bodies, where their velocity is $v_{\infty}>0$.
Setting $r\sim b$ gives the time shift $\Delta t_{\rm in}$ of the point of closest approach. The coefficient of the logarithmic divergence does not depend on angular momentum, which enters only in the suppressed terms in \eqref{geodesic large r}.

The $\GN M$ term in \eqref{t matter} predicts a negative time shift at nonrelativistic velocities (i.e. a time advance), which is easy to understand from Newton's law
since the test particle always moves slightly faster than its asymptotic velocity due to the attractive nature of gravity. At ultrarelativisitic velocities the effect flips sign and gives a time delay consistent with \eqref{Shapiro}.

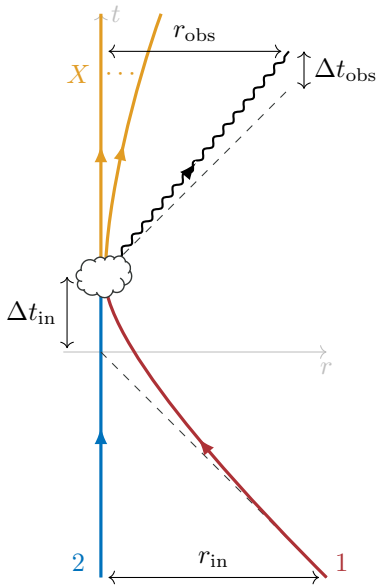
\begin{figure}
\centering
\begin{tikzpicture}
\tikzset{ar/.style={decoration={
				markings,
				mark=at position 0.5 with {\arrow{latex}}},postaction={decorate}}};
\draw[lightgray, ->] (-0.5,0) -- (3,0) node[below] {$r$};
\draw[lightgray, ->] (0,-3) -- (0,4.5) node[right] {$t$};
\draw[RoyalBlue, very thick, ar] (0,-3) -- (0,1);
\draw[charcoal, dashed] (0,0) -- (3,-3);
\draw[<->] (0.1,-3) -- (2.9,-3) node[midway, above] {$r_{\rm in}$};
\draw[Maroon, very thick, ar] (3,-3) to[out=135,in=-90,distance=1cm] (0.05,1);
\draw[yellowish, very thick, ar] (0.06,1) to[out=90,in=-110,distance=1cm] (0.8,4.5);
\draw[yellowish] (-0.3,3.7) node {$X$};
\draw[yellowish] (0.29,3.7) node {$\cdots$};
\draw[RoyalBlue] (-0.3,-2.8) node {$2$};
\draw[Maroon] (3.2,-2.8) node {$1$};
\draw[yellowish, very thick, ar] (0,1) -- (0,4.5);
\draw[charcoal, dashed] (0,1) -- (2.5,3.5);
\draw[photon, thick] (0,1) to[] (2.5,4);
\draw[<->] (0.1,4) -- (2.4,4) node[midway, above] {$r_{\rm obs}$};
\draw[white, very thick, decoration={
		markings,
		mark=at position 0.5 with {\arrow[black]{latex}}},postaction={decorate}] (1.25, 2.5) -- (1.25+0.025, 2.5+0.03); 
\node[scale=0.7] at (0,1) {\cloud};
\draw[<->] (-0.45,0.05) -- (-0.45,1) node[midway,left] {$\Delta t_{\rm in}$};
\draw[<->] (2.7,3.5) -- (2.7,4) node[midway,right] {$\Delta t_{\rm obs}$};
\end{tikzpicture}
\caption{\label{fig:delay}Illustration of the total time delay $\Delta t_{\rm tot} = \Delta t_{\rm in} + \Delta t_{\rm obs}$, given in \eqref{Delta t tot}, for the inclusive radiation produced in a two-body collision whose initial conditions are set in the far past. Dashed lines are tangent to asymptotic velocities. For
non-relativistic objects, $\Delta t_{\rm in}$ is a time advance.}
\end{figure}

It is easy to covariantize the above result
by writing the velocity in terms of rapidity,
$v_\infty=\sqrt{1-\bar{y}^{-2}}$,
and by summing up the delays of each body treated as a test particle in the field of the other, $\Delta t_{\rm tot} = \Delta t_{\rm in} + \Delta t_{\rm obs}$ (which should be complete since this is essentially linear physics).
Combined with \eqref{Shapiro}, this gives the total infrared-divergent offset in the retarded time of observed radiation:
\begin{equation} \label{Delta t tot}
    \Delta t_{\rm tot} = 
   -\GN \, \hat{k}{\cdot}(\bar{p}_1+\bar{p}_2)
   \times\left(
   2\log\frac{r_{\rm obs}}{b}
   +\frac{2-3\bar{y}^{-2}}{(1-\bar{y}^{-2})^{\frac32}}
   \log\frac{r_{\rm in}}{b}\right) + \mbox{finite} \,.
\end{equation}
To be clear, the physical setup here is that the bodies are sent with prescribed initial velocity when they are separated by a distance $r_{\rm in}$, at a time such their naive (straight line) trajectories would reach closest approach near the origin.  The actual point of closest approach, around which radiation is emitted, is then displaced by a logarithmically large amount $\Delta t_{\rm in}$.
The radiation observed at $r_{\rm obs}$ is further displaced by its Shapiro time delay $\Delta t_{\rm obs}$. The spacetime picture of this scattering event is illustrated in Fig.~\ref{fig:delay}.

The $r_{\rm in}$-dependent term in \eqref{Delta t tot} is of course specific to the scattering problem and we expect it to cancel out when extracting an effective theory for generic orbits.
This is consistent
with its absence for bound orbits (see \cite{Goldberger:2009qd,Porto:2012as}).
While astrophysical waveform measurements are only sensitive to time differences, 
the above time delay is physically meaningful if one knows at which time the bodies were sent in from afar, which is by construction the case
for the scattering problem set up by the KMOC formalism \cite{Kosower:2018adc,Cristofoli:2021vyo}.
Therefore, we believe that \eqref{Delta t tot} is the physically correct answer for the quantity defined by \eqref{KMOC}.

We can use the classical physics result \eqref{Delta t tot} to predict the infrared-divergent part of the 1-loop in-in expectation value \eqref{KMOC} to be, up to $\hbar$-suppressed effects (by equating the time delay to the $k^0$ derivative of the phase):
\begin{equation} \label{prediction IR divergences}
   \underbracket[0.4pt]{\Exp_k^{(1\text{-loop})}\Big\vert_{\rm div}
    =  \Exp_k^{({\rm tree})}\times -iG k{\cdot}(\bar{p}_1+\bar{p}_2)
 \left( 2+ \frac{2-3\bar{y}^{-2}}{(1-\bar{y}^{-2})^{\frac32}}\right)
    \log\frac{\Lambda}{\mu_{\rm IR}}}_{\mbox{\small (classical prediction)}}\,.
\end{equation}
We do not expect any further infrared divergence
at this order, in particular no real divergences, since the inclusive waveform should not be suppressed by the small ``probability to not radiate'' that is familiar for exclusive processes.
The prediction \eqref{prediction IR divergences} is supported nontrivially in App.~\ref{app:SK soft} where we compute the soft function for Wilson lines that follow the appropriate Schwinger-Keldysh time-fold.
In this picture it is also clear that this effect exponentiates to give a phase, which is simply the action of a classical configuration.
The discrepancy with \cite{Brandhuber:2023hhy,Herderschee:2023fxh,Elkhidir:2023dco,Georgoudis:2023lgf} is likely related to the question of time-ordered (principal-value) versus retarded propagators noted below \eqref{M and Exp from Geff}.

\subsection{Summary and comparison with the literature}

Equations \eqref{KMOC integrals} constitute the main result of the analysis in this section: the contributions from each eikonal master integral to the in-in expectation value. These can be directly compared with the literature. For the time-ordered amplitude (the part outside square brackets), we find perfect agreement with \cite[Eq.~(6.4)]{Brandhuber:2023hhy} and \cite[Sec.~6, App.~B]{Herderschee:2023fxh}, where the authors computed integrals with time-ordered propagators (which led to principal-values) in accordance with the lower-left entry of \eqref{M and Exp from Geff}. 
In~\cite{Elkhidir:2023dco}, the authors used a different organization principle for their computation: they use Cutkosky rules to highlight different ``Compton'' cut contributions that emphasize instead the separation of the inclusive observable into real and imaginary parts. Later, the authors of~\cite{Georgoudis:2023lgf}, used yet another method, where the waveform was related to the logarithm of
S-matrix. Since the two references~\cite{Elkhidir:2023dco,Georgoudis:2023lgf} had different intermediate results than the master integrals computed here, we did not directly compare our master integrals with theirs.

The agreement with the time-ordered amplitudes of \cite{Brandhuber:2023hhy,Herderschee:2023fxh} confirms the Regge-soft limits recorded in \eqref{eikonal MRK}; this is perhaps not too surprising since time-ordered amplitudes can be computed using any number of standard techniques, such as Feynman parameters.
However, we stress that this is \emph{not} the complete contribution to the inclusive waveform master integrals, which must also account for the terms in the square brackets of~\eqref{KMOC integrals} originating from the $\cut_{1'2'}$ term in \eqref{inclusive KMOC}.  More concretely, this cut term modifies the principal-value propagator that was appropriate for time-ordered amplitudes (after summing over all topologies)
to the retarded propagator displayed in the lower-right entry of ~\eqref{M and Exp from Geff}. As pointed out below that equation, this outcome has a clear physical interpretation and, beyond cancelling the superclassical terms, it contributes non-trivially to classical physics by ensuring the correct causal propagation (this is also seen in impulse calculations, see, e.g., \cite{Kosower:2018adc,Damgaard:2021ipf,Herrmann:2021tct}).

An important question, which is beyond the scope of this work, is whether its inclusion affects the final waveform as opposed to just individual master integrals as considered here.
From the compelling classical picture of infrared divergences presented in Sec.~\ref{ssec:IR}, supported by the Wilson line calculation in \ref{app:SK soft},
we expect this to be the case.
More broadly, the formalism introduced here provides a uniform method for computing scattering amplitudes and inclusive observables by directly incorporating cut contributions through boundary values of differential equations.

\section{Conclusions}
\label{sec:conclusions}

In this paper, we discussed \emph{asymptotic observables}, which represent measurements that can be made from past and future infinity, in situations where an S-matrix makes sense.  In the most straightforward terms, they are defined as vacuum expectation of ``in'' and ``out'' creation/annihilation operators with arbitrary operator ordering. These generalized amplitudes are still functions of on-shell momenta and the novelty is to relax the time-ordering.

Many generalized amplitudes compute interesting observables with standard names, such as inclusive production cross-sections and inclusive waveforms, or, more exotically,
out-of-time-order correlators in a scattering background, as discussed in Sec.~\ref{ssec:asympt}.
Many observables usually thought of as quadratic in the S-matrix are simply linear in generalized amplitudes.
We anticipate that this unified viewpoint will facilitate the application of amplitude techniques to the computation of these more general observables, or even inspire the discovery of new ones. 
The choice of operator ordering can have dramatic physical effects, such as changing the size of the amplitude from being exponentially small to being entirely unsuppressed. As such,  we find it particularly intriguing to understand all possible relations.
In Sec.~\ref{sec:asympt_measurements} we classified all asymptotic measurements that can be obtained by stringing together chains of asymptotic creation and annihilation operators.
This is equivalent to stringing together alternating products of $S$ and $S^\dag$ operators (see \eqref{eq:measurement-type}).  We would like to stress that this
by no means exhausts all possible ways of sewing S-matrices.
While blob diagrams that are not simple chains do not have an immediate Hilbert-space interpretation, we suspect they may be important in understanding the web of connections between different asymptotic measurements.

We presented two main methods to compute asymptotic observables.
The blobology approach in Sec.~\ref{sec:blobology} expresses them as products of ordinary connected (anti-)time-ordered amplitudes (the elementary \say{blobs}) sewn by sums over on-shell intermediate states.  The reduction formula approach in Sec.~\ref{sec:localFields} instead relates generalized amplitudes to the on-shell limit
of amputated correlation functions, with various operator orderings, that are defined within the in-in or Schwinger-Keldysh formalism.  Both techniques lead to identical cutting rules that can be applied in perturbation theory on a diagram-by-diagram basis. Diagrams simply get decorated by cuts that set some propagators on-shell (see the rules in Sec.~\ref{ssec:SK pert}),
not too differently from the Cutkosky rules.

In the context of the differential equation method for Feynman integrals, different choices of operator orderings simply amount to setting different boundary conditions.  We demonstrated this for one-loop integrals contributing to gravitational waveforms in Sec.~\ref{sec:waveforms}, where we obtained all the relevant boundary values in the eikonal limit and also highlighted the physical importance of the ``cut'' terms required by the formalism (see Sec.~\ref{ssec:masterEik}).
In Sec.~\ref{ssec:IR} we further highlighted the differences between the infrared physics of exclusive and inclusive amplitudes.

There are several natural directions for further inquiry.
Clearly, it would be useful to develop further techniques to calculate in-in type Feynman integrals. For example, do these admit convenient Schwinger parameter representations?
In string perturbation theory, can generalized amplitudes be computed by considering strings on a Schwinger-Keldysh target space?

Another question is that of \emph{local analyticity} \cite{Bros:1972jh,Hannesdottir:2022bmo}:
which combinations of generalized amplitudes
are boundary values of a single analytic function at real momenta, as opposed to piecewise analytic? 
An example is the distinction between an ordinary amplitude $\cM$, which can be defined with argument $s+i\varepsilon$, versus ${\rm Im}\,\cM$, whose arguments cannot be deformed by $i\varepsilon$ globally for all $s$; at tree level $\frac{1}{-s + M^2}$ is a single analytic function while $\pi \delta(-s + M^2)$ is not. Analyticity is a powerful tool. When it holds, it alleviates the need to separately consider many kinematic regions and thresholds; instead they can be continuously connected e.g., by differential equations.

The observation that amplitudes and inclusive observables satisfy the same differential equation opens up the possibility of relating them by analytic continuations in kinematic invariants.
Identifying local-analytic amplitudes should be highly relevant to understand crossing of $n$-particle amplitudes, since these are the natural endpoints of analytic continuations in complex kinematics. Does a spacetime picture of Landau singularities a la Coleman-Norton apply to generalized amplitudes, where excitations on some timefolds propagate backward in time? We leave these questions for future work.

\acknowledgments
We would like to thank
Gang Chen, Alessandro Georgoudis, Aidan Herderschee, Harald Ita, Julio Parra-Martinez, 
Radu Roiban, and Fei Teng for helpful discussions.
S.C.H. and M.G.’s work is supported in parts by the National Science and Engineering Council of Canada (NSERC) and by the Canada Research
Chair program, reference number CRC-2022-00421.
S.C.H.'s work is additionally supported by a Simons Fellowships in Theoretical Physics
and by the Simons Collaboration on the Nonperturbative Bootstrap. H.S.H. gratefully acknowledges support from the Simons Foundation (816048). The work of S.M. is supported by the Sivian Fund and the U.S. Department of Energy, Office of Science, Office of High Energy Physics under Award
Number DE-SC0009988. 
S.C.H. thanks the Institute for Advanced Study for hospitality during a sabbatical semester where this work got started.

\appendix
\section{Differential equations and boundary conditions for the eikonal integrals}
\label{sec:eikonalDiffEq}
\allowdisplaybreaks

In this appendix, we provide a subset of the differential equations, along with the boundary conditions in the Regge-soft regime,  satisfied by the basis of master integrals
(\ref{eq:Itilde2}-\ref{remaining ints}) for the family defined in \eqref{eik topology}, relevant to the gravity waveform calculation discussed in Sec.~\ref{sec:waveforms}. For example, we have found that
\def\dlog#1{\left[#1\right]}
\begin{equation} \label{eik first EOM}
    \text{d}\begin{pmatrix} \tilde{\mathcal{I}}_1\\ \tilde{\mathcal{I}}_2
    \\ \mathcal{I}_0 \\ \mathcal{I}_1 \\ \mathcal{I}_2 \end{pmatrix}
    = \epsilon\begin{pmatrix}
    \dlog{\barw_1^{-2}} &0&0&0&0 \\ \dlog{\e^{-2\eta}} & \dlog{\tfrac{\sinh^2\eta}{\barw_1^2}} &0&0&0\\
    0&0&\dlog{q_2^{-2}}&0&0 \\
    2\dlog{\tfrac{\sqrt{q_2^2+\barw_1^2}+\barw_1}{\sqrt{q_2^2+\barw_1^2}-\barw_1}} &0 &
   {-}\!\dlog{\tfrac{\sqrt{q_2^2+\barw_1^2}+\barw_1}{\sqrt{q_2^2+\barw_1^2}-\barw_1}} & \dlog{\tfrac{q_2^2+\barw_1^2}{q_2^4}} &0 \\
    0&0&0&0& \dlog{q_1^{-2}}
    \end{pmatrix}{\cdot}
     \begin{pmatrix} \tilde{\mathcal{I}}_1\\ \tilde{\mathcal{I}}_2
    \\ \mathcal{I}_0 \\ \mathcal{I}_1 \\ \mathcal{I}_2 \end{pmatrix}\,.
\end{equation}
Here, $\dlog{x}=\text{d}\log{x}$ and $\eta=\cosh^{-1}y$ is the rapidity. The remaining differential equations (with $\mathcal{I}_6$ omitted since it only starts contributing at $\mathcal{O}(\epsilon)$ in an expansion around $\D=4-2 \epsilon$ dimensions) which are independent up to permutations, are
\begin{align}
   \frac{\text{d}\tilde{\mathcal{I}}_4}{\epsilon} &=
    \dlog{\tfrac{(\barw_1-\e^\eta \barw_2)(\barw_1-\e^{-\eta}\barw_2)}{\barw_1\barw_2}}\left(\tilde{\mathcal{I}}_4
    -\tilde{\mathcal{I}}_1-\tilde{\mathcal{I}}_1^\prime\right)
+\dlog{\tfrac{\barw_1}{\barw_2}} \left(\tilde{\mathcal{I}}_1-\tilde{\mathcal{I}}_1^\prime\right)
\nonumber\\ &\phantom{=} 
+\dlog{\tfrac{\barw_1-\e^\eta \barw_2}{\barw_1-\e^{-\eta}\barw_2}} \tilde{\mathcal{I}}_2
+\dlog{\tfrac{\barw_2-\e^\eta \barw_1}{\barw_2-\e^{-\eta}\barw_1}} \tilde{\mathcal{I}}_2^\prime
+\dlog{\tfrac{1}{\barw_1\barw_2}}\tilde{\mathcal{I}}_4\,,
\\
    \frac{\text{d}\mathcal{I}_3}{\epsilon}  &=  \dlog{e^{-2\eta}}(\tilde{\mathcal{I}}_1-\mathcal{I}_0)
    +\dlog{\tfrac{q_2^2\sinh^2\eta-\barw_1^2}{\barw_1^2}}\tilde{\mathcal{I}}_2
+\dlog{\tfrac{\sqrt{q_2^2+\barw_1^2}\tanh\eta-\barw_1}{\sqrt{q_2^2+\barw_1^2}\tanh\eta+\barw_1}}\mathcal{I}_1
\nonumber\\ &\phantom{=} 
+ \dlog{\tfrac{q_2\sinh\eta-\barw_1}{q_2\sinh\eta+\barw_1}}\mathcal{I}_2^\prime
+ \dlog{\tfrac{q_2^2\sinh^2\eta-\barw_1^2}{q_2^4\sinh^2\eta}} \mathcal{I}_3 \,,
\\
\frac{\text{d}\mathcal{I}_5}{\epsilon} &=\dlog{\tfrac{(q_1^2-q_2^2)^2-4q_1^2\barw_1^2}{(q_1^2-q_2^2)^2}}(\mathcal{I}_0-\mathcal{I}_0^\prime)
+ \dlog{\tfrac{(q_1^2-q_2^2)^2-4q_1^2\barw_1^2}{q_2^4}}\left(\mathcal{I}_5-\tilde{\mathcal{I}}_1\right)
+\dlog{\tfrac{q_2^4}{q_1^4\barw_1^2}}\mathcal{I}_5
\nonumber\\&\phantom{=} 
+\dlog{\tfrac{\sqrt{q_2^2+\barw_1^2}-\barw_1}{\sqrt{q_2^2+\barw_1^2}+\barw_1}\,
\tfrac{(q_1^2{-}q_2^2)\sqrt{q_2^2+\barw_1^2}-(q_1^2+q_2^2)\barw_1}{(q_1^2{-}q_2^2)\sqrt{q_2^2+\barw_1^2}+(q_1^2+q_2^2)\barw_1}}\mathcal{I}_1+\dlog{\tfrac{q_2^2-q_1^2-2\sqrt{q_1^2} \barw_1}{q_2^2-q_1^2+2\sqrt{q_1^2} \barw_1}}
\mathcal{I}_2\,.  \label{eik last EOM}
\end{align}
The differential equations \eqref{eik first EOM}-\eqref{eik last EOM} enable us to expand the integrals in $\epsilon$, given their value at some boundary point as explain in the main text.  

Next, we elaborate on how these boundary values were obtained. In particular, we show how to use the method of regions~\cite{Beneke:1997zp,Jantzen:2011nz,Jantzen:2012mw} to compute the Regge-soft limit of the $\ItA_2$ triangle from~\eqref{eikonal MRK} (with usual time-ordered $i\varepsilon$). After Schwinger-parametrizing and integrating out the loop momenta~\cite[App.~A]{Hannesdottir:2022bmo}, we can write the expression for $\ItA_2$ as,
\begin{equation}
    \ItA_2 = 2 \epsilon^2 \bary \e^{\epsilon \gamma_E} \Gamma(1+\epsilon) \int_0^\infty \frac{\d^3 \alpha_e}{\GL (1)} \frac{\alpha_2^{-1+2\epsilon}}{[\alpha_1^2+\alpha_3^2-2 \alpha_1 (\barw_1 \alpha_2 + \bary \alpha_3)]^{1+\epsilon}}\,.
\end{equation}
To parametrize the Regge-soft limit, we solve for the potentially divergent regions as $\bary \to \infty$ and $\barw_1 \to 0$, with the method of~\cite{Jantzen:2012mw,Heinrich:2021dbf}. In the gauge fixing where $\alpha_3=1$, we find that the following scalings contribute to the Regge soft limit,
\begin{align}
    \rm{scaling \,\, 1:} \quad (\alpha_1,\alpha_2) & \to (\barw_1 \alpha_1,\barw_1^{-2} \alpha_2) \,, \\
    \rm{scaling \,\, 2:} \quad(\alpha_1,\alpha_2) & \to (\barw_1^{-1} \alpha_1,\barw_1^{-2} \alpha_2) \,,
    \label{eq:scalings}
\end{align}
with $\barw_1 \to 0$, and we assume here that $\frac{1}{\bary}$ goes to zero at the same rate as $\barw_1$. (In general the limits should be taken sequentially, with $y\to\infty$ after $\barw_i\to0$ as mentioned in the main text, but this does not affect this particular diagram hence we take both limits in one step here.)
For both scalings, we change variables according to~\eqref{eq:scalings} and expand to leading order in $\barw_1$. For the first one, the leading-order expansion is
\begin{subequations}
\begin{align}
    \ItA_2 \big\vert_{\rm scaling \,\, 1} & = 2 \epsilon^2 \bary \e^{\epsilon \gamma_E} \Gamma(1+\epsilon) \int_0^\infty \d \alpha_1 \d \alpha_2 \frac{\alpha_2^{-1+2\epsilon}}{[1-2 \alpha_1 (\barw_1 \alpha_2 + \bary)]^{1+\epsilon}}
    \\ & = 
    -\frac{c_1}{2}W_1^{-2\epsilon}Y^{2\epsilon} \,,
    \label{eq:scaling1}
\end{align}
\end{subequations}
while for the second one the expansion is
\begin{subequations}
\begin{align}
    \ItA_2 \big\vert_{\rm scaling \,\, 2} & = 2 \epsilon^2 \bary \e^{\epsilon \gamma_E} \Gamma(1+\epsilon) \int_0^\infty \d \alpha_1 \d \alpha_2 \frac{\alpha_2^{-1+2\epsilon}}{[\alpha_1^2-2 \alpha_1 (\barw_1 \alpha_2 + \bary)]^{1+\epsilon}}
    \\ & = 
    r_\Gamma\frac{(2W_1)^{-2\epsilon}}{2\cos(\pi\epsilon)} \,.
    \label{eq:scaling2}
\end{align}
\end{subequations}
Adding the two contributions~\eqref{eq:scaling1} and~\eqref{eq:scaling2} confirms the limit in~\eqref{eikonal MRK} from the main text.

\section{Schwinger-Keldysh soft factors and infrared divergences 
of \texorpdfstring{$\Exp_k$}{Exp k}}\label{app:SK soft}

Here we study the infrared divergences of the inclusive waveform $\Exp_k$ defined in \eqref{KMOC} in the classical limit.
As usual, infrared divergences come from asymptotic regions where all particles follow approximately straight lines,
and interact weakly with each other through exchange of gravitons.
Since gravity is free in the infrared, it suffices to work in the linearized theory. Expanding the metric around Minkowski as
$g_{\mu\nu}=\eta_{\mu\nu}+2h_{\mu\nu}$,
one can write the soft region contribution to a generic amplitude in terms of gravitational Wilson lines
(see for example \cite{Akhoury:2011kq,Melville:2013qca}):
\begin{equation} \label{Expk soft 1}
    \Exp_k^{\rm (soft)}
    \equiv \<0|\,\cC \exp\left[i\sum_i \int \d\tau p_i^\mu p_i^\nu h_{\mu\nu}(\tau p_i) \right]\,|0\>_{h}\,,
\end{equation}
where the average is computed using the path integral of linearized gravity. The Wilson lines meet at the origin.

When using the standard time-ordered product, \eqref{Expk soft 1} leads to the usual infrared divergent Weinberg factor for exclusive amplitudes.
In the classical limit relevant to $1'2'k\ot 12$ graviton emission,
it was shown in \cite{Brandhuber:2023hhy,Herderschee:2023fxh,Elkhidir:2023dco,Georgoudis:2023lgf} to reduce to a purely superclassical term
(argued to cancel out for inclusive observables),
plus the classical term recorded in \eqref{M IR classical} of the main text.  
As we show here, the appropriate Wilson lines for inclusive observables follow different paths which leads to the different result \eqref{prediction IR divergences}.

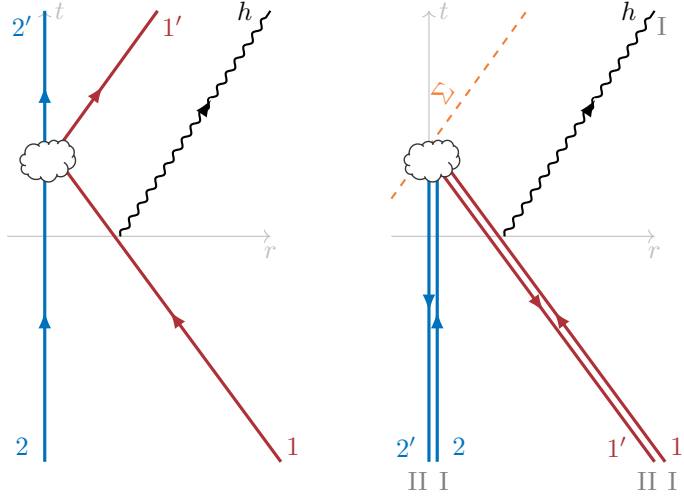
\begin{figure}
	\centering
    \adjustbox{valign=t}{
	\begin{tikzpicture}
		\tikzset{->-/.style={decoration={
					markings,
					mark=at position #1 with {\arrow{>}}},postaction={decorate}}}
		\draw[lightgray, ->] (-0.5,0) -- (3,0) node[below] {$r$};
		\draw[lightgray, ->] (0,-3) -- (0,3) node[right] {$t$};
		\draw[RoyalBlue, very thick, decoration={markings,
			mark=at position 0.5 with {\arrow{latex}},
		},postaction={decorate}] (0,-3) -- (0,1);
		\draw[RoyalBlue, very thick, decoration={markings,
			mark=at position 0.5 with {\arrow{latex}},
		},postaction={decorate}] (0,1) -- (0,3);
		\draw[Maroon, very thick, decoration={markings,
			mark=at position 0.5 with {\arrow{latex}},
		},postaction={decorate}] (0.05,1) -- (1.5,3);
		\draw[Maroon, very thick, decoration={markings,
			mark=at position 0.5 with {\arrow{latex}},
		},postaction={decorate}] (3.14,-3) -- (0.19,1);
		\node[scale=0.7] at (0,1) {\cloud};
		\draw[photon, thick] (1,0) -- (3,3) node[left] {$h\;$};
		\draw[white, very thick, decoration={markings,
		mark=at position 0.5 with {\arrow[black]{latex}},
	},postaction={decorate}] (2.2,1.8) -- (2.22,1.83);
		\draw[RoyalBlue] (-0.3,-2.8) node {$2$};
		\draw[RoyalBlue] (-0.3,2.8) node {$2'$};
		\draw[Maroon] (1.7,2.8) node {$1'$};
		\draw[Maroon] (3.3,-2.8) node {$1$};
	\end{tikzpicture}}
	\qquad
    \adjustbox{valign=t}{
	\begin{tikzpicture}
		\tikzset{->-/.style={decoration={
					markings,
					mark=at position #1 with {\arrow{>}}},postaction={decorate}}}
	\draw[lightgray, ->] (-0.5,0) -- (3,0) node[below] {$r$};
	\draw[lightgray, ->] (0,-3) -- (0,3) node[right] {$t$};
	\draw[RoyalBlue, very thick, decoration={markings,
		mark=at position 0.5 with {\arrow{latex}},
	},postaction={decorate}] (0,1) -- (0,-3);
	\draw[RoyalBlue, very thick, decoration={markings,
		mark=at position 0.5 with {\arrow{latex}},
	},postaction={decorate}] (0.11,-3) -- (0.11,1);
	\draw[Maroon, very thick, decoration={markings,
		mark=at position 0.5 with {\arrow{latex}},
	},postaction={decorate}] (0.05,1) -- (3,-3);
	\draw[Maroon, very thick, decoration={markings,
		mark=at position 0.5 with {\arrow{latex}},
	},postaction={decorate}] (3.14,-3) -- (0.19,1);
	\draw[Orange, thick, dashed] (-0.5,0.5) -- (1.3,3) node[midway, rotate=55,above] {$\Sigma$};
	\node[scale=0.7] at (0,1) {\cloud};
	\draw[photon, thick] (1,0) -- (3,3) node[left] {$h\;$};
	\draw[white, very thick, decoration={markings,
		mark=at position 0.5 with {\arrow[black]{latex}},
	},postaction={decorate}] (2.2,1.8) -- (2.22,1.83);
	\draw[RoyalBlue] (-0.3,-2.8) node {$2'$};
	\draw[RoyalBlue] (0.4,-2.8) node {$2$};
	\draw[Maroon] (2.5,-2.8) node {$1'$};
	\draw[Maroon] (3.3,-2.8) node {$1$};
  	\draw[gray] (-0.15,-3) node [below] {$\small\rtwo$};
	\draw[gray] (0.2,-3) node [below] {$\small\rone$};
    \draw[gray] (2.9,-3) node [below] {$\small\rtwo$};
    \draw[gray] (3.24,-3) node [below] {$\small\rone$};
    \draw[gray] (2.9,2.8) node [right] {$\small\rone$};
  \end{tikzpicture}}
 \caption{\label{fig:Wilson-lines}Example Schwinger-Keldysh contours for the time-ordered (left) and inclusive (right), where the graviton is emitted from the heavy line $1$.
 Zooming out, we have five straight Wilson lines which meet at a common vertex on the first (type-$\rone$) timefold. The folds are glued on an arbitrary spacelike surface $\Sigma$ outside the past lightcone of the observation, after which particles $1'$, $2'$ continue toward the past along the type-$\rtwo$ fold. 
 }
\end{figure}

The novel aspect of the inclusive waveform is that the two heavy bodies after the collision, $1'$ and $2'$, never reach future infinity.
Rather, they return to past infinity along the second Schwinger-Keldysh fold.  The appropriate straight-line trajectories are shown in Fig.~\ref{fig:Wilson-lines}.

We are interested in the classical limit, where
$p_1-p_1' = q_1 \sim \hbar$ and the particles following
the two folds follow approximately the same trajectory.
Moreover, the difference between the gravity
fields is suppressed, $h^\rone-h^\rtwo\sim \hbar$.
Therefore, in the classical limit
it is natural to combine the sources from particles $1$ and $1'$ in \eqref{Expk soft 1} into difference and average fields:
\begin{equation}
 p_1^\mu p_1^\nu h^\rone_{\mu\nu}(\tau p_1)
-p_{1'}^\mu p_{1'}^\nu h^\rtwo_{\mu\nu}(\tau p_{1'})
 = \barp_1^\mu \barp_1^\nu h^\rdiff_{\mu\nu}(\tau \barp_1)
 + q_1{\cdot}\frac{\partial}{\partial p_1}
 \barp_1^\mu \barp_1^\nu h^\ravg_{\mu\nu}(\tau \barp_1) + \mathcal{O}(\hbar^2)\,,
\end{equation}
with $h^\ravg=\frac12(h^\rone+h^\rtwo)$ the average field.
Thus 
\begin{equation}\begin{aligned}
    \Exp_k^{\rm (soft)}
    &= \<0|\,
\cC\exp\Big[i\int_{-\infty}^\Sigma
\d\tau \barp_1^\mu \barp_1^\nu h^\rdiff_{\mu\nu}(\tau \barp_1)
 +i\int_{-\infty}^0 \d\tau q_1{\cdot}\frac{\partial}{\partial p_1}
 \barp_1^\mu \barp_1^\nu h^\ravg_{\mu\nu}(\tau \barp_1)
\\ & \hspace{20mm}+(1\leftrightarrow2) +i\int_0^\infty \d\tau k^\mu k^\nu h^\rone_{\mu\nu}(\tau k)
\Big]\,|0\>_h\,.
\end{aligned}\label{Expk from SK}\end{equation}
Note that all terms are purely classical so there is no need to expand to further orders.
The cutoff $\Sigma$ is the \say{turning point} of the Schwinger-Keldysh contour, which can be an arbitrary spacelike surface outside the past lightcone of the observation point (possibly a surface of constant lightcone time), as depicted in Fig.~\ref{fig:Wilson-lines}.

Taking Wick contractions,
two-point functions of difference fields 
vanish and classical
effects will only arise from the retarded propagator.
Using the familiar combinatorics of Wick contractions, we can thus write the expectation value \eqref{Expk from SK} as an exponential involving retarded two-point functions, up to $\hbar$ suppressed effects:
\begin{equation}\begin{aligned}
   \Exp_k^{\rm (soft)}
    &=
\exp\Big[
\<0|\,\cC
\left(i\int_{-\infty}^\Sigma \d\tau q_2{\cdot}\frac{\partial}{\partial \barp_2}\barp_2^\mu \barp_2^\nu h^\ravg_{\mu\nu}(\tau \barp_2)
+i\int_0^\infty \d\tau k^\mu k^\nu h_{\mu\nu}^\rone(\tau k)\right)
\\
&\hspace{25mm}
\left(i\int_{-\infty}^\delta \d\tau' \barp_1^\rho\barp_1^\sigma h_{\rho\sigma}^\rdiff(\tau' \barp_1)\right)
\,|0\>_h  + (1\leftrightarrow 2)\Big].
\end{aligned}\label{Expk from ret}\end{equation}
Here we have dropped self-energy terms (i.e., the eikonal phase from particle $1$ moving in its own field), which vanish in dimensional regularization.
In harmonic gauge, the retarded two-point function is simply:
\begin{equation}\begin{aligned}
    \< 0|\,\cC\{h_{\mu\nu}^{\ravg}(x)\,h_{\rho\sigma}^{\rdiff}(y)\}\,|0\>_h &= 4\pi G \int \frac{\d^\D p}{(2\pi)^\D} \frac{-i\,e^{ip{\cdot}(x-y)}}{p^2-i\varepsilon p^0}
    \left(
    \eta_{\mu\rho}\eta_{\nu\sigma}+
    \eta_{\mu\sigma}\eta_{\nu\rho}
     -\frac{2\eta_{\mu\nu}\eta_{\rho\sigma}}{\D-2}
\right)
    \\ & \hspace{-2cm}= -2i G \delta_+[(x-y)^2]\,
 \left(\eta_{\mu\rho}\eta_{\nu\sigma}+
    \eta_{\mu\sigma}\eta_{\nu\rho}-\eta_{\mu\nu}\eta_{\rho\sigma}\right)
\quad(\D=4)\,.
\end{aligned}\end{equation}

The representation \eqref{Expk from ret} admits a simple physical interpretation:
the integrals over the difference fields times retarded function simply create the classical Coulomb field of the incoming particles.  Specifically,
\begin{equation}
  \<h^{\mu\nu}(x)\>_i \equiv
 \<0| h_{\mu\nu}^\ravg(x)
\left(i\int_{-\infty}^\Sigma \d\tau' \barp_i^\rho\barp_i^\sigma h_{\rho\sigma}^\rdiff(\tau' \barp_i)\right)|0\>_h
= 
G \barm_i \frac{2\barv_{i}^\mu\barv_i^\nu +\eta^{\mu\nu}}{
\sqrt{x^2+(\barv_i{\cdot}x)^2}}\,.
\end{equation}
This is nothing but a covariant form of the far-field Schwarzschild metric.
Thus the soft factor \eqref{Expk from ret} for the inclusive observable, in the classical limit,
can be written as
\begin{equation} \label{soft near final}
\begin{split}
        \Exp_k^{\rm (soft)}
    =
\exp&\left[
iq_2{\cdot}\frac{\partial}{\partial \barp_2}\int_{-\infty}^\Sigma \!\d\tau \barp_2^\mu \barp_2^\nu
\<h_{\mu\nu}(\tau \barp_2)\>_1
\right.\\&\left.\qquad+i\int_0^\infty \!\d\tau k^\mu k^\nu \<h_{\mu\nu}(\tau k)\>_1
+(1\leftrightarrow 2) \right]\,.
\end{split}
\end{equation}
The integrals are both logarithmically divergent:
\begin{align}
    \int \d\tau\,\barp_2^\mu \barp_2^\nu
\<h_{\mu\nu}(\tau \barp_2)\>_1
 &= G\bar{m}_1\bar{m}_2 \frac{2\bar{y}^2-1}{\sqrt{\bar{y}^2-1}} \int_{-\tau_{\rm max}}^{-\tau_{\rm min}} \frac{\d\tau}{|\tau|}\,,
\\
\int \d\tau\,k^\mu k^\nu
\<h_{\mu\nu}(\tau k)\>_1
&= -2G k{\cdot}\barp_1
\int^{\tau_{\rm max}}_{\tau_{\rm min}} \frac{\d\tau}{\tau}\,.
\end{align}
Plugging back into \eqref{soft near final}
and performing the derivatives, we obtain the expression for the infrared-divergent part of the soft factor:
\begin{equation}
\Exp_k^{\rm (soft)}
    = \exp\left[
-iG k{\cdot}(\barp_1+\barp_2)
\left( 2+ \frac{2-3\bary^{-2}}{(1-\bary^{-2})^{\frac32}}\right)
\log \frac{\tau_{\rm max}}{\tau_{\rm min}} \right]\,.
\end{equation}
This agrees perfectly with the infrared divergence predicted in \eqref{prediction IR divergences} using the classical geodesic
motion of particles in the Schwarzschild metric. In fact, the calculations are easily map into each other since we showed how the expectation value of Wilson lines
in \eqref{Expk from SK} reduces, up to $\hbar$ corrections, to the eikonal phase \eqref{soft near final} of particles moving in the classical field sourced by the others.
The one-loop divergences exponentiate because the graviton field is Gaussian in the infrared.

\bibliographystyle{JHEP}
\bibliography{refs}

\providecommand{\href}[2]{#2}\begingroup\raggedright\begin{thebibliography}{10}

\bibitem{Kosower:2018adc}
D.A.~Kosower, B.~Maybee and D.~O'Connell, \emph{{Amplitudes, Observables, and
  Classical Scattering}},
  \href{https://doi.org/10.1007/JHEP02(2019)137}{\emph{JHEP} {\bfseries 02}
  (2019) 137} [\href{https://arxiv.org/abs/1811.10950}{{\ttfamily
  1811.10950}}].

\bibitem{Cristofoli:2021vyo}
A.~Cristofoli, R.~Gonzo, D.A.~Kosower and D.~O'Connell, \emph{{Waveforms from
  amplitudes}}, \href{https://doi.org/10.1103/PhysRevD.106.056007}{\emph{Phys.
  Rev. D} {\bfseries 106} (2022) 056007}
  [\href{https://arxiv.org/abs/2107.10193}{{\ttfamily 2107.10193}}].

\bibitem{LIGOScientific:2016aoc}
{\scshape LIGO Scientific, Virgo} collaboration, \emph{{Observation of
  Gravitational Waves from a Binary Black Hole Merger}},
  \href{https://doi.org/10.1103/PhysRevLett.116.061102}{\emph{Phys. Rev. Lett.}
  {\bfseries 116} (2016) 061102}
  [\href{https://arxiv.org/abs/1602.03837}{{\ttfamily 1602.03837}}].

\bibitem{LIGOScientific:2017vwq}
{\scshape LIGO Scientific, Virgo} collaboration, \emph{{GW170817: Observation
  of Gravitational Waves from a Binary Neutron Star Inspiral}},
  \href{https://doi.org/10.1103/PhysRevLett.119.161101}{\emph{Phys. Rev. Lett.}
  {\bfseries 119} (2017) 161101}
  [\href{https://arxiv.org/abs/1710.05832}{{\ttfamily 1710.05832}}].

\bibitem{LIGOScientific:2017ync}
{\scshape LIGO Scientific, and others} collaboration, \emph{{Multi-messenger
  Observations of a Binary Neutron Star Merger}},
  \href{https://doi.org/10.3847/2041-8213/aa91c9}{\emph{Astrophys. J. Lett.}
  {\bfseries 848} (2017) L12}
  [\href{https://arxiv.org/abs/1710.05833}{{\ttfamily 1710.05833}}].

\bibitem{Shenker:2013pqa}
S.H.~Shenker and D.~Stanford, \emph{{Black holes and the butterfly effect}},
  \href{https://doi.org/10.1007/JHEP03(2014)067}{\emph{JHEP} {\bfseries 03}
  (2014) 067} [\href{https://arxiv.org/abs/1306.0622}{{\ttfamily 1306.0622}}].

\bibitem{Maldacena:2015waa}
J.~Maldacena, S.H.~Shenker and D.~Stanford, \emph{{A bound on chaos}},
  \href{https://doi.org/10.1007/JHEP08(2016)106}{\emph{JHEP} {\bfseries 08}
  (2016) 106} [\href{https://arxiv.org/abs/1503.01409}{{\ttfamily
  1503.01409}}].

\bibitem{Caron-Huot:2022lff}
S.~Caron-Huot, \emph{{Holographic cameras: an eye for the bulk}},
  \href{https://doi.org/10.1007/JHEP03(2023)047}{\emph{JHEP} {\bfseries 03}
  (2023) 047} [\href{https://arxiv.org/abs/2211.11791}{{\ttfamily
  2211.11791}}].

\bibitem{Bros:1965kbd}
J.~Bros, H.~Epstein and V.~Glaser, \emph{{A proof of the crossing property for
  two-particle amplitudes in general quantum field theory}},
  \href{https://doi.org/10.1007/BF01646307}{\emph{Commun. Math. Phys.}
  {\bfseries 1} (1965) 240}.

\bibitem{crossing}
S.~Caron-Huot, M.~Giroux, H.S.~Hannesdottir and S.~Mizera, \emph{{Crossing
  beyond scattering amplitudes}},
  \href{https://arxiv.org/abs/2310.12199}{{\ttfamily 2310.12199}}.

\bibitem{Anastasiou:2002yz}
C.~Anastasiou and K.~Melnikov, \emph{{Higgs boson production at hadron
  colliders in NNLO QCD}},
  \href{https://doi.org/10.1016/S0550-3213(02)00837-4}{\emph{Nucl. Phys. B}
  {\bfseries 646} (2002) 220}
  [\href{https://arxiv.org/abs/hep-ph/0207004}{{\ttfamily hep-ph/0207004}}].

\bibitem{Jain:2014nza}
S.~Jain, M.~Mandlik, S.~Minwalla, T.~Takimi, S.R.~Wadia and S.~Yokoyama,
  \emph{{Unitarity, Crossing Symmetry and Duality of the S-matrix in large N
  Chern-Simons theories with fundamental matter}},
  \href{https://doi.org/10.1007/JHEP04(2015)129}{\emph{JHEP} {\bfseries 04}
  (2015) 129} [\href{https://arxiv.org/abs/1404.6373}{{\ttfamily 1404.6373}}].

\bibitem{Mehta:2022lgq}
U.~Mehta, S.~Minwalla, C.~Patel, S.~Prakash and K.~Sharma, \emph{{Crossing
  Symmetry in Matter Chern-Simons Theories at finite $N$ and $k$}},
  \href{https://arxiv.org/abs/2210.07272}{{\ttfamily 2210.07272}}.

\bibitem{Csaki:2022tvb}
C.~Cs\'aki, Z.-Y.~Dong, O.~Telem, J.~Terning and S.~Yankielowicz,
  \emph{{Dressed vs. pairwise states, and the geometric phase of monopoles and
  charges}}, \href{https://doi.org/10.1007/JHEP02(2023)211}{\emph{JHEP}
  {\bfseries 02} (2023) 211}
  [\href{https://arxiv.org/abs/2209.03369}{{\ttfamily 2209.03369}}].

\bibitem{Anastasiou:2015vya}
C.~Anastasiou, C.~Duhr, F.~Dulat, F.~Herzog and B.~Mistlberger, \emph{{Higgs
  Boson Gluon-Fusion Production in QCD at Three Loops}},
  \href{https://doi.org/10.1103/PhysRevLett.114.212001}{\emph{Phys. Rev. Lett.}
  {\bfseries 114} (2015) 212001}
  [\href{https://arxiv.org/abs/1503.06056}{{\ttfamily 1503.06056}}].

\bibitem{Anastasiou:2016cez}
C.~Anastasiou, C.~Duhr, F.~Dulat, E.~Furlan, T.~Gehrmann, F.~Herzog et~al.,
  \emph{{High precision determination of the gluon fusion Higgs boson
  cross-section at the LHC}},
  \href{https://doi.org/10.1007/JHEP05(2016)058}{\emph{JHEP} {\bfseries 05}
  (2016) 058} [\href{https://arxiv.org/abs/1602.00695}{{\ttfamily
  1602.00695}}].

\bibitem{Dreyer:2016oyx}
F.A.~Dreyer and A.~Karlberg, \emph{{Vector-Boson Fusion Higgs Production at
  Three Loops in QCD}},
  \href{https://doi.org/10.1103/PhysRevLett.117.072001}{\emph{Phys. Rev. Lett.}
  {\bfseries 117} (2016) 072001}
  [\href{https://arxiv.org/abs/1606.00840}{{\ttfamily 1606.00840}}].

\bibitem{Chen:2021isd}
X.~Chen, T.~Gehrmann, E.W.N.~Glover, A.~Huss, B.~Mistlberger and A.~Pelloni,
  \emph{{Fully Differential Higgs Boson Production to Third Order in QCD}},
  \href{https://doi.org/10.1103/PhysRevLett.127.072002}{\emph{Phys. Rev. Lett.}
  {\bfseries 127} (2021) 072002}
  [\href{https://arxiv.org/abs/2102.07607}{{\ttfamily 2102.07607}}].

\bibitem{EventHorizonTelescope:2019dse}
{\scshape Event Horizon Telescope} collaboration, \emph{{First M87 Event
  Horizon Telescope Results. I. The Shadow of the Supermassive Black Hole}},
  \href{https://doi.org/10.3847/2041-8213/ab0ec7}{\emph{Astrophys. J. Lett.}
  {\bfseries 875} (2019) L1}
  [\href{https://arxiv.org/abs/1906.11238}{{\ttfamily 1906.11238}}].

\bibitem{Diehl:2017wew}
M.~Diehl and J.R.~Gaunt, \emph{{Double parton scattering theory overview}},
  \href{https://doi.org/10.1142/9789813227767_0002}{\emph{Adv. Ser. Direct.
  High Energy Phys.} {\bfseries 29} (2018) 7}
  [\href{https://arxiv.org/abs/1710.04408}{{\ttfamily 1710.04408}}].

\bibitem{Eden:1966dnq}
R.J.~Eden, P.V.~Landshoff, D.I.~Olive and J.C.~Polkinghorne, \emph{{The
  Analytic S-Matrix}}, Cambridge Univ. Press, Cambridge (1966).

\bibitem{Stapp:1976mx}
H.P.~Stapp, \emph{{Discontinuity Formulas for Multiparticle Amplitudes}},  in
  \emph{{Ecole d'Ete de Physique Theorique - Methods in Field Theory}}, 3,
  1976.

\bibitem{Iagolnitzer:1994xv}
D.~Iagolnitzer, \emph{{Scattering in Quantum Field Theories: The Axiomatic and
  Constructive Approaches}}, vol.~48, Princeton University Press (2014).

\bibitem{Peskin:1995ev}
M.E.~Peskin and D.V.~Schroeder, \emph{{An Introduction to quantum field
  theory}}, Addison-Wesley, Reading, USA (1995).

\bibitem{Cutkosky:1960sp}
R.E.~Cutkosky, \emph{{Singularities and discontinuities of Feynman
  amplitudes}}, \href{https://doi.org/10.1063/1.1703676}{\emph{J. Math. Phys.}
  {\bfseries 1} (1960) 429}.

\bibitem{Campbell:2017hsr}
J.~Campbell, J.~Huston and F.~Krauss, \emph{{The Black Book of Quantum
  Chromodynamics : a Primer for the LHC Era}}, Oxford University Press (2018),
  \href{https://doi.org/10.1093/oso/9780199652747.001.0001}{10.1093/oso/9780199652747.001.0001}.

\bibitem{ruelle1961}
D.~Ruelle, \emph{Connection between wightman functions and green functions
  inp-space}, {\emph{Il Nuovo Cimento (1955-1965)} {\bfseries 19} (1961) 356}.

\bibitem{araki1960properties}
H.~Araki and N.~Burgoyne, \emph{{Properties of the Momentum Space Analytic
  Function}}, \href{https://doi.org/10.1007/BF02725943}{\emph{Nuovo Cim.}
  {\bfseries 18} (1960) 342}.

\bibitem{Das:1997gg}
A.K.~Das, \emph{{Finite Temperature Field Theory}}, World Scientific, New York
  (1997).

\bibitem{Meltzer:2020qbr}
D.~Meltzer and A.~Sivaramakrishnan, \emph{{CFT unitarity and the AdS Cutkosky
  rules}}, \href{https://doi.org/10.1007/JHEP11(2020)073}{\emph{JHEP}
  {\bfseries 11} (2020) 073}
  [\href{https://arxiv.org/abs/2008.11730}{{\ttfamily 2008.11730}}].

\bibitem{Stanford:2015owe}
D.~Stanford, \emph{{Many-body chaos at weak coupling}},
  \href{https://doi.org/10.1007/JHEP10(2016)009}{\emph{JHEP} {\bfseries 10}
  (2016) 009} [\href{https://arxiv.org/abs/1512.07687}{{\ttfamily
  1512.07687}}].

\bibitem{Haehl:2017qfl}
F.M.~Haehl, R.~Loganayagam, P.~Narayan and M.~Rangamani, \emph{{Classification
  of out-of-time-order correlators}},
  \href{https://doi.org/10.21468/SciPostPhys.6.1.001}{\emph{SciPost Phys.}
  {\bfseries 6} (2019) 001} [\href{https://arxiv.org/abs/1701.02820}{{\ttfamily
  1701.02820}}].

\bibitem{Caron-Huot:2007zhp}
S.~Caron-Huot, ``{Heavy quark energy losses in the quark-gluon plasma: beyond
  leading order}.''
  \url{https://escholarship.mcgill.ca/concern/theses/5h73q0751}, MSc.~thesis,
  McGill, 2007.

\bibitem{Srednicki:2007qs}
M.~Srednicki, \emph{{Quantum field theory}}, Cambridge University Press (1,
  2007).

\bibitem{Lehmann:1954rq}
H.~Lehmann, K.~Symanzik and W.~Zimmermann, \emph{{On the formulation of
  quantized field theories}},
  \href{https://doi.org/10.1007/BF02731765}{\emph{Nuovo Cim.} {\bfseries 1}
  (1955) 205}.

\bibitem{Henning:2022xlj}
B.~Henning, H.~Murayama, F.~Riva, J.O.~Thompson and M.T.~Walters,
  \emph{{Towards a nonperturbative construction of the S-matrix}},
  \href{https://doi.org/10.1007/JHEP05(2023)197}{\emph{JHEP} {\bfseries 05}
  (2023) 197} [\href{https://arxiv.org/abs/2209.14306}{{\ttfamily
  2209.14306}}].

\bibitem{Lehmann:1957zz}
H.~Lehmann, K.~Symanzik and W.~Zimmermann, \emph{{On the formulation of
  quantized field theories. II}},
  \href{https://doi.org/10.1007/BF02832508}{\emph{Nuovo Cim.} {\bfseries 6}
  (1957) 319}.

\bibitem{Bros:1964iho}
J.~Bros, H.~Epstein and V.J.~Glaser, \emph{{Some rigorous analyticity
  properties of the four-point function in momentum space}},
  \href{https://doi.org/10.1007/BF02733596}{\emph{Nuovo Cim.} {\bfseries 31}
  (1964) 1265}.

\bibitem{Sommer:1970mr}
G.~Sommer, \emph{{Present state of rigorous analytic properties of scattering
  amplitudes}}, \href{https://doi.org/10.1002/prop.19700181102}{\emph{Fortsch.
  Phys.} {\bfseries 18} (1970) 577}.

\bibitem{Bros:1972jh}
J.~Bros, V.~Glaser and H.~Epstein, \emph{{Local analyticity properties of the n
  particle scattering amplitude}},
  \href{https://doi.org/10.5169/seals-114374}{\emph{Helv. Phys. Acta}
  {\bfseries 45} (1972) 149}.

\bibitem{Steinmann1960a}
O.~Steinmann, \emph{{{\"U}ber den Zusammenhang zwischen den Wightmanfunktionen
  und den retardierten Kommutatoren}},
  \href{https://doi.org/10.5169/seals-113076}{\emph{Helv. Phys. Acta}
  {\bfseries 33} (1960) 257}.

\bibitem{Steinmann1960b}
O.~Steinmann, \emph{{Wightman-Funktionen und retardierte Kommutatoren. II}},
  \href{https://doi.org/10.5169/seals-113079}{\emph{Helv. Phys. Acta}
  {\bfseries 33} (1960) 347}.

\bibitem{Bartels:2008ce}
J.~Bartels, L.N.~Lipatov and A.~Sabio~Vera, \emph{{BFKL Pomeron, Reggeized
  gluons and Bern-Dixon-Smirnov amplitudes}},
  \href{https://doi.org/10.1103/PhysRevD.80.045002}{\emph{Phys. Rev. D}
  {\bfseries 80} (2009) 045002}
  [\href{https://arxiv.org/abs/0802.2065}{{\ttfamily 0802.2065}}].

\bibitem{Caron-Huot:2016owq}
S.~Caron-Huot, L.J.~Dixon, A.~McLeod and M.~von Hippel, \emph{{Bootstrapping a
  Five-Loop Amplitude Using Steinmann Relations}},
  \href{https://doi.org/10.1103/PhysRevLett.117.241601}{\emph{Phys. Rev. Lett.}
  {\bfseries 117} (2016) 241601}
  [\href{https://arxiv.org/abs/1609.00669}{{\ttfamily 1609.00669}}].

\bibitem{Bartels:2014jya}
J.~Bartels, A.~Kormilitzin and L.N.~Lipatov, \emph{{Analytic structure of the
  $n=7$ scattering amplitude in $\mathcal{N}=4$ theory in multi-Regge
  kinematics: Conformal Regge cut contribution}},
  \href{https://doi.org/10.1103/PhysRevD.91.045005}{\emph{Phys. Rev. D}
  {\bfseries 91} (2015) 045005}
  [\href{https://arxiv.org/abs/1411.2294}{{\ttfamily 1411.2294}}].

\bibitem{Brower:2008nm}
R.C.~Brower, H.~Nastase, H.J.~Schnitzer and C.-I.~Tan, \emph{{Implications of
  multi-Regge limits for the Bern-Dixon-Smirnov conjecture}},
  \href{https://doi.org/10.1016/j.nuclphysb.2009.02.009}{\emph{Nucl. Phys. B}
  {\bfseries 814} (2009) 293}
  [\href{https://arxiv.org/abs/0801.3891}{{\ttfamily 0801.3891}}].

\bibitem{Chou:1984es}
K.-c.~Chou, Z.-b.~Su, B.-l.~Hao and L.~Yu, \emph{{Equilibrium and
  Nonequilibrium Formalisms Made Unified}},
  \href{https://doi.org/10.1016/0370-1573(85)90136-X}{\emph{Phys. Rept.}
  {\bfseries 118} (1985) 1}.

\bibitem{PolkinhorneDisp}
J.C.~Polkinhorne, \emph{General dispersion relations},
  \href{https://doi.org/10.1007/BF02745445}{\emph{Il Nuovo Cimento (1955-1965)}
  {\bfseries 4} (1956) 216}.

\bibitem{tHooft:1973wag}
G.~'t~Hooft and M.J.G.~Veltman, \emph{{Diagrammar}},
  \href{https://doi.org/10.1007/978-1-4684-2826-1_5}{\emph{NATO Sci. Ser. B}
  {\bfseries 4} (1974) 177}.

\bibitem{Damgaard:2023vnx}
P.H.~Damgaard, E.R.~Hansen, L.~Plant\'e and P.~Vanhove, \emph{{The Relation
  Between KMOC and Worldline Formalisms for Classical Gravity}},
  \href{https://arxiv.org/abs/2306.11454}{{\ttfamily 2306.11454}}.

\bibitem{Mueller:2002gd}
A.H.~Mueller and D.T.~Son, \emph{{On the Equivalence between the Boltzmann
  equation and classical field theory at large occupation numbers}},
  \href{https://doi.org/10.1016/j.physletb.2003.12.047}{\emph{Phys. Lett. B}
  {\bfseries 582} (2004) 279}
  [\href{https://arxiv.org/abs/hep-ph/0212198}{{\ttfamily hep-ph/0212198}}].

\bibitem{Caron-Huot:2010fvq}
S.~Caron-Huot, \emph{{Loops and trees}},
  \href{https://doi.org/10.1007/JHEP05(2011)080}{\emph{JHEP} {\bfseries 05}
  (2011) 080} [\href{https://arxiv.org/abs/1007.3224}{{\ttfamily 1007.3224}}].

\bibitem{Damgaard:2021ipf}
P.H.~Damgaard, L.~Plante and P.~Vanhove, \emph{{On an exponential
  representation of the gravitational S-matrix}},
  \href{https://doi.org/10.1007/JHEP11(2021)213}{\emph{JHEP} {\bfseries 11}
  (2021) 213} [\href{https://arxiv.org/abs/2107.12891}{{\ttfamily
  2107.12891}}].

\bibitem{Damgaard:2023ttc}
P.H.~Damgaard, E.R.~Hansen, L.~Plant\'e and P.~Vanhove, \emph{{Classical
  Observables from the Exponential Representation of the Gravitational
  S-Matrix}},  \href{https://arxiv.org/abs/2307.04746}{{\ttfamily 2307.04746}}.

\bibitem{Bern:2019nnu}
Z.~Bern, C.~Cheung, R.~Roiban, C.-H.~Shen, M.P.~Solon and M.~Zeng,
  \emph{{Scattering Amplitudes and the Conservative Hamiltonian for Binary
  Systems at Third Post-Minkowskian Order}},
  \href{https://doi.org/10.1103/PhysRevLett.122.201603}{\emph{Phys. Rev. Lett.}
  {\bfseries 122} (2019) 201603}
  [\href{https://arxiv.org/abs/1901.04424}{{\ttfamily 1901.04424}}].

\bibitem{Damour:2020tta}
T.~Damour, \emph{{Radiative contribution to classical gravitational scattering
  at the third order in $G$}},
  \href{https://doi.org/10.1103/PhysRevD.102.124008}{\emph{Phys. Rev. D}
  {\bfseries 102} (2020) 124008}
  [\href{https://arxiv.org/abs/2010.01641}{{\ttfamily 2010.01641}}].

\bibitem{DiVecchia:2021bdo}
P.~Di~Vecchia, C.~Heissenberg, R.~Russo and G.~Veneziano, \emph{{The eikonal
  approach to gravitational scattering and radiation at $ \mathcal{O}
  $(G$^{3}$)}}, \href{https://doi.org/10.1007/JHEP07(2021)169}{\emph{JHEP}
  {\bfseries 07} (2021) 169}
  [\href{https://arxiv.org/abs/2104.03256}{{\ttfamily 2104.03256}}].

\bibitem{Bini:2021gat}
D.~Bini, T.~Damour and A.~Geralico, \emph{{Radiative contributions to
  gravitational scattering}},
  \href{https://doi.org/10.1103/PhysRevD.104.084031}{\emph{Phys. Rev. D}
  {\bfseries 104} (2021) 084031}
  [\href{https://arxiv.org/abs/2107.08896}{{\ttfamily 2107.08896}}].

\bibitem{Bern:2021yeh}
Z.~Bern, J.~Parra-Martinez, R.~Roiban, M.S.~Ruf, C.-H.~Shen, M.P.~Solon et~al.,
  \emph{{Scattering Amplitudes, the Tail Effect, and Conservative Binary
  Dynamics at O(G4)}},
  \href{https://doi.org/10.1103/PhysRevLett.128.161103}{\emph{Phys. Rev. Lett.}
  {\bfseries 128} (2022) 161103}
  [\href{https://arxiv.org/abs/2112.10750}{{\ttfamily 2112.10750}}].

\bibitem{Manohar:2022dea}
A.V.~Manohar, A.K.~Ridgway and C.-H.~Shen, \emph{{Radiated Angular Momentum and
  Dissipative Effects in Classical Scattering}},
  \href{https://doi.org/10.1103/PhysRevLett.129.121601}{\emph{Phys. Rev. Lett.}
  {\bfseries 129} (2022) 121601}
  [\href{https://arxiv.org/abs/2203.04283}{{\ttfamily 2203.04283}}].

\bibitem{Streater:1989vi}
R.F.~Streater and A.S.~Wightman, \emph{{PCT, spin and statistics, and all
  that}} (1989).

\bibitem{Witten:2018zxz}
E.~Witten, \emph{{APS Medal for Exceptional Achievement in Research: Invited
  article on entanglement properties of quantum field theory}},
  \href{https://doi.org/10.1103/RevModPhys.90.045003}{\emph{Rev. Mod. Phys.}
  {\bfseries 90} (2018) 045003}
  [\href{https://arxiv.org/abs/1803.04993}{{\ttfamily 1803.04993}}].

\bibitem{Herderschee:2023fxh}
A.~Herderschee, R.~Roiban and F.~Teng, \emph{{The Sub-Leading Scattering
  Waveform from Amplitudes}},
  \href{https://arxiv.org/abs/2303.06112v2}{{\ttfamily 2303.06112v2}}.

\bibitem{Brandhuber:2023hhy}
A.~Brandhuber, G.R.~Brown, G.~Chen, S.~De~Angelis, J.~Gowdy and G.~Travaglini,
  \emph{{One-loop Gravitational Bremsstrahlung and Waveforms from a Heavy-Mass
  Effective Field Theory}},
  \href{https://arxiv.org/abs/2303.06111v3}{{\ttfamily 2303.06111v3}}.

\bibitem{Elkhidir:2023dco}
A.~Elkhidir, D.~O'Connell, M.~Sergola and I.A.~Vazquez-Holm, \emph{{Radiation
  and Reaction at One Loop}},
  \href{https://arxiv.org/abs/2303.06211v1}{{\ttfamily 2303.06211v1}}.

\bibitem{Georgoudis:2023lgf}
A.~Georgoudis, C.~Heissenberg and I.~Vazquez-Holm, \emph{{Inelastic
  Exponentiation and Classical Gravitational Scattering at One Loop}},
  \href{https://arxiv.org/abs/2303.07006v1}{{\ttfamily 2303.07006v1}}.

\bibitem{Beneke:1997zp}
M.~Beneke and V.A.~Smirnov, \emph{{Asymptotic expansion of Feynman integrals
  near threshold}},
  \href{https://doi.org/10.1016/S0550-3213(98)00138-2}{\emph{Nucl. Phys. B}
  {\bfseries 522} (1998) 321}
  [\href{https://arxiv.org/abs/hep-ph/9711391}{{\ttfamily hep-ph/9711391}}].

\bibitem{Jantzen:2011nz}
B.~Jantzen, \emph{{Foundation and generalization of the expansion by regions}},
  \href{https://doi.org/10.1007/JHEP12(2011)076}{\emph{JHEP} {\bfseries 12}
  (2011) 076} [\href{https://arxiv.org/abs/1111.2589}{{\ttfamily 1111.2589}}].

\bibitem{Jantzen:2012mw}
B.~Jantzen, A.V.~Smirnov and V.A.~Smirnov, \emph{{Expansion by regions:
  revealing potential and Glauber regions automatically}},
  \href{https://doi.org/10.1140/epjc/s10052-012-2139-2}{\emph{Eur. Phys. J. C}
  {\bfseries 72} (2012) 2139}
  [\href{https://arxiv.org/abs/1206.0546}{{\ttfamily 1206.0546}}].

\bibitem{Heinrich:2021dbf}
G.~Heinrich, S.~Jahn, S.P.~Jones, M.~Kerner, F.~Langer, V.~Magerya et~al.,
  \emph{{Expansion by regions with pySecDec}},
  \href{https://doi.org/10.1016/j.cpc.2021.108267}{\emph{Comput. Phys. Commun.}
  {\bfseries 273} (2022) 108267}
  [\href{https://arxiv.org/abs/2108.10807}{{\ttfamily 2108.10807}}].

\bibitem{Parra-Martinez:2020dzs}
J.~Parra-Martinez, M.S.~Ruf and M.~Zeng, \emph{{Extremal black hole scattering
  at $\mathcal{O}(G^3)$: graviton dominance, eikonal exponentiation, and
  differential equations}},
  \href{https://doi.org/10.1007/JHEP11(2020)023}{\emph{JHEP} {\bfseries 11}
  (2020) 023} [\href{https://arxiv.org/abs/2005.04236}{{\ttfamily
  2005.04236}}].

\bibitem{Brandhuber:2021bsf}
A.~Brandhuber, G.~Chen, H.~Johansson, G.~Travaglini and C.~Wen,
  \emph{{Kinematic Hopf Algebra for Bern-Carrasco-Johansson Numerators in
  Heavy-Mass Effective Field Theory and Yang-Mills Theory}},
  \href{https://doi.org/10.1103/PhysRevLett.128.121601}{\emph{Phys. Rev. Lett.}
  {\bfseries 128} (2022) 121601}
  [\href{https://arxiv.org/abs/2111.15649}{{\ttfamily 2111.15649}}].

\bibitem{Brandhuber:2021eyq}
A.~Brandhuber, G.~Chen, G.~Travaglini and C.~Wen, \emph{{Classical
  gravitational scattering from a gauge-invariant double copy}},
  \href{https://doi.org/10.1007/JHEP10(2021)118}{\emph{JHEP} {\bfseries 10}
  (2021) 118} [\href{https://arxiv.org/abs/2108.04216}{{\ttfamily
  2108.04216}}].

\bibitem{Henn:2013pwa}
J.M.~Henn, \emph{{Multiloop integrals in dimensional regularization made
  simple}}, \href{https://doi.org/10.1103/PhysRevLett.110.251601}{\emph{Phys.
  Rev. Lett.} {\bfseries 110} (2013) 251601}
  [\href{https://arxiv.org/abs/1304.1806}{{\ttfamily 1304.1806}}].

\bibitem{Smirnov:2019qkx}
A.V.~Smirnov and F.S.~Chuharev, \emph{{FIRE6: Feynman Integral REduction with
  Modular Arithmetic}},
  \href{https://doi.org/10.1016/j.cpc.2019.106877}{\emph{Comput. Phys. Commun.}
  {\bfseries 247} (2020) 106877}
  [\href{https://arxiv.org/abs/1901.07808}{{\ttfamily 1901.07808}}].

\bibitem{Klappert:2020nbg}
J.~Klappert, F.~Lange, P.~Maierh\"ofer and J.~Usovitsch, \emph{{Integral
  reduction with Kira 2.0 and finite field methods}},
  \href{https://doi.org/10.1016/j.cpc.2021.108024}{\emph{Comput. Phys. Commun.}
  {\bfseries 266} (2021) 108024}
  [\href{https://arxiv.org/abs/2008.06494}{{\ttfamily 2008.06494}}].

\bibitem{Herrmann:2021lqe}
E.~Herrmann, J.~Parra-Martinez, M.S.~Ruf and M.~Zeng, \emph{{Gravitational
  Bremsstrahlung from Reverse Unitarity}},
  \href{https://doi.org/10.1103/PhysRevLett.126.201602}{\emph{Phys. Rev. Lett.}
  {\bfseries 126} (2021) 201602}
  [\href{https://arxiv.org/abs/2101.07255}{{\ttfamily 2101.07255}}].

\bibitem{Goldberger:2009qd}
W.D.~Goldberger and A.~Ross, \emph{{Gravitational radiative corrections from
  effective field theory}},
  \href{https://doi.org/10.1103/PhysRevD.81.124015}{\emph{Phys. Rev. D}
  {\bfseries 81} (2010) 124015}
  [\href{https://arxiv.org/abs/0912.4254}{{\ttfamily 0912.4254}}].

\bibitem{Porto:2012as}
R.A.~Porto, A.~Ross and I.Z.~Rothstein, \emph{{Spin induced multipole moments
  for the gravitational wave amplitude from binary inspirals to 2.5
  Post-Newtonian order}},
  \href{https://doi.org/10.1088/1475-7516/2012/09/028}{\emph{JCAP} {\bfseries
  09} (2012) 028} [\href{https://arxiv.org/abs/1203.2962}{{\ttfamily
  1203.2962}}].

\bibitem{Herrmann:2021tct}
E.~Herrmann, J.~Parra-Martinez, M.S.~Ruf and M.~Zeng, \emph{{Radiative
  classical gravitational observables at $ \mathcal{O} $(G$^{3}$) from
  scattering amplitudes}},
  \href{https://doi.org/10.1007/JHEP10(2021)148}{\emph{JHEP} {\bfseries 10}
  (2021) 148} [\href{https://arxiv.org/abs/2104.03957}{{\ttfamily
  2104.03957}}].

\bibitem{Hannesdottir:2022bmo}
H.S.~Hannesdottir and S.~Mizera, \emph{{What is the i\ensuremath{\varepsilon}
  for the S-matrix?}}, SpringerBriefs in Physics, Springer (1, 2023),
  \href{https://doi.org/10.1007/978-3-031-18258-7}{10.1007/978-3-031-18258-7},
  [\href{https://arxiv.org/abs/2204.02988}{{\ttfamily 2204.02988}}].

\bibitem{Akhoury:2011kq}
R.~Akhoury, R.~Saotome and G.~Sterman, \emph{{Collinear and Soft Divergences in
  Perturbative Quantum Gravity}},
  \href{https://doi.org/10.1103/PhysRevD.84.104040}{\emph{Phys. Rev. D}
  {\bfseries 84} (2011) 104040}
  [\href{https://arxiv.org/abs/1109.0270}{{\ttfamily 1109.0270}}].

\bibitem{Melville:2013qca}
S.~Melville, S.G.~Naculich, H.J.~Schnitzer and C.D.~White, \emph{{Wilson line
  approach to gravity in the high energy limit}},
  \href{https://doi.org/10.1103/PhysRevD.89.025009}{\emph{Phys. Rev. D}
  {\bfseries 89} (2014) 025009}
  [\href{https://arxiv.org/abs/1306.6019}{{\ttfamily 1306.6019}}].

\end{thebibliography}\endgroup

\end{document}